\documentclass[11pt]{article}
\usepackage[utf8]{inputenc} 

\usepackage{geometry} 
\usepackage{graphics}
\usepackage{graphicx}
\usepackage{amssymb}
\usepackage{array}
\usepackage{fancyhdr}
\usepackage{subfigure}
\usepackage{graphicx}
\usepackage{mathrsfs}
\usepackage{slashed}
\usepackage{stmaryrd}
\usepackage{epsfig}
\usepackage{dcolumn}
\usepackage{bm}
\usepackage{amsmath}
\usepackage{wrapfig}
\usepackage{makeidx}
\usepackage{wasysym}
 \usepackage{relsize}
\usepackage{la}
\usepackage[utf8]{inputenc}

\usepackage{array}
\usepackage{fancyhdr}
\usepackage{graphicx}
\usepackage{mathrsfs}
\usepackage{slashed}
\usepackage{stmaryrd}
\usepackage{simplewick}
\allowdisplaybreaks[4]

\begin{document}
%
\title{\bf  Field Theory in the Imaginary-time Formulation}
\author{Yi-Cheng Huang\\
E-mail: {\tt ychuang1109@msn.com}
}                     
%
%
\date{}
\maketitle

%
%
{\bf Abstract}
\vskip.2cm

{
The imaginary-time formulation is investigated in the field theory. The Dirac and photon's Lagrangians are treated in the imaginary-time and space and are invariant under scale transformations, which are motivated by the Tolman-Ehrenfest relation and Wilson's approach in the renormalization group. With the functional approaches, the propagators of leptons and photons can be derived both for the real- and the imaginary-time. The theory is free of UV divergence in the integration range of the imaginary-time, $(0,\beta\,]$, which is required by the KMS condition. The results of one-loop radiative corrections are proved to be consistent with those in QED. The renormalization group equations with respect to vacuum's temperature give no Landau pole and show identical renormalization coefficients as those from the $\overline{MS}$ scheme of the renormalization in the limit of zero temperature.
%
%
\newpage
\section{Introduction}
\label{intro}
A field theory of  imaginary-time and space based on the assumption that vacuum is a thermodynamical system is constructed. It is figured to be filled with off-shell particles, and may communicate with physical ones as being the so-called virtual particles. The macroscopic observable of the system, the temperature, is introduced  through the imaginary-time, and the theory is built according to Matsubara's formalism \cite{matsubara} for either a fermionic or a bosonic  many-particle system. The quantizations of the field operators are achieved in terms of the Matsubara frequency, instead of the energy, and 3-momentum. The propagators are derived for the real-time and the imaginary-time from the same partition functions through the path integral approach \cite{shankar93}; 
the loop corrections of QED according to the corresponding Feynman rules are calculated respectively for the real-time and the imaginary-time. The  propagators of the real-time are obtained by summing over all of the Matsubara frequencies, $\omega_n$, then the analytical continuation is performed from the imaginary-time to its real axis. They become the traditional propagators in the limit of zero temperature, which is parametrized by the variable $\beta$ $(=\frac{1}{k_{\rm B}T})$, as it gets close to infinity. This is because that the additional factors in the two-point correlation functions besides the conventional components are the density functions of  fermions and bosons, $1\mp n_{\rm F,B}(\xi_{\bf p})$, where $n_{\rm F,B}(\xi_{\bf p})=\frac{1}{e^{\beta\xi_{\bf p}}\pm 1}$ and $\xi_{\bf p}$ is the energy carried by the particle. 
If the temperature of the hypothetical vacuum is comparable to that of the microwave background radiations (CMB) \cite{ryden}, which is around a few Kelvins, the factor $\beta$ is about $10^4 \,\, ( {\rm eV}^{-1} )$, or equivalently $10^{16} \,\,(  {\rm TeV}^{-1} )$ so that the exponential function is vastly large in the denominator and the density function becomes unity. As particles are created in the modern accelerators, this condition could be easily fulfilled. Although it appears that the density functions violate the Lorentz invariance, for a very tiny temperature it would, on the contrary, explains why the Lorentz invariance holds in the many-particle field theory. For a typical loop integral in this formalism, the calculations compose of a traditional Feynman integral containing the density functions and the residues of the poles from the density functions. 
The loop corrections of the real-time, for the real part of the physical observables, such as self-energies, etc., give  negligible contribution, therefore they are free of the ultraviolet divergences (UV). Beyond the threshold of the ingoing momenta, the same imaginary part as in the field theory is generated from the branch cut.\par
The loop calculations for the imaginary-time, $\tau$, are similar to the traditional ways; only the zeroth-component of the four-momentum for a particle is replaced by $i\omega_n$, where $\omega_n$ is equal to $\frac{2n\pi}{\beta}$ or $\frac{(2n+1)\pi}{\beta}$ for a boson or a fermion. Instead of integrating $\tau$ from 0 to $\beta$, the lower bound of the imaginary-time is replaced by an infinitesimal $\beta_0$ in the imaginary-time evolution operator for the perturbation theory. The loop integrals at $\beta_0$ provide a reference point for radiative corrections, and the UV divergences for the imaginary-time could be removed automatically 
without introducing any counter term to absorb them. The radiative corrections derived from QED, such as the anomalous magnetic dipole moment, $g-2$, of the electron are consistent with the results in this formalism. Moreover, the renormalization group equations can be also derived as functions of $\beta$, which plays the role of the renormalization scale $\mu$ in field theory. They are consistent with the results of those in the $\overline{MS}$ scheme of renormalization \cite{grozin}. One thing related to this idea is the Tolman-Ehrenfest relation \cite{tolman30}, and from the studies in the general relativity it implies that the physical time is proportional to the proper time by the formula $t=\beta\tau_p$, where the ratio $\beta$ is often regarded as the "speed of time", and $t$ is then called the thermal time \cite{rovelli93}. Both of them imply that the temperature of the vacuum could determine the physical scale. It will be shown that they are related to the sacle transformations applied in the imaginary-time theory. On the other hand, there have been many efforts in studying the many-particle  theory of the relativistic quantum fields \cite{landsman87}. Some of the approaches toward a QED or a QCD plasma \cite{bellac96} are similar to those presented here; instead of studying an on-shell many-particle system,  some of the differences from their works are to treat the vacuum as a macroscopic system, which only off-shell particles are filling in, and to seek regenerating the known results in loop computations in the conventional field theories. In the work of \cite{huang13b} that follows this one, the relations between the imaginary-time hamiltonian with various vacuum effects are discussed, such as  the Casimir effect \cite{casimir} and the van der Waals forces \cite{vanderwaals}. For the two effects, the thermal theory of vacuum not only generates consistent results with the conventional calculations but also yields the cutoff functions to automatically regulate the divergences, while in the precedent approaches the regularization functions are added intentionally. Other effects, like Unruh effect \cite{unruh} and the Hawking radiation \cite{hawking}, also exhibit many agreements. In short, the proposed thermal vacuum provides a solid thermal bath for the uniformly accelerated observer in the Unruh's thought experiment to observe the black-body radiation. Meanwhile, in the general relativity, one theoretical source of the black-body radiation is the black hole, which establishes a unique environment, the event horizon, for the virtual photons to radiate and make the black hole evaporate in a very slow pace, and agreements between the imaginary-time formalism and the viewpoint from Hawking's approach are explained in the same paper. From the last two effects, they correspond an acceleration or a surface gravity, $g$, to an effective temperature, $T=\frac{\hbar g}{2\pi c k_B}$. To estimate the temperature of the vacuum, it is about $4\times 10^{-20}K$ for the surface gravity on earth or $\sim6\times10^{-8}K$ for a black hole of a sun's size. This supports  the assumption that the propagators of the particles deduced from the imaginary-time formalism in an infinitesimal temperature would become the conventional ones, and thus secure the Lorentz invariance.  To extend and show the usefulness of the imaginary-time field theory, the cosmological constant can be derived through the approaches of the DeWitt-Schwinger representation \cite{dewitt75} and the Casimir effect in ref. \cite{huang13c}. One of the unique features for the cosmological constant, the ratio $w=-1$, in the equation of state, $p=w \rho$, can be obtained without any trouble from the divergence. In a recent article \cite{huang13d}, an application on the quantization of the weak gravitational field is discussed.
\par
In the next section, the free and the interaction Lagrangian of electrons and photons in the imaginary-time and space are discussed, as well as their relations to the scale invariance. The interaction range of the imaginary-time in the $S$-matrix is given a nonzero lower bound, which leads to the cancellations of the UV divergences. The Green functions of both kinds of particles are derived for the real-time and the imaginary-time in Section \ref{greenfunction}. In the following section, the radiative corrections, such as the self-energy, are performed and the comparisons with the results from the field theory will be checked. Then the renormalization group equations with respective to the variation of the vacuum temperature are presented in Section \ref{rgeq}. In the end, a conclusion will be given. In the appendices, some details of the calculations are provided for reader's convenience.

\section{Lagrangian in imaginary-time}

\label{lagimag}

\subsection{Fermion Lagrangian}\label{fermionL}
Here we may start with the path integral approach for fermions, and what in the following is basically generalized from the derivations in ref. \cite{shankar93}.  Let's consider a  partition function over an imaginary-time variable, ${{\tau}}$,
\begin{eqnarray*}
Z&\equiv&{\rm Tr}\,e^{-\beta \mathcal{K}}=\prod_{\{{\tau}\}}{\rm Tr}\, e^{-\beta({\tau}) \mathcal{\mathcal{K}}({\bf \tau})\Delta {\tau }},
\end{eqnarray*}
where $\mathcal{\mathcal{K}}({\tau})$ $(=\hat{H}-\mu_{\rm ch} \hat{N})$ is a normal ordered operator, $\mathcal{K}(\psi^\dagger({\tau}),\psi({\tau}))$. The operator $\hat{H}$ and $\hat{N}$ are the Hamiltonian and the number of the particle, and $\mu_{\rm ch}$ is the chemical potential. A notation, which is used throughout the paper, is the boldface that indicates a 3-dimensional vector, such as the position vector, ${\bf x}$, and 3-momentum, ${\bf p}$. The fermion fields, which are generalized to four dimensions of the imaginary-time and space, $\psi(\tau, \bf x)$ and  $\psi^\dagger (\tau, \bf x)$, are the so-called grassmann numbers in the path integral formalism. For the first step,  the exponential function is divided into products of infinitesimal changes with respect to the variation of the imaginary time, $\tau$. After summing over all of the functional changes of the fields,  we may obtain from the Hamiltonian density of Dirac particles, $ \mathcal{K}(\tau, {\bf x})=\psi^\dagger(\tau, {\bf x})(-i\gamma^0\vec{\gamma}\cdot   \vec{\nabla}+m_{\rm f}\gamma^0-\mu_{\rm f})\psi(\tau, {\bf x})$:
\begin{eqnarray}
Z&=&\int e^{\int^\beta_0 d\tau\int d^3{\bf x} \psi^\dagger(\tau,{\bf x})\left(-\frac{\partial}{\partial \tau}+i\gamma^0\vec{\gamma}\cdot\vec{\nabla}-m_{\rm f}\gamma^0+\mu_{\rm f} \right)\psi(\tau,\bf{x})}[d\psi^\dagger(\tau,{\bf x}) d\psi(\tau,\bf{x})],\label{partitionZ2}
\end{eqnarray}
where $\mu_{\rm f}$ is the chemical potential of fermions. The notation for a vector, $\vec{v}$, means a 3 dimensional vector. In Appendix \ref{appA}, the details of the derivation are provided. Here introduce a rescaling factor $e^{\mu_{\rm f}\tau}$ for a transformation of the field operator, $\psi_r(\tau,\vec{\bf x})=e^{\mu_{\rm f}\tau}\psi(\tau,\vec{\bf x})$. With the inclusion of this factor, the reference point of the energy of a fermion is shifted to the Fermi surface, since the term of the chemical potential $\mu_{\rm f}$ is removed from the new Lagrangian. 
From above, the rescaled imaginary-time, space and mass are
\begin{eqnarray*}
\tau_r=\frac{3}{2\mu_{\rm f}}\left(e^{\frac{2}{3}\mu_{\rm f}\tau}-1\right),\hspace{.3cm}
\vec{\bf x}_r= e^{\frac{2}{3}\mu_{\rm f}\tau}\vec{\bf x}, \hspace{.3cm}
m_r= e^{-\frac{2}{3}\mu_{\rm f}\tau}m\hspace{.3cm}{\rm and}\hspace{.3cm}
\beta_r=\frac{3}{2\mu_{\rm f}}\left(e^{\frac{2}{3}\mu_{\rm f}\beta}-1\right),
\end{eqnarray*}
so that the new partition function becomes
\begin{eqnarray}
Z&=&\int e^{\int^{\beta}_0 d\tau_r \int d^3{\bf x}_ r\psi^\dagger_r(\tau,{\bf x})\left(-\frac{\partial}{\partial \tau_r}+i\gamma^0\vec{\gamma}\cdot\vec{\nabla}_r-m_r\gamma^0\right)\psi_r(\tau,\bf{x})}[d\psi^\dagger_r(\tau,{\bf x}) d\psi_r(\tau,\bf{x})].\label{partitionF}
\end{eqnarray}
The subscript, $r$, will be dropped hereafter without causing any ambiguity. In the following, besides fermion's propagator of the imaginary-time is derived from this partition function, that of the real-time can also be obtained from it through the summation of the Matsubara frequency and the analytic continuation from the imaginary-time, $\tau$, to the real-time, $t$.


\subsection{Interaction Lagrangian in QED} \label{IntQED}
The QED Lagrangian in the imaginary-time and space, $(\tau, \vec{\bf x})$, can be identified from the partition function, eq. (\ref{partitionF}). After replaced with the covariant derivative, it becomes 
\begin{eqnarray*}
\mathcal{L}_{\rm Dirac}+\mathcal{L}_{\rm int}&=&
\left.\bar{\psi} (i \slashed{D}_\tau-m )\psi
,\right.
\end{eqnarray*}
where the covariant derivative is ${D}^\mu_\tau= \left(\frac{\partial }{i\partial \tau},{\vec{\nabla}}\right)+ie {A}^\mu(\tau,\vec{\bf x}).$ The Lagrangian of the imaginary-time is invariant  under the following gauge transformations 
\begin{eqnarray*}
&&\hspace{1.7cm}\psi(\tau, \vec{\bf x})\rightarrow e^{i\Lambda(\tau, \vec{\bf x})} \psi(\tau, \vec{\bf x}),\\
&&A_0\rightarrow A_0+\frac{i}{e}\partial_{\tau}\Lambda(\tau,\vec{\bf x}),\,\,\,
\vec{A}\rightarrow \vec{A}-\frac{1}{e}\vec{\nabla}\Lambda(\tau,\vec{\bf x}),
\end{eqnarray*}
where $A_0$ is the time component of the vector potential and $\vec{A}$ is for the spatial dimensions.
So the action of the interaction Lagrangian, $\mathcal{S}_{\rm int}$, in the generating function is 
\begin{eqnarray}
i\mathcal{S}_{\rm int}&=&i\int d(i\tau) d^3{\bf x}\,\mathcal{L}_{\rm int}(\tau, \vec{\bf x})=e\int d\tau d^3{\bf x}\,
\bar{\psi} \slashed{{A}}\psi. \label{Linteraction}
\end{eqnarray}
The only difference from the usual interaction action is an extra imaginary number $i$, therefore as 
we apply the corresponding Feynman rules, the corresponding factor for each vertex is $e\gamma^\mu$, instead of $-ie\gamma^\mu$.

\subsection{Scale invariance and thermal time}\label{scaleinv}
The scale invariance \cite{difrancesco97} is one of the important features in diverse fields of science. In statistical mechanics, it is used to 
study phase transitions, and is found that near the critical point the fluctuations happen at all length scales \cite{goldenfeld92}. In the study under the imaginary-time and space, the same feature can also been found. The Lagrangian $\mathcal{L}_{\rm Dirac}$ in eq. (\ref{partitionF})  for fermions is 
\begin{eqnarray}
\mathcal{L}_{\rm Dirac}(\psi,\psi^\dagger,\tau,{\bf x})&=&\psi^\dagger(\tau,{\bf x})\left(-\frac{\partial}{\partial \tau}+i\gamma^0\vec{\gamma}\cdot\vec{\nabla}-m_{\rm f}\gamma^0\right)\psi(\tau,\bf{x}).\label{Lfermion}
\end{eqnarray}
 As for the photons the corresponding Lagrangian, $\mathcal{L}_{\rm Maxwell}(A_\mu, \tau,\vec{\bf x})$, including the term of the chemical potential, $\mu_\gamma$, could be obtained by replacing the time, $t$, with the imaginary-time, $-i\tau$; combined with the gauge fixing term, $\mathcal{L}_{\rm fix}$, they are:
\begin{eqnarray}
&&\hspace{-0.5cm}\mathcal{L}_{\rm Maxwell}+\mathcal{L}_{\rm fix}
=-\frac{1}{4}F_{\mu\nu}F^{\mu\nu}+\mu_\gamma A_\mu A^\mu-\frac{\zeta}{2} (\partial_\mu A^\mu)^2\nonumber\\ \label{Lphoton}
\hspace{-1cm}&=&\frac{1}{2} A_\nu\left(-\frac{\partial^2}{\partial \tau^2}+\partial_i\partial^i \right)A^\nu+\frac{\zeta-1}{2}\left(-A_0 \frac{\partial^2}{\partial\tau^2} A^0+A_i \partial^i\partial_k A^k\right) +\mu_\gamma A_\mu A^\mu,\nonumber\\
\end{eqnarray}
where the indices, 0, and $i\,\,(=1,2,3)$ refer to the components of the imaginary-time variable, $\tau$, and 3-dimensional space, ${\bf x}$. The factor $\zeta$ is the gauge parameter. The integration by parts has been used to derive from the first to the second line in eq. (\ref{Lphoton}). 
eq. (\ref{Lfermion}) and (\ref{Lphoton}) are invariant under the transformations
\begin{eqnarray}
\psi(\tau,\bf{x})&\rightarrow& 
s^{3/2}\psi_{0}(\tau,{\bf x}),\hspace{.5cm} \tau\rightarrow
s^{-1}\tau_0,\hspace{.5cm}  {\bf x}\rightarrow 
s^{-1}{\bf x}_0 \nonumber\\
A^\mu(\tau,\bf{x})&\rightarrow&
s A^\mu_{0}(\tau,{\bf x}),\hspace{.5cm}\mu_\gamma\rightarrow s^2\, \mu_{\gamma,0} \hspace{.5cm} {\rm and}\,\,\,\, m\rightarrow s\, m_0, \label{scaletrans}
\end{eqnarray}
where $s$ is a scale factor.  In the representation of the momentum space, $(\omega_n,{\bf p})$, in expansions like eq. (\ref{psix}) and (\ref{photonex}), the transformation are 
\begin{eqnarray*}
\psi(\omega_n,\bf{p})&\rightarrow&
s^{{5}/{2}}\psi_{0}(\omega_n,{\bf p}),\hspace{.5cm}\omega_n\rightarrow
s\,\omega_{0,n},\\
A^\mu(\tau,\bf{x})&\rightarrow&
s^3\hspace{.2cm}A^\mu_{0}(\tau_r,{\bf x}),\hspace{.9cm}
{\bf p}\rightarrow
s\,{\bf p}_0.
\end{eqnarray*} 
Here a temperature dependent cutoff may be introduced, the maximal number of the Matsubara frequency, $N_{\rm max}$, is related to it by:  
\begin{eqnarray} 
\omega_{\rm cutoff}=\frac{2\pi N_{\rm max}}{\beta},\label{cutoff1}
\end{eqnarray} 
The number $N_{\rm max}$ is a constant and is only constrained by the total number of particles in the system. Obviously, it is meaningless that the maximal number of mode is larger than the total number of the particles in a many-particle system. This assumption will be useful as the UV divergences are considered in Section \ref{oneloop}. 
 The cutoff frequencies between two different temperatures implies a scale factor,
\begin{eqnarray} 
s=\frac{\beta_0}{\beta},\label{sfactor}
\end{eqnarray} 
with the assumption that $N_{\rm max}$ is the same regardless of different temperatures.  As for the loop integrals in Section \ref{oneloop}, the 3-momentum integration and  frequency summation are performed separately. In the field theory, no matter what regularization of divergence is used, the cutoffs for each of four dimensions of  the momentum are equal quantities. The same fashion is adopted in the following calculations, so, in a similar way, we may define a cutoff for the 3-momentum phase space
\begin{eqnarray}
\Lambda&=&\frac{2\pi N_{\rm max}}{\beta}.\label{cutoff2}
\end{eqnarray} 
The change of the scale leads to the theory of the renormalization group; this is known from the efforts made by Kadanoff \cite{kadanoff66} in 1966, Wilson \cite{wilson75}  et al. in 1975\,. An example proposed is the spins in a solid; the renormalization group describes the couplings' variation as observed in different sizes of blocks. The effective Lagrangian may be obtained after the scale transformation and integrating out the momentum phase space between different cutoffs, $\Lambda$ and $s\Lambda$. The coupling constants and masses are re-defined in the new scale. Another example is the Hawking radiation \cite{hawking}, for the field theory in the curved space-time, the ground states are defined separately near the event horizon and in the distance, the match of the conformally invariant wave functions  for incident rays and outgoing rays leads to the discovery of the Hawking temperature.  In the formalism presented in this paper, the Lagrangian densities possesses the same features as those in their works; in addition, the cutoffs, in eq. (\ref{cutoff1}) and (\ref{cutoff2}), obey the transformation law, $\Lambda=s \Lambda_0$, without the need of integration between two scales. As a result, the action in the partition function is scale invariant. Another important concept developed over decades is the thermal time \cite{rovelli93}, especially in the discussions of the possibilities for a quantum gravity theory. The connection between the thermodynamics and the general relativity has been discussed with intense literatures. It may be started from the Tomita-Takesaki theorem \cite{takesaki70} to derive a time flow from a generic thermal physical state, and the Unruh effect and  the Hawking radiation are shown to relate to this idea.  The thermal time, $t$, is related to the geometrical time, $\tau_p$, or say the proper time, by a simple formula: $t=\beta\tau_p$. This agrees with the scale transformations for the imaginary-time in eq. (\ref{scaletrans}) for different thermal times. As early as the 1930s, it was figured in a stationary spacetime with a time-like Killing vector field $\xi$, a temperature of the vacuum satisfies the Tolman-Ehrenfest relation \cite{tolman30}
  \begin{eqnarray}
T||\xi||=const.\,,
\end{eqnarray}
where $||\xi||=\sqrt{g_{ab}\xi^a\xi^b}$ is the norm of $\xi$. A thermal equilibrium point of view was adopted regarding the gravity, and in the Newtonian limit, the gravitational field was related to the gravitational field by 
  \begin{eqnarray}
\frac{\nabla T}{T}=\frac{\vec{g}}{c^2},
\end{eqnarray}
where $\vec{g}$ is the strength of the gravitational field. As for the spatial dimensions, in Section 5 the temperature is shown to play the same role to vary the scale of the dimensions as the renormalization scale $\mu$ in renormalization.  The Tolman-Ehrenfest relation, the Hawking radiation and the result of the renormalization group from the imaginary-time theory all imply that the temperature of  vacuum determines the physical scales. In conclusion, the transformations from eq. (\ref{scaletrans}) and (\ref{sfactor}) can be regarded as those between coordinates of flat space-time to hold the Lagrangian invariant under scale changes.
More discussions are given in Section \ref{oneloop}.

\begin{figure}[t]
\begin{center}
   \subfigure[]{\includegraphics[width=5cm]{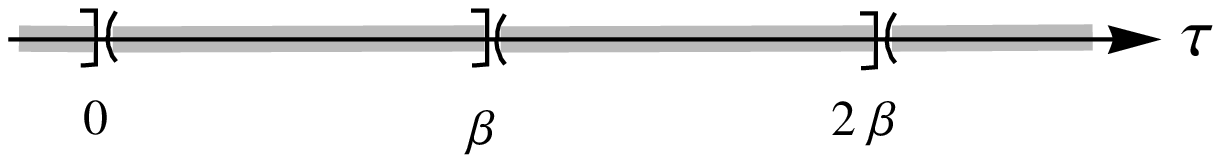}}
   \hspace*{0.07\textwidth}
   \subfigure[]{\includegraphics[width=5cm]{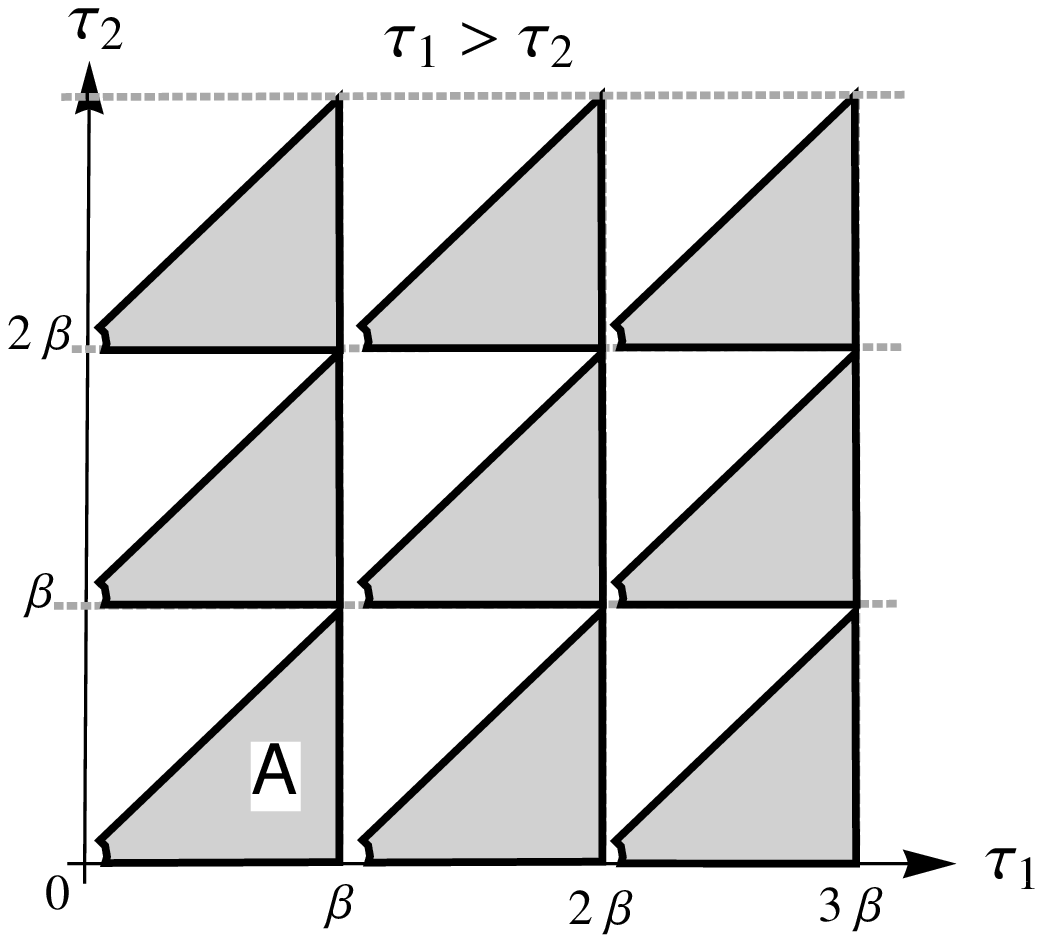}}
   \hspace*{0.0\textwidth}
\caption{\small  From the KMS condition, the Green function is cyclic with respect to the imaginary-time $\tau$. Therefore, for $n=1$, in eq. (\ref{perturbative}) the intergation domain of $\tau$ is an open-closed interval $(0,\beta]$ in order to avoid the overlap of the periodic domains at points $-\beta$, 0, $\beta$, $2\beta$...\, as shown in (a). Similarly, for a specific time order $\tau_1>\tau_2$ in the case of $n=2$, 
the periodic domains in the plane of $\tau_1$-$\tau_2$, the origin has to be taken out from the gray triangular domain A to prevent the confliction.  The same reason and result can also apply on and be obtained for the cases of the other time order $\tau_2>\tau_1$ and the higher dimensions of $\tau$, $n>2$. 
  } 
  \label{Fig:Cyctau}
\end{center}
\end{figure}

\subsection{Perturbation theory for imaginary-time}\label{perturbation}
Similar to the perturbation for the real-time, the $S$-matrix is the imaginary-time evolution operator from 0 to $\beta$:
\begin{eqnarray*}
\langle {\bf p}_1,\cdots,{\bf p}_n | \,S\, | {\bf k}_1,\cdots,{\bf k}_m \rangle=\langle {\bf p}_1,\cdots,{\bf p}_n | \,T_\tau\,e^{-\int^\beta_0 d\tau H_{int}(\tau)}\, | {\bf k}_1,\cdots,{\bf k}_m \rangle,
\end{eqnarray*}
where $H_{int}(\tau)=\int d^3{\bf x} \,\mathcal{H}_{int}(x)$ and $T_\tau$ is the operator of the imaginary-time ordering. According to the KMS condition \cite{kms}, the Green functions of imaginary-time are cyclic in the interval from zero to $\beta$. As illustrated in Figure \ref{Fig:Cyctau} (a), the domain has to be an open-closed interval. Therefore we have to slightly modify the above definition by replacing the lower bound of the imaginary-time, zero, with $\beta_0$:  
\begin{eqnarray*}
\lim_{\beta_0\rightarrow 0^+}\hspace{.3cm}\langle {\bf p}_1,\cdots,{\bf p}_n | \,T_\tau \,e^{-\int^\beta_{\beta_0} d\tau H_{int}(\tau)}\, | {\bf k}_1,\cdots,{\bf k}_m \rangle,
\end{eqnarray*}
\begin{figure}[t]
\begin{center}
{\includegraphics[height=5cm,width=5cm]{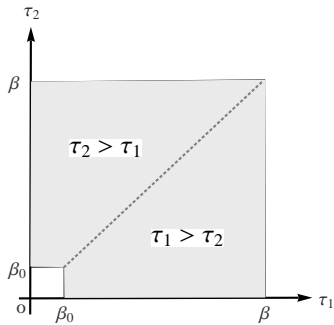}}
   \hspace*{0.0\textwidth}
\caption{\small After combined with the contributions from two different  time orders, $\tau_1>\tau_2$ and $\tau_2>\tau_1$, as shown in Figure \ref{Fig:Cyctau} (b), the integration domain for two imaginary-time variables, $\tau_1$ and $\tau_2$, is illustrated as above. The gray area is the new integration domain with the infinitesimal white square removed from the multi-dimensional integration of $\tau$. }
  \label{Fig:taudomain}
\end{center}
\end{figure}
where $\beta_0$ is an infinitesimal positive number. It may not appear to make a difference from the above expression. As we will find out later, the infinitesimal shift of the lower integration bound of $\tau$ gives a new reference point for the radiative corrections, and it automatically removes the UV divergences without the need to introduce any counter term. The expansion of the imaginary-time evolution operator
can be written as
\begin{eqnarray}\label{newIT}
&&\hspace{-1cm}\lim_{\beta_0\rightarrow 0^+} e^{-\int^\beta_{\beta_0} d\tau H_{int}(\tau)}= \lim_{\beta_0\rightarrow 0^+} \sum_{n=0}^\infty \frac{(-1)^n}{n!} \int^\beta_{\beta_0} d\tau_1\dots d\tau_n T_\tau\left\{ H_{int} (\tau_1)\dots  H_{int} (\tau_n)\right\}\nonumber\\
&&\hspace{-1.0cm}= \sum_{n=0}^\infty \frac{(-1)^n}{n!} \left(\int^\beta_{0} d\tau_1\dots d\tau_n T_\tau\left\{ H_{int} (\tau_1)\dots  H_{int} (\tau_n)\right\}\right.\nonumber\\
&&
\left.\hspace{2.5cm}-\lim_{\beta_0\rightarrow 0^+}  \int^{\beta_0}_0 d\tau_1\dots d\tau_n T_\tau\left\{ H_{int} (\tau_1)\dots  H_{int} (\tau_n)\right\}\right).\label{perturbative}
\end{eqnarray}
As for the multi-dimensional integration of the imaginary-time in the above expression, a 2-dimensional case is illustrated in Figure \ref{Fig:Cyctau} (b), a small neighborhood around the origin in the $\tau_1$-$\tau_2$ plane has to be removed from the integration domain.  From the new definition, the radiative corrections computed from this imaginary-time evolution operator have to subtract the contributions of $\tau$ from 0 to $\beta_0$. In the two dimensional case,  after combined with all possible time orders, as shown in Figure \ref{Fig:taudomain}, a small white square, which is the area from 0 to $\beta_0$ for $\tau_1$ and $\tau_2$, is excluded from the integration domain. Even though $\beta_0$ is infinitely close to zero, the contributions are divergent if the integral of $\beta$ also has UV divergence. In this definition, the UV divergences are canceled spontaneously along with the consideration of the consistency in scale, which will be depicted in Section \ref{phi3}. In Section \ref{oneloop}, all of the radiative corrections that are computed for the imaginary-time have to be subtracted from the contributions from 0 to $\beta_0$, as shown as the second term in the parenthesis of eq. (\ref{newIT}). The integral of $\beta_0$ plays a similar role to a counter term as in the renormalization, and the computed radiative corrections that are shifted to the reference point at $\beta_0$ will be called renormalized radiative corrections throughout the paper for convenience and also to emphasize their correspondences in the field theory. 

\subsubsection{Examples from  $\phi^3$- and $\phi^4$-theory}
\label{phi3}
Here to provide some examples to manifestly explain how the UV divergences are canceled by a closed-open domain of the imaginary-time $\tau$, $(0,\beta]$, where the lower bound will be denoted as $\beta_0(=0^+)$. The action of the interaction hamiltonian of the $\phi^3$-theory is 
\begin{eqnarray*}
S_I=\int^\beta_0d\tau \int d^3{\bf x}\,H_I(\tau, {\bf x}),\,\,{\rm where}\,\,H_I(\tau, {\bf x})=\frac{\lambda}{3!}\, \phi^3(\tau,{\bf x}),
\end{eqnarray*}
where $\lambda$ is the coupling constant. The coupling $\lambda$ has dimension [mass], and it is a function of $\beta$ because of the scale invariance of the action, namely $\lambda(\beta)=\frac{\beta_1}{\beta}\lambda_1$, where $\lambda_1$ is the coupling constant at another temperature $\beta_1$.  They are assumed to be massless particles here. Consider the correlation function 
\begin{eqnarray*}
\langle 0|T_\tau\{ \phi(x)\phi(y)e^{ \int^\beta_0d\tau \int d^3{\bf x}\,H_I(\tau, {\bf x})}\}|0\rangle
\end{eqnarray*}
to its one-loop level 
\begin{eqnarray*}
\langle 0|T_\tau\left\{ \phi(x)\phi(y){ \int^\beta_{\beta_0}d\tau_{z_1} \int d^3{\bf z}_1\,H_I(\tau_{z_1}, {\bf z}_1)}
{ \int^\beta_{\beta_0}d\tau_{z_2} \int d^3{\bf z}_2\,H_I(\tau_{z_2}, {\bf z}_2)}\right\}|0\rangle.
\end{eqnarray*}
According to the difinition given in eq. (\ref{newIT}), the above expression can be separated into two counterparts
\begin{eqnarray}
&&\hspace{-0.8cm}=\langle 0|T_\tau\left\{ \phi(x)\phi(y){ \int^\beta_{0}d\tau_{z_1} \int d^3{\bf z}_1\,H_I(\tau_{z_1}, {\bf z}_1)}
{ \int^\beta_{0}d\tau_{z_2} \int d^3{\bf z}_2\,H_I(\tau_{z_2}, {\bf z}_2)}\right\}|0\rangle\nonumber\\
&&\hspace{9cm}-\{\beta\rightarrow\beta_0\}.\hspace{.8cm}
\label{ham1}
\end{eqnarray}
The first term is well known, and the second one is supposed to play the role of counter term. As $\beta_0\rightarrow 0^+$, it would be interesting to see what happens to the second one, especially two of the Green functions are expressed in different scales, or say different temperatures  $\beta$ and $\beta_0$. For example, the coordinates $x$ and $z_{0,1}$ in $\acontraction[0.5ex]{}{\phi}{(x)}{\phi} \phi(x)\phi_0(z_{0,1})$ are of different scales, where the subscript-0  corresponds to the temperature parameter $\beta_0$ for the the coordinate variables or fields. In order to compare the difference of the first and the second term in eq. (\ref{ham1}),
we may perform the scale transformation on the coordinate variables and the field operators by 
\begin{eqnarray*}
\tau_{z_{0,i}}=\frac{\beta_0}{\beta}\tau_{z_i},\,\, {\bf z}_{0,i}=\frac{\beta_0}{\beta} {\bf z}_i,\,\,{\rm and}\,\,\phi_0 (z_{0,i})=\frac{\beta}{\beta_0}\phi (z_i),\,\,{\rm where}\,\, i=1,2.
\end{eqnarray*}
We may consider the self-energy diagram, $\raisebox{-3.mm}{\psfig{figure=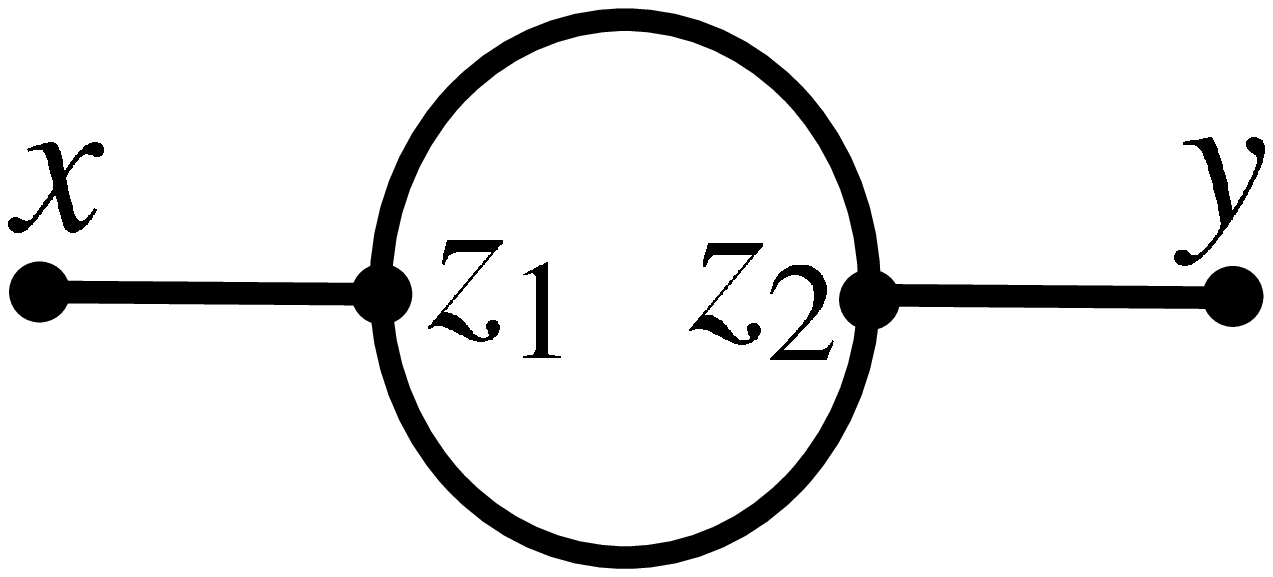, width=2cm,height=0.8cm}}$, for the second term in eq. (\ref{ham1}) as below. Since we don't know what is the Green function of different scales, like $\acontraction[0.5ex]{}{\phi}{(x)}{\phi}
\phi(x)\phi_0(z_{0,i})$,  scale transformations for the external field operators from $\phi(x)\,\,\left(=\frac{\beta_0}{\beta}\phi_0(x_0)\right)$ and $\phi(y)\,\,\left(=\frac{\beta_0}{\beta}\phi_0(y_0)\right)$ and  needed. Thus, 
\begin{eqnarray*}
&&\hspace{-2cm}\lambda_0^2\int^{\beta_0}_{0}d\tau_{z_{0,1}}  d^3{\bf z}_{0,1}\, \int^{\beta_0}_{0}d\tau_{z_{0,2}}  d^3{\bf z}_{0,2}\,\langle 0|\left\{
\acontraction[1ex]{}{\phi}{(x)}{\phi}
\phi(x)\phi_0(z_{0,1})
\acontraction[1ex]{}{\phi}{_0(z_{0,1})}{\phi}
\phi_0(z_{0,1})\phi_0(z_{0,2})
\acontraction[1ex]{}{\phi}{_0(z_{0,2})}{\phi}
\phi_0(z_{0,2})\phi_0(z_{0,1})
\acontraction[1ex]{}{\phi}{_0(z_{0,2})}{\phi}
\phi_0(z_{0,2})\phi(y)
\right\}|0\rangle,\\
&&\hspace{-1cm}=\lambda_0^2 \left(\frac{\beta_0}{\beta}\right)^2\frac{1}{\beta_0}\sum_n\int d^3{\bf p}_0\frac{1}{p_{0,n}^2}\left(\frac{1}{\beta_0}\sum_m\int d^3{\bf k}_0 \frac{1}{k^2_{0,m}}\frac{1}{(k_{0,m}+p_{0,n})^2}\right)\frac{1}{p_{0,n}^2}e^{-ip_{0,n}\cdot (x_0-y_0)},\\
&&\hspace{-1cm}=\frac{1}{\beta}\sum_n\int d^3{\bf p}\frac{1}{p_{n}^2}\left(\frac{\lambda^2}{\beta_0}\sum_m\int d^3{\bf k}_0 \frac{1}{k^2_{0,m}}\frac{1}{(k_{0,m}+p_{0,n})^2}\right)\frac{1}{p_{n}^2}e^{-ip_{n}\cdot (x-y)},
\end{eqnarray*}
where $p_n=(i\omega_n,{\bf p})$ and  $k_{0,n}=(i\omega_{0,m},{\bf k}_0)$.
The factor $\left(\frac{\beta_0}{\beta}\right)^2$ in the second line is from the scale transformations of $\phi (x)$ and $\phi (y)$, and in the third line the scale transformations on $p_{0,n}$ and $\lambda_0=\frac{\beta}{\beta_0}\lambda$ are performed. The loop integral now is in the parenthesis of the third line. We may explain the idea of cancellation with some examples. Consider the imaginary-time Feynman integral without  coupling constants,
\begin{eqnarray*}
I_1(\Delta)&=&  
\frac{1}{\beta}\sum_{n=1}^{N_{\rm max}}\frac{1}{\sqrt{\big(\frac{ n}{\beta}\big)^2+\Delta}}\,\,\left(\simeq \int^\Lambda_0 dx \frac{1}{\sqrt{x^2+\Delta}},\,\,{\rm for}\,\,\beta\gg 1 \right),\\
I_2(\Delta)&\equiv&\lim_{\beta_0\rightarrow0}\frac{1}{\beta_0}\sum_{n=1}^{N_{\rm max}}\frac{1}{\sqrt{\big(\frac{n}{\beta_0}\big)^2+\Delta_0}},
\end{eqnarray*}
where the continuous variable $x\,(=n/\beta)$ has a dimension of [mass], $\Delta$ is the function of the external momenta or masses and $\Lambda=\frac{N_{\rm max}}{\beta}$. To simplify the expressions, the factor of $(2\pi)$ in the Matsubara frequency is ignored. The above sums are from the loop integral, $\frac{1}{\beta}\sum_n^{N_{\rm max}}\int\frac {d^3{\bf k}}{(k^2_n+\Delta)}$, after integrating over the 3-momentum. As $\beta\gg 1$, $I_1$ becomes an integral over $x$. The term $\Delta_0$ in $I_2$ is negligible after taking the limit of $\beta_0\rightarrow 0$. In this case, there are $\beta_0$ coming  from outside the loop integral and they happen to be canceled out, but we should pay attention that this definition just include $\beta_0$ from inside the loop integral.  We obtain
\begin{eqnarray*}
\lambda^2 I_1(\Delta)&=&\lambda^2\log\left(\mathsmaller{\frac{\Lambda+\sqrt{\Lambda^2+\Delta^2}}{\sqrt{\Delta}}}\right)\simeq
\lambda^2\log\Lambda+...\,,\\
\lambda^2  I_2(\Delta_0)&=&\lambda^2\sum_{n=1}^{N_{\rm max}}\frac{1}{n}=\lambda^2(\log N_{\rm max}+\gamma_{\rm E}).
\end{eqnarray*}
After substituting $\Lambda=\frac{N_{\rm max}}{\beta}$ into $I_1$, it is obviously that the terms with $N_{\rm max}$ are the same in $I_1$ and $I_2$. Thus a clean cancellation, $\lambda^2 I_1(\Delta)-\lambda^2 I_2(\Delta)$, can be fulfilled. As a double check for the integrals without UV divergence, consider an example in the vertex diagram, $\raisebox{-3.mm}{\psfig{figure=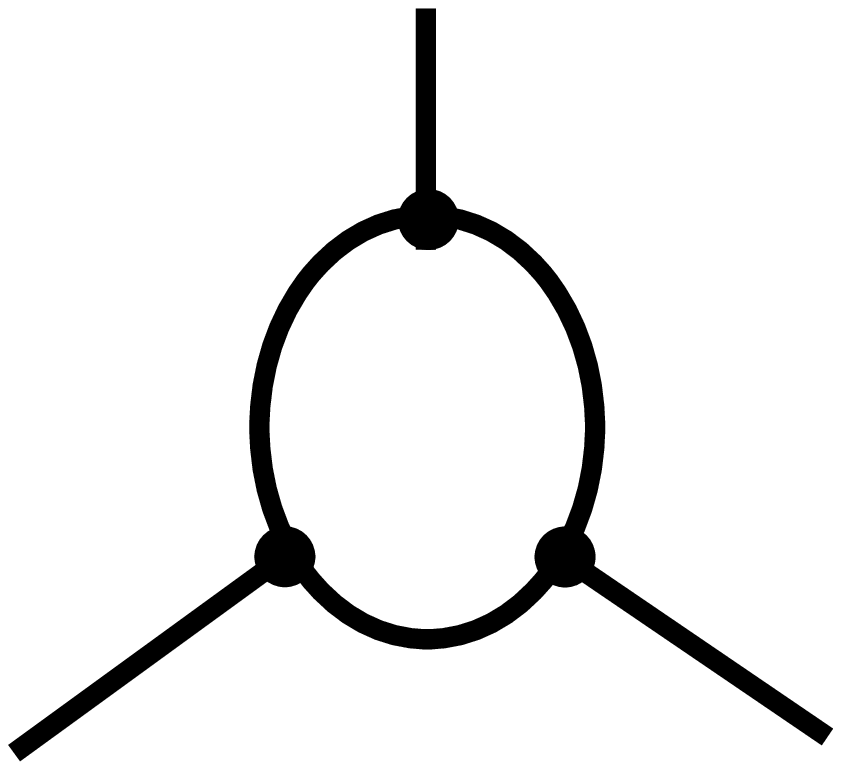, width=2cm,height=1cm}}$, similar to eq. (\ref{vertexqed}):
\begin{eqnarray*}
I_3(\Delta)&=&
\frac{1}{\beta}\sum_{n=1}^{N_{\rm max}}\frac{1}{\left(\big(\frac{n}{\beta}\big)^2+\Delta^2\right)^{\frac{3}{2}}}\,\,\left(\simeq \int^\Lambda_0 dx \frac{1}{({x^2+\Delta^2})^{\frac{3}{2}}} ,\,\,{\rm for}\,\,\beta\gg 1 \right),\\
I_4(\Delta_0)&=&\lim_{\beta_0\rightarrow 0}\frac{1}{\beta_0}\sum_{n=1}^{N_{\rm max}}\frac{1}{\left(\big(\frac{n}{\beta_0}\big)^2+\Delta_0^2\right)^{\frac{3}{2}}}=0.
\end{eqnarray*}
For $\beta_0$ inside the loop and with its limit to zero, we will have $I_4$ as a reference point for $I_3$, and $I_4(\Delta_0)= \lim_{\beta_0\rightarrow 0} \sum\frac{\beta_0^2}{n^3}= 0$. The triangular Feynman integral   is then
$\lambda^2 I_3(\Delta)$, as the same as the traditional integral.  In the summations of the Matsubara frequency in $I_2$ and $I_4$, the limiting process of $\beta_0$ inside the loop is treated as a definition.  As we may remember from the conventional calculations of the Casimir effect \cite{casimir}, the difference of the continuous and discrete potential functions is calculated. The discrete potential is due to the discrete mode number  of the electromagnetic standing waves between the two plates, as the momentum in the normal direction of the plates is $\frac{2\pi n}{L}$, where $L$ is a small distance, similar to the case of $\beta_0\rightarrow 0$ in the Matsubara frequency $\frac{2\pi n}{\beta_0}$.  This could be an analogy for  such an  assumption and condition. Therefore, it could be tricky to give those reference integrals certain kinds of conditions, but it is also important to see if it is universal to all perturbative theories.\par
Let's also see an example from $\phi^4$-theory. The corresponding interaction Lagrangian is $\mathcal{L}_I=\frac{\lambda}{4!}\phi^4$, where the coupling constant $\lambda$ is dimensionless. For a self-energy diagram, $\raisebox{-3.mm}{\psfig{figure=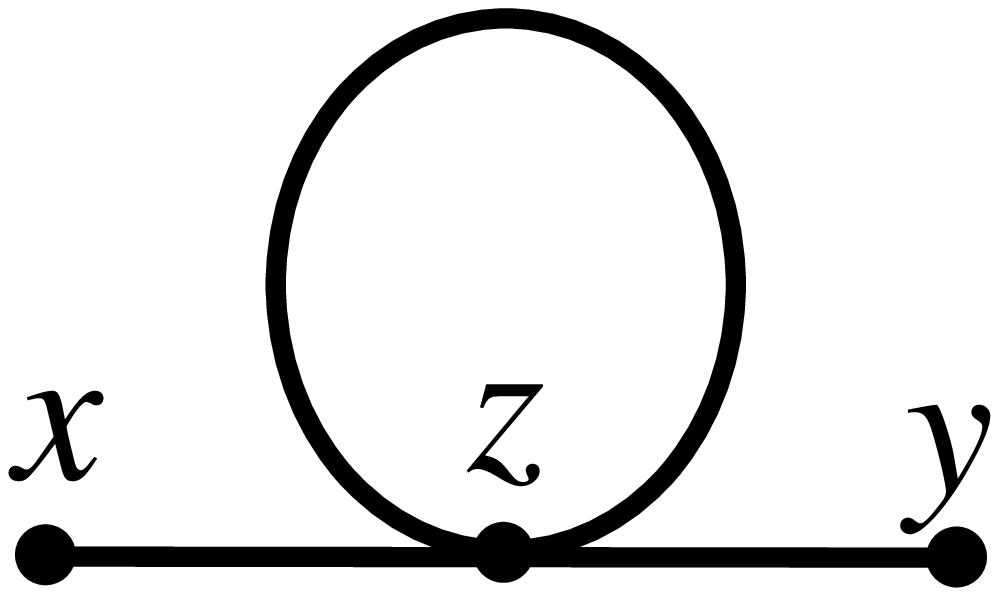, width=1.6cm,height=1cm}}$, similarly perform the scale transformation on $\phi(x)$ and $\phi(y)$ first, we have 
\begin{eqnarray*}
&&\hspace{-0cm}\lambda\int^{\beta_0}_{0}d\tau_{z_{0}}  d^3{\bf z}_{0}\,\,\langle 0|\left\{
\acontraction[1ex]{}{\phi}{(x)}{\phi}
\phi(x)\phi_0(z_{0})
\acontraction[1ex]{}{\phi}{_0(z_{0})}{\phi}
\phi_0(z_{0})\phi_0(z_{0})
\acontraction[1ex]{}{\phi}{_0(z_{0})}{\phi}
\phi_0(z_{0})\phi(y)
\right\}|0\rangle,\\
&&\hspace{-0cm}=\lambda \left(\frac{\beta_0}{\beta}\right)^2\frac{1}{\beta_0}\sum_n\int d^3{\bf p}_0\frac{1}{p_{0,n}^2}\left(\frac{1}{\beta_0}\sum_m\int d^3{\bf k}_0 \frac{1}{k^2_{0,m}}\right)\frac{1}{p_{0,n}^2}e^{-ip_{0,n}\cdot (x_0-y_0)},\\
&&\hspace{-0cm}=\left(\frac{\beta_0}{\beta}\right)^2\frac{1}{\beta}\sum_n\int d^3{\bf p}\frac{1}{p_{n}^2}\left\{\frac{\lambda}{\beta_0}\sum_m\int d^3{\bf k}_0 \frac{1}{k^2_{0,m}}\right\}\frac{1}{p_{n}^2}e^{-ip_{n}\cdot (x-y)}.
\end{eqnarray*}
The part in the curly bracket is the imaginary-time Feynman integral for $\beta_0$. We may notice that there is an extra factor 
$\big(\frac{\beta_0}{\beta}\big)^2$ from outside the loop, and it has to be taken into account in the result. 
From the above two examples. we may generalize the above the derivation by considering 
\begin{eqnarray*}
\langle 0|\phi(x_1)...\phi(x_E)\prod_{i=1}^V \int d^dz_iH_I(z_i)|0\rangle,
\end{eqnarray*}
where $E$ is the number of the external legs and $V$ is the number of the vertices. The extra factors of $\big(\frac{\beta}{\beta_0}\big)$ are given by
\begin{eqnarray*}
\left(\frac{\beta}{\beta_0}\right)^{V\cdot \lambda_d} \cdot \left(\frac{\beta_0}{\beta}\right)^{E} \cdot
\left(\frac{\beta}{\beta_0}\right)^{4E-4} \cdot \left(\frac{\beta}{\beta_0}\right)^{-2E},
\end{eqnarray*}
where $\lambda_d$ is the mass dimension of the coupling constant.
The first is from all of the coupling constants, and the second is from the scale transformation of $E$ external field operators $\phi(x)=\frac{\beta_0}{\beta}\phi_0(x_{0})$. The third is due to the integrations of external momentum space and $-4$ in the exponent is from the $\delta$-function to ensure  momentum and frequency conservations; the external propagators give  the last ratio factor. To conclude, the extra factor that has to be taken into account for $\beta_0$ is  
\begin{eqnarray*}
\left(\frac{\beta}{\beta_0}\right)^{V\cdot \lambda_d+E-4}.
\end{eqnarray*}
Thus, for the diagram of $\phi^3$-theory   $\,\raisebox{-3.mm}{\psfig{figure=selfephi3a.eps, width=2cm,height=0.8cm}}$, we have $V=2$, $\lambda_d=1$ and $E=2$, the  factor is one. For the one $\raisebox{-3.mm}{\psfig{figure=vertexphi3.eps, width=2cm,height=1cm}}$, $V=3$, $\lambda_d=1$ and $E=3$, and the factor is $\big(\frac{\beta}{\beta_0}\big)^2$. As for the diagram of $\phi^4$-theory, $\raisebox{-3.mm}{\psfig{figure=selfephi4.eps, width=1.6cm,height=1cm}}$, $V=1$, $\lambda_d=0$ and $E=2$, the factor is $\big(\frac{\beta_0}{\beta}\big)^2$.


\section{Green's function}
\label{greenfunction}
\subsection{Propagator of fermions}
\subsubsection{Real-time}
In terms of the real-time, it is obtained according to an analytic continuation from the imaginary-time $\tau$, after the Matsubara frequencies are summed. As for the quantization of the fields, the creation and annihilation operators  $a^s_{\omega_n,{{\bf p}}}$, $b^s_{\omega_n,{{\bf p}}}$ are quantized for the respective Matsubara frequency, $\omega_n$, and 3-momentum ${\bf p}$, and they will be treated as the grassmann numbers latter in the functional formalism. 
The field operators are expanded by the Fourier expansion and transform for the respective discrete and continuous phase spaces  as 
\begin{eqnarray}
\psi(\tau,x)=\frac{1}{\beta}\sum_{n}\int\frac{d^3{\bf p}}{(2\pi)^3}\,\psi(\omega_n,{\bf \vec{p}})e^{-i\omega_n\tau+i\vec{\bf p}\cdot \vec{\bf x}},\nonumber\\
{\rm and }\hspace{.5cm}\psi^\dagger(\tau,x)=\frac{1}{\beta}\sum_{n}\int\frac{d^3{\bf p}}{(2\pi)^3}\,\psi^\dagger(\omega_n,{\bf \vec{p}})e^{i\omega_n\tau-i\vec{\bf p}\cdot \vec{\bf x}},\label{psix}
\end{eqnarray}
where in the momentum phase space: 
\begin{eqnarray}
\psi(\omega_n,{\bf p})=\frac{1}{\sqrt{2\xi_{\bf p}}}\sum_s\left(a^s_{\omega_n,{\bf{p}}}u^s({\bf p})+{b^s}^\dagger_{-\omega_n,\bf -p}{{v}^s}({\bf -p})\right),\nonumber\\
{\rm and }\hspace{.5cm}\psi^\dagger(\omega_n,{\bf p})=\frac{1}{\sqrt{2\xi_{\bf p}}}\sum_s\left(b^s_{-\omega_n,\bf -p}{{v}^s}^\dagger({\bf -p})+{a^s}^\dagger_{\omega_n,{\bf{p}}}{u^s}^\dagger({\bf p})\right).
\label{psip}
\end{eqnarray}
The superscript $s$ indicates the spin state of the spinor $u^s({\bf p})$ or $v^s(-{\bf p})$. The spinors satisfy the Dirac equation, such as $(\slashed{p}-m_{\rm f})u({\bf p})=0$ and so on. 
In the traditional way, the factor $1/\sqrt{2\xi_{\bf p}}$ is inserted  in the expansions, eq. (\ref{psix}), to ensure the Lorentz Invariance. In fact, even without this factor, it can be shown that the Lorentz invariance is still hold for the propagators in this Matsubara frequency expansion and the resultant 2-point correlation function in eq. (\ref{diracprop}), will not be modified. The action, $\mathcal{A}$, from the partition function $Z$ is  
\begin{eqnarray*}
\mathcal{A}=\int^{\beta}_0 d\tau\int d^3 \vec{\bf x}\,\,\mathcal{L}_{\rm Dirac}(\tau, \vec{\bf x_r})
= \frac{1}{\beta}\sum_{\omega_n}\int d^3 \vec{\bf p}\,\,\mathcal{L}_{\rm Dirac}(\omega_n, \vec{\bf p}),
\end{eqnarray*}
where the Lagrangians in  the respective representations are  
\begin{eqnarray}
\mathcal{L}_{\rm Dirac}(\tau, \vec{\bf x})&=&\psi^\dagger(\tau,{\bf x})\left(-\frac{\partial}{\partial \tau}+i\gamma^0\vec{\gamma}\cdot\vec{\nabla}-m_{\rm f}\gamma^0\right)\psi(\tau,\bf{x}),\nonumber\\
\mathcal{L}_{\rm Dirac}(\omega_n, \vec{\bf p})&=&\psi^\dagger(\omega_n,{\bf p})\left(i\omega_n-\gamma^0\vec{\gamma}\cdot\vec{\bf p}-m_{\rm f}\gamma^0\right)\psi(\omega_n,\bf{p}),\label{Lagrangian}
\\
\hspace{-2.cm}&=&\hspace{-.3cm}\sum_{s}\left.\left(i\omega_n-\xi_{\bf p}\right){a^s}^\dagger_{\omega_n,{\bf{p}}}a^s_{\omega_n,{\bf{p}}}\right.
-\left.\sum_{s}\left(i\omega_n+\xi_{\bf p}\right)b^s_{\omega_n,\bf p}{b^s}^\dagger_{\omega_n,\bf p}\right. .\label{Lagrangian2}
\end{eqnarray}
The lines from eq.  (\ref{Lagrangian}) to (\ref{Lagrangian2}) are derived by using the Dirac equation, and the factor, $2\xi_{\bf p}$, coming from the spinor products,  ${u^r}^\dagger({\bf p})u^s({\bf p})$ and ${v^s}^\dagger({\bf p})v^s({\bf p})$, are canceled by those in eq.  (\ref{psip}).
The correlation function for two different imaginary-time and space points, $(\tau_x,{\bf x})$ and $(\tau_y,{\bf y})$, is related to the one for the Matsubara frequency and  3-momentum as follows
\begin{eqnarray*}
&&\langle\bar{\psi}(\tau_x,\vec{\bf x}){\psi}(\tau_y,\vec{\bf y})\rangle
=\frac{1}{\beta^2}\sum_{n,m}\int\frac{d^3{\bf p}}{(2\pi)^3}\frac{d^3{\bf k}}{(2\pi)^3}
\langle \bar{\psi}(\omega_n,{\bf p})\psi(\omega_m,{\bf k})\rangle
e^{-i\omega_n\tau_x+i\omega_m\tau_y+i{\bf p}\cdot {\bf x}-i{\bf k}\cdot {\bf y}}.
\end{eqnarray*}
As the creation and annihilation operators are regarded as grassmann numbers, the partition function is treated as a functional of them; the correlation function for the Matsubara frequency and  3-momentum can be obtained by inserting eq. (\ref{psip}) into:
\begin{eqnarray*}
&&\langle {\psi}(\omega_n,{\bf p})\bar{\psi}(\omega_m,{\bf k})\rangle \\
&=&
-\sum_{s}u^s({\bf p})\bar{u}^s({\bf p})\frac{\beta\delta_{mn}(2\pi)^3\delta^3({\bf p-k})}{2\xi_{\bf p}(i\omega_n-\xi_{\bf p})}-\sum_{s}v^s({\bf -p})\bar{v}^s({\bf -p})\frac{\beta\delta_{mn}(2\pi)^3\delta^3({\bf p-k})}{2\xi_{\bf p}(i\omega_n+\xi_{\bf p})}.
\end{eqnarray*}
As for the case of $\tau_x>\tau_y$, we may obtain the retarded propagator by summing over the Matsubara frequency with the help of eq. (\ref{sumfermi}), and choose the semicircle below the real axis of $p_0$ for the contour. This is achievable as the analytic continuation of the variable $\tau= it$ is applied. The retarded fermion propagator is derived from 
\begin{eqnarray}
&&\hspace{-1.5cm}\langle{\psi}(\tau_x,\vec{\bf x})\bar{\psi}(\tau_y,\vec{\bf y})\rangle_{\rm Ret}
=\int_{\otimes}\frac{d^4 p}{(2\pi)^4}\frac{i}{\slashed{p}-m_{\rm f}}e^{-p_0(\tau_x-\tau_y)+i{\bf p}\cdot ({\bf x-y})}(1-n_{\rm F}(p_0)),\label{diracpropR}
\end{eqnarray}
where we have used the relations 
\begin{eqnarray*}
\sum_s u^s({\bf p})\bar{u}^s({\bf p})&=&\xi_{\bf p}\gamma_0-\vec{\bf p}\cdot \vec{\gamma}+m_{\rm f} =\slashed{p}+m_{\rm f},\\
\sum_s v^s({\bf p})\bar{v}^s({\bf p})&=&\xi_{\bf p}\gamma_0-\vec{\bf p}\cdot \vec{\gamma}-m_{\rm f}  =\slashed{p}-m_{\rm f},
\end{eqnarray*}
and the formulas in Appendix \ref{appD} for the sum of Matsubara frequencies for fermions. The notation, $\int_\otimes$, indicates that the contributions from the poles of the density function have to be excluded by either adjusting the contour off them, such as illustrated in fig. \ref{Fig:threshold} (b),  or removing the contributions from the enclosed residues.   For  fermions, there are poles at $p_0=\pm\frac{\pi}{\beta}$, $\pm\frac{3\pi}{\beta}$... . For the other imaginary-time ordering of the field operators,  $\tau_y>\tau_x$,
\begin{eqnarray*}
&&\langle \bar{\psi}_{\alpha}(\omega_m,{\bf k}){\psi}_{\beta}(\omega_n,{\bf p})\rangle \\
&=&
\sum_{s}u^s_\beta({\bf p})\bar{u}^s_\alpha({\bf p})\frac{\beta\delta_{mn}(2\pi)^3\delta^3({\bf p-k})}{2\xi_{\bf p}(i\omega_n-\xi_{\bf p})}+\sum_{s}v^s_\beta({\bf -p})\bar{v}^s_\alpha({\bf -p})\frac{\beta\delta_{mn}(2\pi)^3\delta^3({\bf p-k})}{2\xi_{\bf p}(i\omega_n+\xi_{\bf p})}.
\end{eqnarray*}
The chosen contour is the upper semicircle of the $p_0$-complex plane.
\begin{eqnarray}
&&\hspace{-1cm}\langle\bar{\psi}_{\alpha}(\tau_y,\vec{\bf y}){\psi}_{\beta}(\tau_x,\vec{\bf x})\rangle_{\rm Adv}
=\int_{\otimes}\frac{d^4 p}{(2\pi)^4}\frac{i}{\slashed{p}-m_{\rm f}}e^{p_0(\tau_x-\tau_y)+i{\bf p}\cdot ({\bf x-y})}n_{\rm F}(p_0).\label{diracpropA}
\end{eqnarray}
eq. (\ref{diracpropR}) and (\ref{diracpropA}) have shown the retarded and advanced propagators for fermions. 
By choosing the Feynman boundary conditions as in the field theory \cite{peskin95}, we may define the corresponding Feynman propagator as follows
\begin{eqnarray*}
S_F(\tau_x-\tau_y, {\bf x- y})& \equiv &\Theta (\tau_x-\tau_y)\langle0|\psi(\tau_x,{\bf x})\bar{\psi}(\tau_y,{\bf y})|0\rangle-\Theta (\tau_y-\tau_x)\langle0|\bar{\psi}(\tau_y,{\bf y})\psi(\tau_x,{\bf x})|0\rangle\\
&=&\int_{\otimes} \frac{d^4{ p}}{(2\pi)^4}e^{-p_0(\tau_x-\tau_y)+i{\bf p}\cdot ({\bf x}-{\bf y})}\frac{i}{\slashed{p}-m_{\rm f}+i\varepsilon}\left(1-n_F(p_0)\right).
\end{eqnarray*}
 The analytic continuation of the imaginary-time $\tau$ to the real-time $t$ by making $\tau=it$ may be applied as soon as the Matsubara frequencies are summed, there is no confusion that we do the replacement now 
\begin{eqnarray}
S_F(t_x-t_y, {\bf x- y})
&=&\int_\otimes \frac{d^4{ p}}{(2\pi)^4}e^{-i p\cdot (x-y)}\frac{i}{\slashed{p}-m_{\rm f}+i\varepsilon}\left(1-n_F(p_0)\right).\label{diracprop}
\end{eqnarray}
In the limit of $\beta \rightarrow \infty$, the above Feynman propagator becomes the one that we are familiar with. The density function becomes irrelevant as to suppress the violation of the Lorentz invariance. In the tree-level, the four momentum integral $\int_\otimes  \frac{d^4{ p}}{(2\pi)^4} \rightarrow \int  \frac{d^4{ p}}{(2\pi)^4}$ is a usual one, since the poles do not lie in the real axis of $p^0$.

\begin{figure}[t]
\begin{center}
   \subfigure[]{\includegraphics[width=5cm]{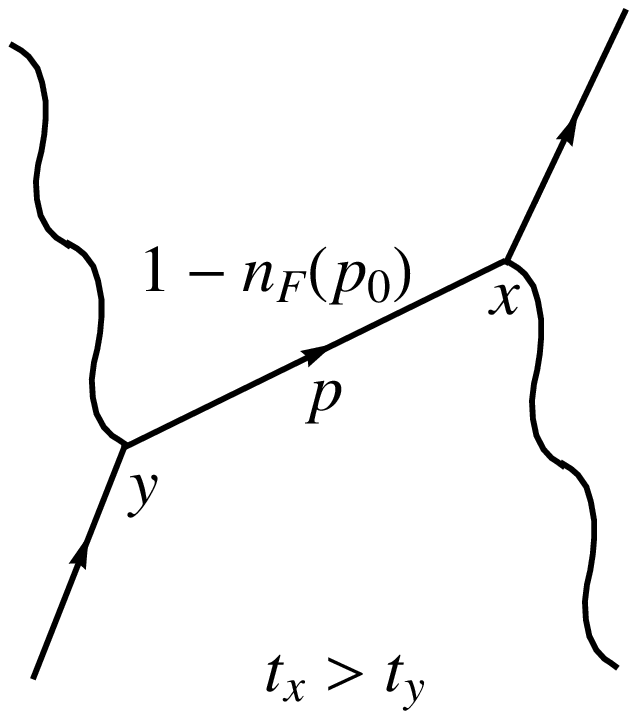}}
   \hspace*{0.07\textwidth}
   \subfigure[]{\includegraphics[width=5cm]{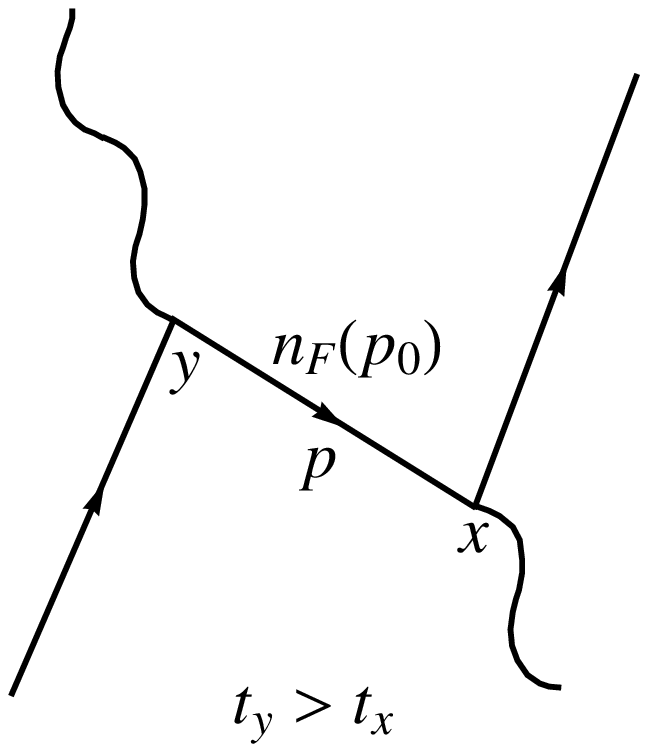}}
   \hspace*{0.0\textwidth}
\caption{\small  The Feynman diagrams are drawn for a fermion being produced and propagating through space between two space-time points, $(t_x, {\bf x} )$ and $(t_y, {\bf y} )$.  (a) for $t_x>t_y$, the chance for a fermion with a momentum $p$ to be created at $y$ and annihilated at $x$ is proportional to $1-n_F(p_0)$, since there is filled with $n_F(p_0)$ of fermions.  (b) for $t_x<t_y$, as a fermion is annihilated at $y$ and created at $x$, the probability is proportional to $n_F(p_0)$, because it cannot be annihilated without the existence in the first place. } 
  \label{Fig:feynDF}
\end{center}
\end{figure}

\subsubsection{Imaginary-time}
From the functional approach, the field operators are treated as grassmann numbers without the use of the spinors. According to the Lagrangian in eq. (\ref{Lagrangian}), the correlation function for the imaginary-time can also be obtained directly: 
\begin{eqnarray*}
&&\hspace{-.5cm}\langle \bar{\psi}(\omega_n,{\bf p})\psi(\omega_m,{\bf k})\rangle 
=\beta \delta_{nm} (2\pi)^3\delta^{(3)}({\bf p-k})\frac{1}{(i\omega_{n}\gamma_0 -\vec{\gamma}\cdot\vec{{\bf p}}-m_{\rm f})}.
\end{eqnarray*}
Followed by the same approaches in the field theory, the Feynman propagator between the two points, $(\tau_x, {\bf x})$ and $(\tau_y, {\bf y})$, in the imaginary-time and space  is 
\begin{eqnarray}
S_{\rm F}(\tau_x-\tau_y, {\bf x- y})
=\frac{1}{\beta}\sum_{n={\rm odd}}\int \frac{d^3{ \bf p}}{(2\pi)^3}e^{-i\omega_n(\tau_x-\tau_y)+i{\bf p}\cdot ({\bf x}-{\bf y})}\frac{1}{(i\omega_{n}\gamma_0 -\vec{\gamma}\cdot\vec{{\bf p}}-m_{\rm f})}.\hspace{.5cm}\label{Imtpropfermi}
\end{eqnarray}

\subsection{Propagator of photons}
From the Lagrangian densities in eq. (\ref{Lphoton}), the partition function of the photon field is known as 
\begin{eqnarray}
Z&=&\int e^{\int^{\beta}_0 d\tau \int d^3{\bf x}\left(-\frac{1}{4}F^{\mu\nu}F_{\mu\nu}-\frac{\zeta}{2}\partial_\mu A^\mu\partial_\nu A^\nu+\mu_\gamma A_\mu A^\mu\right)}[dA_\sigma(\tau,{\bf x}) dA^\sigma(\tau,\bf{x})],\label{PhotonZ}
\end{eqnarray}
where $\zeta$ is the gauge parameter. The choice of the gauge will be $\zeta=1$ in the following calculations. The role that the chemical potential $\mu_\gamma$ plays is similar to the mass squared, $m^2_\gamma=\frac{\mu_\gamma}{2}$. This term will be ignored in the derivation of photon's propagators and will be considered after the chemical potential is computed in Section \ref{chempot}.
\subsubsection{Real-time}
\begin{figure}[t]
\begin{center}
   \subfigure[]{\includegraphics[width=5cm]{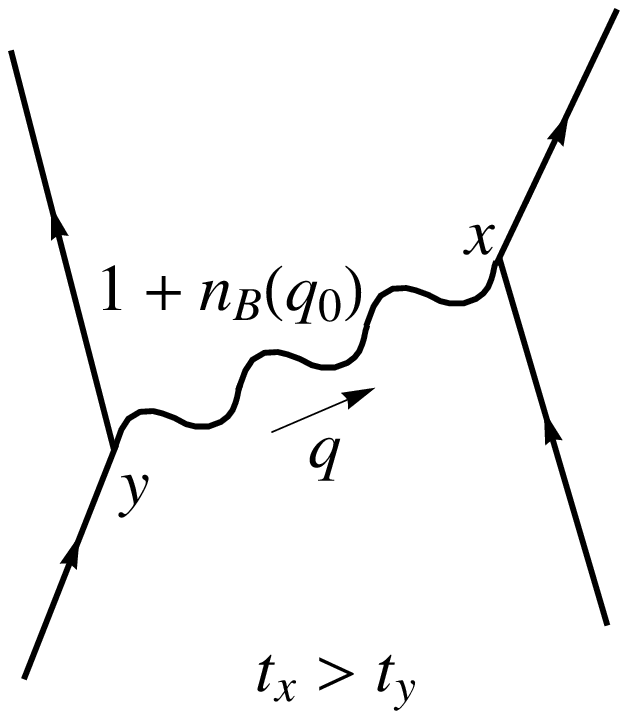}}
   \hspace*{0.07\textwidth}
   \subfigure[]{\includegraphics[width=5cm]{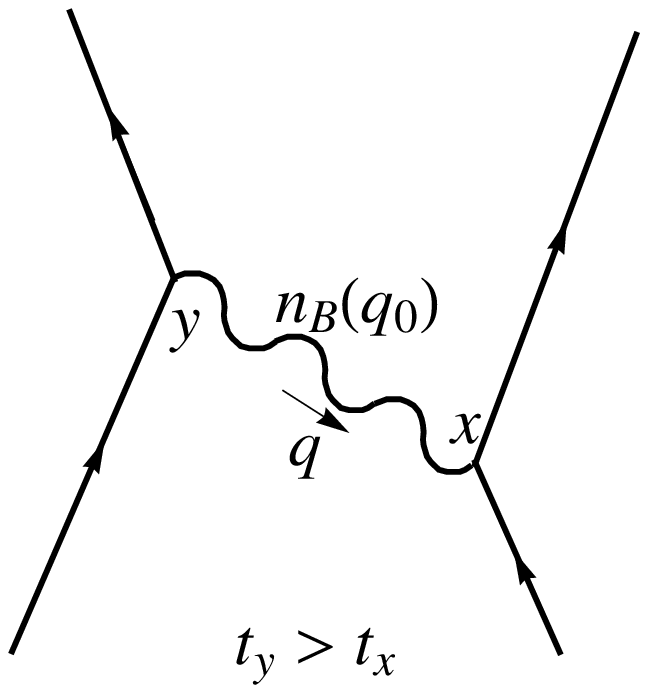}}
   \hspace*{0.0\textwidth}
\caption{\small  Feynman diagrams are illustrated for a boson being produced and propagating through space between two space-time points, $(t_x, {\bf x} )$ and $(t_y, {\bf y} )$. (a) for $t_x>t_y$, the chance for a boson with a momentum $q$ to be annihilated at  a latter time, $t_x$, is proportional to $1+n_B(q_0)$, since there are $1+n_B(q_0)$ of bosons in the vacuum.  (b)  for $t_x<t_y$, as a boson to be annihilated at $t_x$ ahead of the event at $t_y$, the chance is proportional to $n_B(q_0)$, because it can only annihilated $n_B(q_0)$ of bosons before $t_y$. } 
  \label{Fig:feynDF}
\end{center}
\end{figure}
One of the differences in quantizing the photon fields is that it is expanded by the Matsubara frequency, $\omega_n\,\,(=\frac{2\pi n}{\beta})$, instead of the energy. 
The other is that the expansion is the Fourier expansion without intentionally adding the factor $\frac{1}{\sqrt{2E}}$ to ensure the Lorentz Invariance. It can be shown later that the Lorentz invariance is secured as $\beta\rightarrow \infty$, like in the case of fermion's. The field operator of photons is expanded with respective to the Matsubara frequency and the 3-momentum as follows 
\begin{eqnarray}
A_\mu(\tau,{\bf x})
&\equiv&\frac{1}{\beta}\sum_n \int \frac{d^3 {\bf q}}{(2\pi)^3}A_\mu(\omega_n,{\bf q})e^{-i\omega_n\tau+i{\bf q\cdot x}},\label{photonex}
\end{eqnarray}
where the field operator in the momentum representation is
\begin{eqnarray*}
A_\mu(\omega_n,{\bf q})&=&\frac{1}{\sqrt{2|{\bf q}|}}
\sum^3_{\lambda=0}
\left(a^\lambda_{\omega_n,{\bf q}}\epsilon^\lambda_\mu({\bf q})+
a^{\lambda\dagger}_{-\omega_n,{\bf -q}}\epsilon^{\lambda*}_\mu({\bf -q})\right).
\end{eqnarray*}
The action, $i\mathcal{A}$, of the photon fields, is  
\begin{eqnarray*}
i\mathcal{A}=\int^{\beta}_0 d\tau \int d^3 \vec{\bf x}\,\,\mathcal{L}_0(A_\mu,\tau, \vec{\bf x})
= \frac{1}{\beta}\sum_{\omega_n}\int d^3 \vec{\bf q}\,\,\mathcal{L}_0(a^\lambda_{\bf q}, a^{\lambda\dagger}_{\bf q},\omega_n, \vec{\bf q}),
\end{eqnarray*}
where the Lagrangian of free photons, denoted as $\mathcal{L}_0$, includes only the first two terms in eq. (\ref{Lphoton}). The term with the chemical potential density $\mu_\gamma$ is dropped temporarily and will be taken into consideration later in the computation of the photon's self-energy.  We choose the gauge parameter to be $\zeta=1$, the Lagrangian in the momentum space is 
\begin{eqnarray*}
\mathcal{L}_0(a^\lambda_{\bf q}, a^{\lambda\dagger}_{\bf q}, \omega_n,\vec{\bf q})
&=&\frac{1}{4|{\bf q}|}(\omega_n^2+|{\bf q}|^2)\sum_\lambda (-g_{\lambda\lambda})
\left(a^\lambda_{\omega_n,{\bf q}} a^{\lambda\dagger}_{\omega_n,{\bf q}} + a^{\lambda\dagger}_{\omega_n,{\bf q}} a^\lambda_{\omega_n,{\bf q}}\right),
\end{eqnarray*}
where the matrix $g_{\lambda\lambda'}={\rm diag}(1,-1,-1,-1)$. The propagator of photons in two representations can be related by 
\begin{eqnarray*}
\hspace{-0.8cm}\left\langle A_\mu(\tau_x,\vec{\bf x}) A_\nu(\tau_y,\vec{\bf y})\right\rangle
=\frac{1}{\beta^2}\sum_{n,m}\int \frac{d^3 {\bf q}}{(2\pi)^3} \frac{d^3 {\bf k}}{(2\pi)^3}
\left\langle A_\mu(\omega_n,{\bf q}) A_\nu(\omega_m,{\bf k})\right\rangle e^{-i\omega_n\tau_x-i\omega_m\tau_y+i{\bf q\cdot x}+i{\bf k\cdot y}}.
\end{eqnarray*}
The result of the summation for the polarization vectors is $\sum_{\lambda=0}^{3}g_{\lambda\lambda}\epsilon^\lambda_\mu \epsilon^\lambda_\nu=g_{\mu\nu}$.
The two-point correlation function for the momentum representation can be derived as follows:
\begin{eqnarray}
\hspace{-1cm}\left\langle A_\mu(\omega_n,{\bf q}) A_\nu(\omega_m,{\bf k})\right\rangle
&=&\beta\delta_{-n,m}(2\pi)^3\delta^3({\bf q+k })\frac{-g_{\mu\nu}}{2|{\bf q}|}\left\{\frac{1}{i\omega_n+|{\bf q}|}-
\frac{1}{i\omega_n-|{\bf q}|} \right\}.\nonumber\\
\label{PhotonProp1}
\end{eqnarray}
The sum of the polarization vectors is replaced by a negative metric tensor. When computing the propagator in the imaginary-time and space representation, assume $\tau_x>\tau_y$ and sum over the Matsubara frequency, then we may obtain a two-point correlation function of a 4-momentum integral by introducing a complex integral for a variable $q_0$ with a semi-circle contour in the lower $q_0$-complex plane. We use the formulas in eq. (\ref{sumbose}) in Appendix \ref{appD} to sum over the Matsubara frequencies for bosons. 
The retarded propagator is
\begin{eqnarray*}
\left\langle A_\mu(\tau_x,\vec{\bf x}) A_\nu(\tau_y,\vec{\bf y})\right\rangle_{\rm Ret}
&=&\int_\otimes \frac{d^4 { q}}{(2\pi)^4} \frac{-ig_{\mu\nu}}{q^2}
\left(1+\frac{1}{e^{\beta q_0}-1}\right)
e^{-q_0(\tau_x-\tau_y)+i{\bf q\cdot( x-y)}}.
\end{eqnarray*}
The contour is chosen to enclose the  two residues of the poles at $q_0=-|{\bf q}|$ and $|{\bf q}|$. The resultant expression is similar to the one in the field theory except the factor  $1+n_B(q_0)$. In the limit of large value of $\beta$, the density function becomes unity as in the case of fermions. The integral sign, $\int_\otimes$, remind us of that the poles of the density function  have not to be enclosed by the contour or their residues inside the contour have to be excluded; this will be taken into account when the radiative corrections are computed in the following sections.  Analytically continuation from the imaginary-time to the real-time is made by letting $\tau=i t$; the corresponding Feynman propagator can be obtained in a similar manner to the retarded one: 
\begin{eqnarray*}
D_{{\rm F }}^{\mu\nu}(t_x-t_y, {\bf x- y})
&=&\int_\otimes \frac{d^4 { q}}{(2\pi)^4} \frac{-ig^{\mu\nu}}{q^2+i\varepsilon}
\left(1+n_B(p_0)\right)
e^{-i{ q\cdot( x-y)}}.
\end{eqnarray*}


\subsubsection{Imaginary-time}
Like fermion's propagator of the imaginary-time in eq. (\ref{Imtpropfermi}), the photon's imaginary-time propagator is derived from the Lagrangian of photons in eq. (\ref{Lphoton}):
\begin{eqnarray}
D_{{\rm }}^{\mu\nu}(\tau_x-\tau_y, {\bf x- y})
=\frac{1}{\beta}\sum_n\int \frac{d^3{ \bf q}}{(2\pi)^3}e^{-i\omega_n(\tau_x-\tau_y)+i{\bf q}\cdot ({\bf x}-{\bf y})}\frac{-1}{q^2_n}\left(g^{\mu\nu}-(\zeta-1)\frac{q_n^\mu q_n^\nu}{q^2_n}\right),\hspace{.6cm}\label{}
\end{eqnarray}
where $q_n=(i\omega_n, {\bf q})$. The corresponding Matsubara frequency, $\omega_n=\frac{2\pi n}{\beta}$, and $n$ is an integer. As mentioned, the chemical potential has not been included in the above, and will be included in the computations of radiative corrections.

\section{One-loop radiative corrections}
\label{oneloop}

\subsection{Self-energy of photons}
\subsubsection{Real-time}\label{photonreal}

\begin{figure}[t]
\begin{center}
   \subfigure[]{\includegraphics[width=5cm]{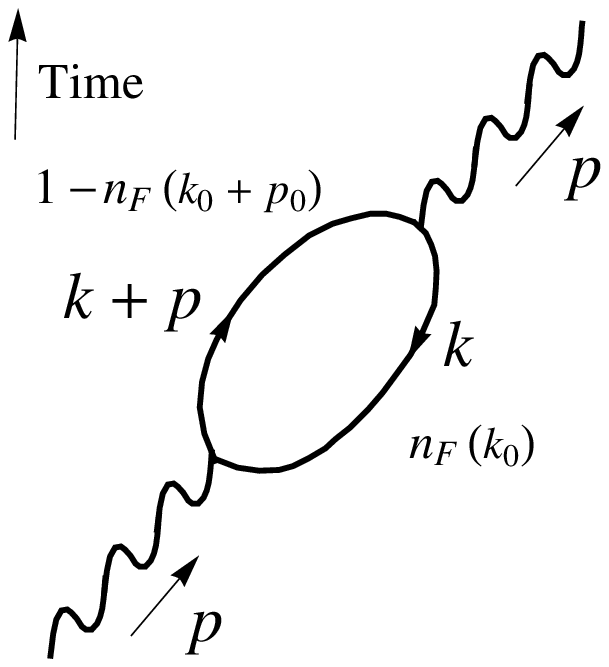}}
   \hspace*{0.07\textwidth}
   \subfigure[]{\includegraphics[width=5cm]{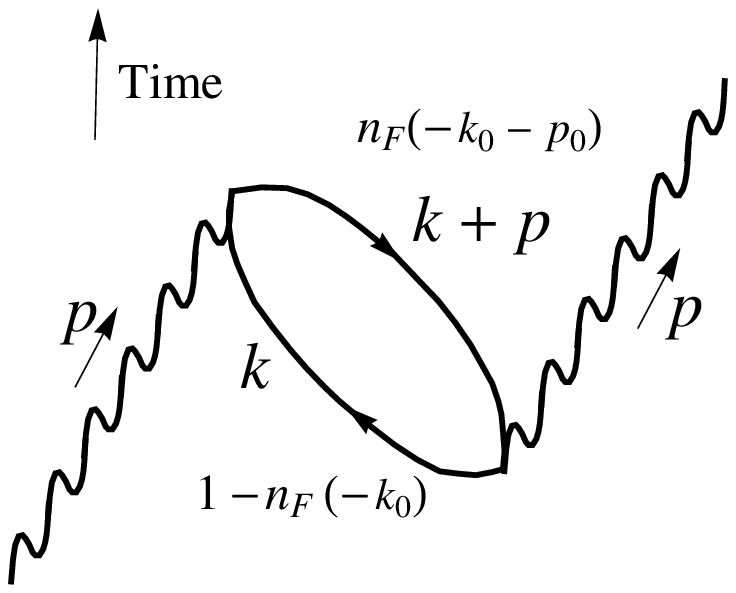}}
   \hspace*{0.0\textwidth}
\caption{\small    One-loop self-energy Feynman diagrams of a photon for different time ordering of the interaction vertices: (a) A photon annihilates first and creates an electron and a positron, later the two annihilate and a photon is created, the probability for this process is proportional to the product of $1-n_{\rm F}(k_0+p_0)$ and $n_{\rm F}(k_0)$. (b) A photon disappears due to the annihilation of an electron and a positron in the background, and its momentum has been carried away by another photon that is produced earlier. The probability is proportional to the product of $1-n_{\rm F}(-k_0)$ and $n_{\rm F}(-k_0-p_0)$.    } 
  \label{Fig:feynSEphoton}
\end{center}
\end{figure}
The radiative corrections in QED with the propagators that are formulated in the previous section will be applied on loop calculations. Feynman rules are similar except the extra density functions that are attached to each propagating particle and the redundant residues from the same density functions that have to be carefully dealt with.  As a photon carries a momentum, $p^\mu$, an electron and a positron are created and annihilates with momenta $k^\mu+p^\mu$ and $k^\mu$, as shown in fig. \ref{Fig:feynSEphoton}. The 4-velocity of the whole statistical system measured by an observer is denoted as $u^\mu$.  In the theory of relativity, the energy of a particle, with a 4-momentum $p^\mu$, measured by the system is written as a scalar product, $(p\cdot u)$. As the observer is at rest with respective to the system, $u^\mu=(1,0,0,0)$, the energy of the particle then happens to be $p^0$. In this section, the calculations for the self-energy of a photon may be proceeded in a similar way from the those in ref. \cite{peskin95}, 
with $\Delta=m^2_{\rm f}-x(1-x)p^2 $, after the Feynman parametrization is applied and the loop momentum is shifted from $q^\mu$ to $l^\mu$:
\begin{eqnarray}
i\Pi_2^{\mu\nu}(p)=-4e^2\int^1_0 dx\int_\otimes\frac{d^4 l}{(2\pi)^4}\frac{2l^\mu l^\nu -g^{\mu\nu}l^2-2x(1-x)p^\mu p^\nu+g^{\mu\nu}(m^2_{\rm f}+x(1-x)p^2)}{(l^2-\Delta)^2}\nonumber\\
\hspace{0cm}\times 
\frac{1}{e^{-\beta l\cdot u-\log b}+1}\frac{1}{e^{\beta l\cdot u-\log a}+1},\hspace{.5cm}
\label{SEofphoton}
\end{eqnarray}
where $l^\mu=k^\mu+x p^\mu$, $a=e^{\beta x p\cdot u}$, $b=e^{\beta(1- x)p\cdot u}$ and $x$ is the Feynman parameter. In general, the integrals with different powers of the denominator after the Feynman parameterization are in the form:
\begin{eqnarray}
&&\hspace{-1.5cm}\int_\otimes \frac{d^4 l}{(l^2-\Delta)^\lambda}\left\{1-n_{\rm F}((k+p)\cdot u)\right\}\left\{1-n_{\rm F}(-k\cdot u)\right\}\nonumber\\
&&\hspace{-1.5cm}
=\int_\otimes \frac{d^4 l}{(l^2_0-{\bf l}^2-\Delta)^\lambda}\frac{1}{e^{-\beta l\cdot u-\log b}+1}\frac{1}{e^{\beta l\cdot u-\log a}+1}.\label{photonInt}
\end{eqnarray}
 After applying the Wick rotation and  taking into account the shifting of the poles from the density functions, the integral becomes
\begin{eqnarray*}
&&\hspace{-1.cm}i\oint_{\rm RC,LC} \frac{(-1)^\lambda d l_{\rm E}d^3 {\bf l}}{(l^2_{\rm E}+{\bf l}^2+\Delta)^\lambda}\frac{1}{e^{i\beta l_{\rm E}-\log a}+1}\frac{1}{e^{-i\beta l_{\rm E}-\log b}+1}-({\rm residues\,\,of\,\,\,2\,\,n_F }),\\
&&\hspace{-1.5cm}=
\begin{cases}
i\oint_{\rm RC} d l_{\rm E}\int d^3 {\bf l} \frac{(-1)^\lambda}{(l^2_{\rm E}+{\bf l}^2+\Delta)^\lambda}\frac{1}{e^{i\beta l_{\rm E}-\log a}+1}\frac{1}{e^{-i\beta l_{\rm E}-\log b}+1}+2\pi i\frac{ab}{(1-ab)\beta}\sum_{n\in {\rm I}}\int d^3{\bf l}\frac{(-1)^\lambda}{\left[\Delta+{\rm l}^2+(\frac{(2n+1)\pi}{\beta}-i\frac{\log a}{\beta})^2\right]^\lambda}\\
i\oint_{\rm LC} d l_{\rm E}\int d^3 {\bf l} \frac{(-1)^\lambda}{(l^2_{\rm E}+{\bf l}^2+\Delta)^\lambda}\frac{1}{e^{i\beta l_{\rm E}-\log a}+1}\frac{1}{e^{-i\beta l_{\rm E}-\log b}+1}-2\pi i\frac{ab}{(ab-1)\beta}\sum_{n\in {\rm I}}\int d^3{\bf l}\frac{(-1)^\lambda}{\left[\Delta+{\rm l}^2+(\frac{(2n+1)\pi}{\beta}-i\frac{\log b}{\beta})^2\right]^\lambda},
\end{cases}
\end{eqnarray*}
\begin{figure}[t]
\begin{center}
   \subfigure[]{\includegraphics[width=5.5cm]{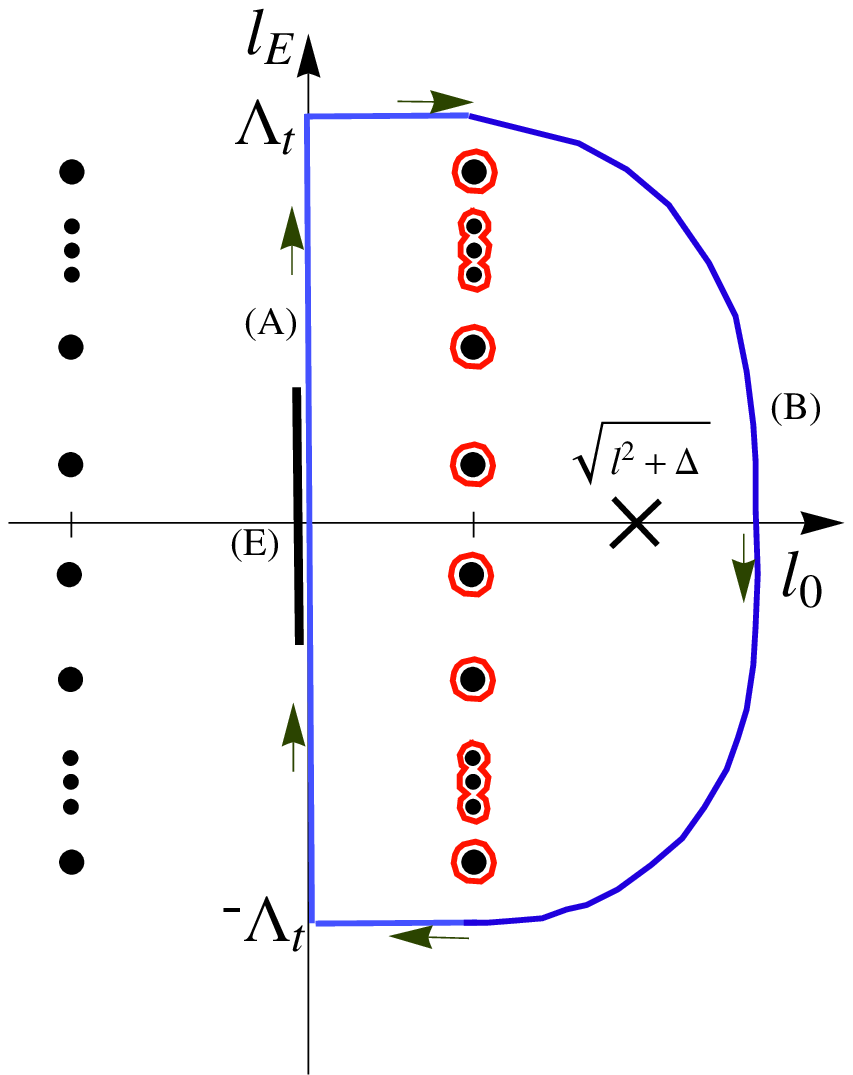}}
   \hspace*{0.0\textwidth}
   \subfigure[]{\includegraphics[width=5.5cm]{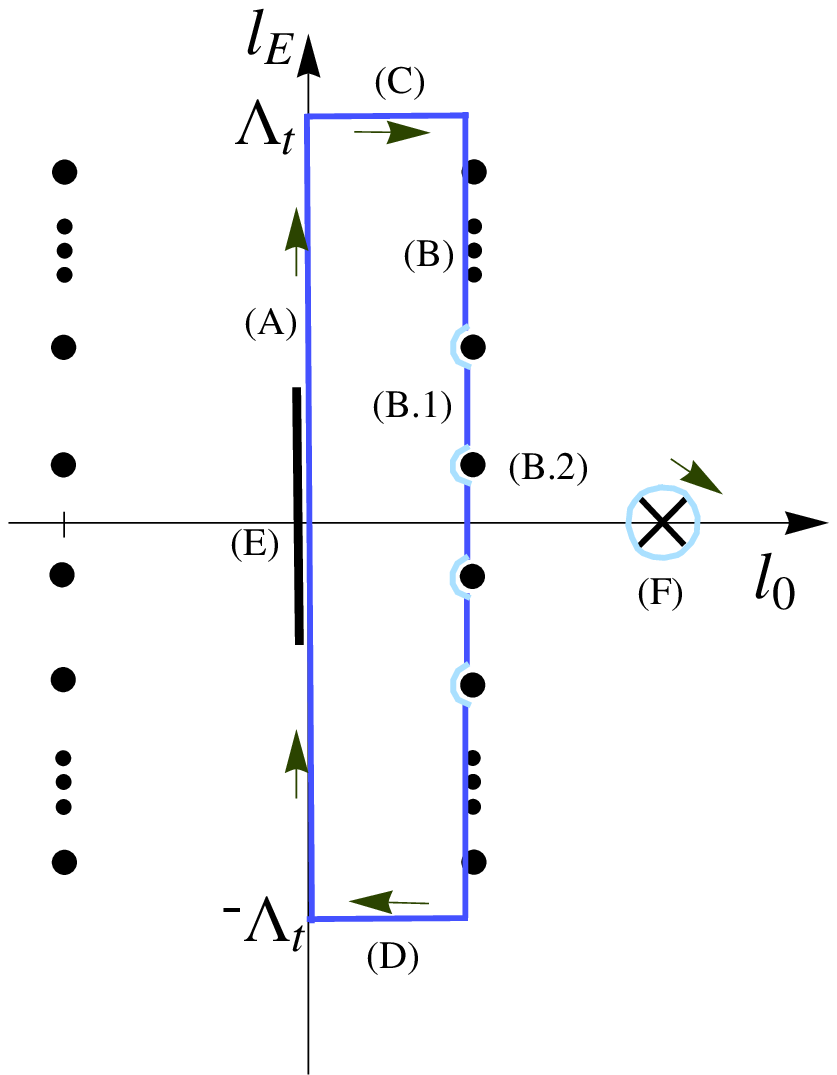}}
   \hspace*{0.0\textwidth}
\caption{\small  (a) The traditional contour are drawn after the Wick rotation is applied. The contour in blue, including (A) and (B), is running clockwise and those in red for the poles from one of the density function are counterclockwise. (b) The closed contour is represented by the blue lines with four segments (A), (B), (C) and (D). The integral in eq. (\ref{Eq:integralL})  is for the contours (A) and (B). The thick line of (E) is the branch cut of the integrand in the complex plane of $l_0$ as $\Delta<0$. The value $\Lambda_t$ is the cutoff for the energy. } 
  \label{Fig:threshold}
\end{center}
\end{figure}
where we have changed the variable by making $l_0=il_{\rm E}$. Before the Wick rotation being applied, the residues of the poles from the density functions are either above or below the real axis of the complex variable $l_0$, as illustrated in fig. \ref{Fig:threshold}. After the rotation, the poles that have to be taken into account are either in the right- or  left-hand side of the imaginary axis, depending on which contour is adopted. As $\beta \gg 1$, for the integration along the imaginary axis, the density functions in the first integral of both contours become unity, it changes to a traditional Feynman integral used in field theory. In practical calculations, the formulas below eq. (\ref{photonInt}) include an integration over a closed contour and a series of residues, and the contour can be chosen as the outlines of the rectangular box, (A), (B), (C) and (D), as shown in fig. \ref{Fig:threshold}, where the contour (B) runs down the complex plane and gets around the poles. The integral for the contour (B) comprises two parts, one is for the vertical lines between two poles, (B.1), and the other is the semicircles to avoid the poles, (B.2), whose resides happen to be half values of the full poles. For the first, take the right contour as an example, the parametrization of the contour is $l_0=il_{E}+(\log a/\beta)$, where $l_E$ is from $(2n-1)\pi/\beta$ to the next pole $(2n+1)\pi/\beta$. The factor in the integrand of eq. (\ref{photonInt}), $1/(e^{\beta l_0-\log a}+1)$, following the contour is  
\begin{eqnarray*}
\frac{1}{e^{i\beta l_E}+1}=\frac{1}{2}-i\frac{\sin\beta l_E}{2+2\cos\beta l_E},
\end{eqnarray*}
while other density function along the contour (B.1), $1/(e^{i\beta l_E-\log a-\log b}+1)$ becomes unity for large value of $\beta$.
The imaginary part are canceled by the upper and the lower halves of the contour (B), and can be ignored from now. Therefore, the sum of the integral ng all of the vertical lines, (B.1), is half size of the value for that of the contour (A) due to the factor $\frac{1}{2}$ in the real part.  For the semicircles, the density function gives a  factor $-2\pi/\beta$ to the residue for each pole, and it can be regarded as $-\Delta l_E$ as $\beta\gg 1$. The contribution from the half residues happens to be same  as that from the first part, therefore the combined result for the integral along the contour (B) can be expressed similar to the that along the imaginary axis as it runs through the contour, $l_0=il_E+\log a/\beta$. As we choose the new contour, the pole of the factor $1/(l_0^2-{\bf l}^2-\Delta)$, as indicated by (F) in the above figure, could be outside the rectangular box, and its residue is proportional to 
\begin{eqnarray}
\frac{1}{2\sqrt{{\bf l}^2+\Delta}}\frac{1}{e^{-\beta \sqrt{{\bf l}^2+\Delta}-\log b}+1}\frac{1}{e^{\beta \sqrt{{\bf l}^2+\Delta}-\log b}+1}.\label{pole1}
\end{eqnarray}
The residue is vanishing due to the last factor in eq. (\ref{pole1}) for large loop momentum ${\bf l}^2$ and large $\beta$; the same can also be applied for the LC. Based on the above discussions, the result of eq. (\ref{photonInt}) for the contours (A) and (B) can be written as 
\begin{eqnarray}\label{Eq:integralL}
&&\hspace{-1cm}(A)+(B)=
i\int^{\Lambda_t}_{-\Lambda_t} d l_{\rm E}\int d^3 {\bf l} \frac{(-1)^\lambda}{(l^2_{\rm E}+{\bf l}^2+\Delta)^\lambda}+i\int^{\frac{2\pi N_m}{\beta}}_{-\frac{2\pi N_m}{\beta}}d l_{\rm E}\int d^3{\bf l}\frac{(-1)^{\lambda+1}}{\left[\Delta+{\bf l}^2+\left(l_{\rm E}-i\frac{\log a}{\beta} \right)^2\right]^\lambda}.\nonumber
\end{eqnarray}
The integrals along the contours, (B.1) and (B.2), are combined into the second integral in the above.
The integrations of the loop momentum that we will apply are separated  into two steps, the first is to integrate the 3-momentum, then  the energy, instead of being treated  equivalently in the field theory. 
For the case of $\lambda=1$, and use the cutoffs, $\Lambda_t=\Lambda=\frac{2\pi N_{\rm max}}{\beta}$, which are introduced in Section \ref{scaleinv}, the two integrals of (A) and (B) become
\begin{eqnarray}
&&\hspace{-1cm}(A)+(B)=
-i\int^{\Lambda_t}_{-\Lambda_t} d l_{\rm E}\int d^3 {\bf l} \frac{1}{(l^2_{\rm E}+{\bf l}^2+\Delta)}+i\int^{\frac{2\pi N_m}{\beta}}_{-\frac{2\pi N_m}{\beta}}d l_{\rm E}\int d^3{\bf l}\frac{1}{\left[\Delta+{\bf l}^2+\left(l_{\rm E}-i\frac{\log a}{\beta} \right)^2\right]}\nonumber\\
\label{Eq:IntL1}
&&\hspace{-.0cm}=
-4\pi i\left[\left(\frac{\Lambda \Lambda_t }{(\Delta+\Lambda^2+\Lambda_t^2)}-\frac{\Lambda_t}{\sqrt{\Delta +\Lambda_t^2}}\tan^{-1}\left(\frac{\Lambda}{\sqrt{\Delta+\Lambda^2_t}}\right)
\right)\frac{\log^2 a}{\beta^2}+\cdots\right].
\end{eqnarray}
As the cutoff $\Lambda_t$ and $\Lambda$ are of the same size, from power countings, the coefficient of the order $\frac{\log^4 a}{\beta^4}$ is proportional to $\frac{1}{\Lambda^2}$, so it is negligible, as well as those of higher orders. The correction to the order of $p^2_0\,(\propto\frac{\log^2 a}{\beta^2})$ gives a non-negligible contribution; the value in the parenthesis of order $p_0^2$ is $-0.285398$ when $\Delta$ is neglected. The mass of the photon is expected to be $<10^{-18}{\rm eV}$ \cite{tu05}, so  this non-trivial corrections needs carefully examinations to see if it can be removed. As we discussed the contour in fig. \ref{Fig:threshold}, the segmants of (C) and (D) are not included in the discussion so far, and it can be proved that the contributions from these two are of the same size as (A)+(B) up to a minus sign. For the contours of (C) and (D), the integrals are
\begin{eqnarray*}
(C)+(D)&=&-4\pi\int^{-\frac{\log a}{\beta}}_0dx\sqrt{\Delta+(\Lambda_t+ix)^2}\tan^{-1}\left(\frac{\Lambda}{\sqrt{\Delta+(\Lambda_t+i x)^2}}\right)\\
&&\hspace{0.5cm}-4\pi\int^0_{-\frac{\log a}{\beta}}dx\sqrt{\Delta+(-\Lambda_t+ix)^2}\tan^{-1}\left(\frac{\Lambda}{\sqrt{\Delta+(-\Lambda_t+i x)^2}}\right)\\
&&\hspace{-2cm}=\,\,4\pi i\left[ \left(\frac{\Lambda_t\Lambda}{\Delta+\Lambda^2+\Lambda_t^2}-
\frac{\Lambda_t}{\sqrt{\Delta +\Lambda_t^2}}\tan^{-1}\left(\frac{\Lambda}{\sqrt{\Delta+\Lambda^2_t}}\right)
\right)\frac{\log^2 a}{\beta^2}
+O\left(\frac{\log^4 a}{\beta^4}\right)\right],
\end{eqnarray*}
which happens to cancel the correction in eq. (\ref{Eq:IntL1}). As it applies to the self-energy of the photon, the combined results contribute zero to the next-leading order. As for $\lambda=2$, the corresponding integrals for the contours (A) and (B), are
\begin{eqnarray*}
&&\hspace{-1cm}(A)+(B)=
i\int^{\Lambda_t}_{-\Lambda_t} d l_{\rm E}\int d^3 {\bf l} \frac{1}{(l^2_{\rm E}+{\bf l}^2+\Delta)^2}-i\int^{\frac{2\pi N_m}{\beta}}_{-\frac{2\pi N_{\rm max}}{\beta}}d l_{\rm E}\int d^3{\bf l}\frac{1}{\left[\Delta+{\bf l}^2+\left(l_{\rm E}-i\frac{\log a}{\beta} \right)^2\right]^2}\\
&&\hspace{-1cm}
=2\pi i\left[\left(-\frac{1}{\Lambda\Lambda_t}+\frac{2\Lambda\Lambda_t}{(\Lambda^2+\Lambda^2_t)^2}+\frac{\Lambda_t}{\Lambda(\Lambda^2+\Lambda^2_t)}-\frac{1}{\Lambda_t^2}\tan^{-1}\left(\frac{\Lambda}{\Lambda_t}\right)\right)\frac{\log^2a}{\beta^2}
+O\left(\frac{\log^4 a}{\beta^4}\right)\right].
\end{eqnarray*}
The result of the above approaches zero once the cutoffs are taken to infinity, and that of (C) and (D) can be proved to be vanished in the same manner. Therefore, the whole integral for different $\lambda$ contributes nothing to the self-energy as they are used in eq. (\ref{SEofphoton}). The only nonzero contribution to the self-energy happens when $\Delta < 0$ and the branch cut appears along the contour (A). 
Since the integral along contour (A) is identical to the traditional loop integral as $\beta\gg 1$, it gives the same imaginary part over the threshold as that from the field theory.

\subsubsection{Imaginary-time}\label{photonimg}
The one-loop radiative corrections to the self-energy of photon are computed according to the Lagrangian in eq. (\ref{Lphoton}) with the gauge $\zeta=1$. From the corresponding Feynman rules, the self-energy to the next-leading order, $\Omega_{2,\beta}^{\mu\nu}$, is 
\begin{eqnarray}
\Omega_{2,\beta}^{\mu\nu}(p_n{\bf })=(-1)e^2\frac{1}{\beta}\sum_{m}\int\frac{d^3{\bf q}}{(2\pi)^3}
{\rm Tr}\left[\gamma^\mu\frac{1}{\slashed{q}_m-m_{\rm f}}\gamma^\nu\frac{1}{(\slashed{p}_n-\slashed{q}_m)-m_{\rm f}}\right],\label{omega}
\nonumber
\end{eqnarray}
where $p^0_m=i\frac{\omega_m}{\beta}$ and $\omega_m=\frac{2\pi}{\beta}(m+\frac{1}{2})$. The minus sign is added for a closed fermion loop. The self-energy of the photon is taking a general form 
\begin{eqnarray}
&&\hspace{-0.7cm}\Omega_{2,\beta}^{\mu\nu}(p_n)\equiv\left(g^{\mu\nu}-\frac{p^\mu_n p^\nu_n}{p^2_n}\right)\left(p^2_n\Omega_T(p^2_n)-M(p^2_n)\right)+\frac{p^\mu_np^\nu_n}{p^2_n}\left(p^2_n\Omega_L(p^2_n)-M(p^2_n)\right),
\hspace{1cm}\label{SEphoton}
\end{eqnarray}
We make a shorthand for the factor
$\Delta(p_n^2)=m^2_{\rm f}-x(1-x){p}_n^2$. 
The contribution to the correction mass, $M(p^2_n)$, is from the term of the metric tensor, $g^{\mu\nu}$, therefore it is the same for the longitudinal and the transverse parts. 
The integration over 3-momentum  
 are taken first while the cutoff $\Lambda$ is kept finite until the whole integration is finished. 
One the other hand, in the limit of $\beta_0\rightarrow 0$, according to Section \ref{phi3}, an extra scaling factor $(\frac{\beta_0}{\beta})^2$ has to be added. The dimensionless quantities, like $\Omega_{0T}
$ and $\Omega_{0L}$, are free from the extra factor, only $M_0$ has to be corrected. It is noticed that there is in fact no quadratic divergence for the photon mass in QED due to  gauge- and Lorentz-invariance. However the Lorentz invariance is only secured for $\beta\gg 1$ in the imaginary-time formalism, there would be no wonder that the quadratic divergence will appear in the following calculation.  
The arguments of the self-energy  functions can be dropped, since they do not depend on them, and we obtain
\begin{eqnarray}
\Omega_{0T}
=
-\left.\frac{2\alpha}{3\pi}(\log N_{\rm max}+C_E)+\frac{\alpha}{3\pi^2}\left(G+\frac{\pi}{4}\right),
 \right. \hspace{1.3cm}&&
\hspace{-1cm}\Omega_{0L}=\frac{\alpha}{3\pi^2}\left(G+\frac{\pi}{4}\right),\nonumber\\
M_0
=\,\,\frac{2\alpha}{\pi}\left(\frac{\beta_0}{\beta}\right)^2
\left\{ \frac{4\pi N_{\rm max}^2}{\beta_0^2}+\left(G+\frac{\pi}{4}\right)\frac{m^2_{\rm  f }}{\pi}-\frac{\pi^2}{12\beta_0^2}\right\},
 \label{betaInftyTps}
\end{eqnarray}
where $C_E=-0.036489...$ for $\sum_{m=1}^{N_{\rm max}}\frac{1}{m-\frac{1}{2}}=\log N_{\rm max}+C_E$.  Besides, we have used $\sum_{m=1}^{N_{\rm max}}\left({m-\frac{1}{2}}\right)= \frac{N^2_{\rm max}}{2}$, $\sum_{m=1}^{N_{\rm max}}$\hspace{.5cm}$\left({m-\frac{1}{2}}\right)^2=\frac{N_{\rm max}}{3}\left( \mathsmaller{N^2_{\rm max}}-\frac{1}{4}\right)$, etc.\,.
The factor $G\,\,(=0.915966...)$ is Catalan's constant. It is obtained from the sum, $\sum_{n=0}^\infty \frac{(-1)^n n}{(2n-1)^2}=\frac{G}{2}+\frac{\pi}{8}$, as the integral is expanded in powers of $1/\Lambda$. \par
As for the other limiting case, for $\beta \gg 1$, the sum of the fermionic frequency becomes an integral over the variable $\omega$, that is  $\frac{2\pi}{\beta}\sum\rightarrow\int d\omega$. 
 When integration over $d\omega$, as shown in Appendix \ref{appB}:  eq. (\ref{omegaT2}) and (\ref{omegaL2}), the following treatment for the inverse tangent function is adopted by making 
\begin{eqnarray*}
\tan^{-1}\left(\frac{\Lambda}{\sqrt{\omega^2+\Delta}}\right)=\frac{\pi}{2}-
\tan^{-1}\left(\frac{\sqrt{\omega^2+\Delta}}{\Lambda}\right).
\end{eqnarray*}
This will make a good expansion of $1/\Lambda$ when the cutoff is taken to infinity. 
The corrections to the photon mass is independent from the incoming momentum. We will drop its arguments hereafter. We may set $\Lambda_t=\Lambda=\frac{2\pi N_{\rm max}}{\beta}$ and obtain
\begin{eqnarray}
\Omega_T(p^2_n)&=&-\frac{2\alpha}{\pi}\left(\frac{G}{6\pi}+\frac{1}{3}\log N_{\rm max}-\frac{1}{3}\log \beta+\frac{1}{3}\log (4\pi)-\int^1_0dxx(1-x)\log\Delta(p^2_n)\right),\nonumber \\
\Omega_L(p^2_n)&=&\frac{G\alpha}{3\pi^2},\hspace{.5cm}{\rm and}\hspace{.5cm}M=\frac{2\alpha}{\pi}\left( {\frac{4\pi}{\beta^2}N_{\rm max}^2}+\frac{Gm^2_{\rm f}}{\pi} \right).\label{beta0ps}
\end{eqnarray}
In Section \ref{perturbation}, the radiative corrections, similar to the renormalized conditions, are defined as the differences between $\beta$ and $\beta_0\,\, (=0^+)$ in eq. (\ref{newIT}); they can be written for the respective quantities such as follows
\begin{eqnarray*}
&&\hat{\Omega}_T(p^2_n)\equiv{\Omega}_T(p^2_n)-{\Omega}_{0T},\hspace{.3cm}
\hat{M}\equiv{M}-{M}_{0},\,\,
 {\rm and}\hspace{.5cm} \hat{\Omega}_L(p^2_n)\equiv{\Omega}_L(p^2_n)-{\Omega}_{0L}.
\end{eqnarray*}
The renormalized results for transverse and longitudinal parts are
\begin{eqnarray}
\hat{\Omega}_T(p^2_n)&=&-\frac{2\alpha}{\pi}\left(-\frac{1}{3}C_E+\frac{\alpha}{24}+\frac{1}{3}\log 4\pi -\frac{1}{3}\log \beta-\int^1_0dxx(1-x)\log\Delta(p_n^2)\right),\nonumber\\
\hat{\Omega}_L(p^2_n)&=&-\frac{\alpha}{12\pi}, \hspace{.2cm}{\rm and}\hspace{.2cm}
\hat{M}
=\,\,\frac{2\alpha}{\pi}
\left\{
\frac{Gm^2_{\rm  f }}{\pi}+\frac{\pi^2}{12\beta^2}\right\}.\hspace{.3cm}
\label{pitpil}
\end{eqnarray}
For the corrected photon mass squared, $\hat{M}$, looks non-trivial, we will leave it temporarily and be back on this later.  It  can be compared with the self-energy of the gluon in a QCD plasma \cite{landsman87}; its Debye mass is 
\begin{eqnarray}
m_D&=&g^2\left(\frac{N_f T^2}{6}+\frac{N_c T^2}{3}+\frac{N_f \mu_q^2}{2\pi^2}\right),
\end{eqnarray}
where $\mu_q$ is the quark chemical potential and $g$ is the QCD coupling constant. We know that in a sense the $\mu_q$ is equivalent to a mass term in the Lagrangian, therefore the correction to the  Debye mass of a gluon in QCD plasma is similar to that of a photon in the imaginary-time formalism of vacuum. They all depend on the temperature and fermion masses or chemical potentials in a similar form.

 We now combine the tree-level and the one-loop contribution for the photon propagator, as we assume it carries a mass $m_\gamma$:
\begin{eqnarray}
&&\frac{-g^{\mu\nu}}{(i\omega_n)^2-|{\bf q}|^2-m^2_{\gamma}}+\frac{-g^{\mu\alpha}}{(i\omega_n)^2-|{\bf q}|^2-m^2_{\gamma}}
(\hat{\Omega}_{\alpha\beta})\frac{-g^{\beta\nu}}{(i\omega_n)^2-|{\bf q}|^2-m^2_{\gamma}}+\dots\nonumber
\\
\hspace{-3.8cm}&=&
-\frac{g^{\mu\nu}(1-\hat{\Omega}_T)}{(i\omega_n)^2-{\bf q'}^2 }+\frac{p^\mu_ n p^\nu_n(\hat{\Omega}_L-\hat{\Omega}_T)}{(p^2_n-m^2_{\gamma})^2}+\dots, 
\label{propcorr}
\end{eqnarray}
where ${\bf q'}^2={\bf q}^2+(m^2_{\gamma}+\hat{M})(1-\hat{\Omega}_L)$. The role of a photon's intrinsic mass is played by the chemical potential $\mu_{\gamma}$ in the Lagrangian, eq. (\ref{Lphoton}), so  the effective mass of the photon to the leading order can be defined from above as 
\begin{eqnarray}
m^2_{\rm eff,\gamma}=\frac{\mu_{\gamma}}{2}+\hat{M}.\label{meffphoton}
\end{eqnarray}
Unlike the effective mass of the fermion which will be defined in eq. (\ref{effmassF}), the correction from the imaginary-time context will shift the pole of the photon propagator as seen in eq. (\ref{PhotonProp1}).  
To see if the effective mass of the photon is nearly zero or not, the calculation of the chemical potential, which is ignored in the derivation of the photon propagator, is necessary and will be performed below. On the other hand, the renormalized amplitude of the photon is defined as 
\begin{eqnarray}
Z_3\equiv 1-\hat{\Omega}_T.
\end{eqnarray}
Its dependence on $\beta$ results in one of the renormalization group equation, and will be discussed in detail  in Section \ref{rgeq}.



\subsubsection{Running coupling and Lamb shift}
The radiative corrections to the self-energy of photons give rise to the slight change of the electric potential $V({\bf x})$ in QED. In this section, we will see if the same results are concluded for the imaginary-time formalism. 
The correction to the amplitude can approximated by setting $\omega_n=0$, since $\hat{\Omega}_T$ depends weakly on it, especially as $\beta \gg 1$. To Ignore the constants in the expression, the quantity can be redefined in the ${\bf q}^2$ dependence of the effective charge
\begin{eqnarray*}
{\Omega}_2(0,{\bf q})&\equiv& \hat{\Omega}_T(0,{\bf q})-\hat{\Omega}_T(0,{\bf 0})
=-\frac{2\alpha}{\pi}\int^1_0dx (1-x)\log\left(\frac{m^2_{\rm f}}{m^2_{\rm f}+x(1-x){\bf p}^2}\right).
\end{eqnarray*}
The expression is the same  as what is obtained in the field theory \cite{peskin95} in the non-relativistic limit. 
In the non-relativistic limit, the potential $V({\bf x})$ is obtained from the formula
\begin{eqnarray*}
V({\bf x})&=&-e^2\int \frac{d^3 {\bf p}}{(2\pi)^3}\frac{e^{i{\bf p\cdot x}}}{{\bf p}^2(1-\Omega_2(0,{\bf p}))}.
\end{eqnarray*}
From above, we may see that it also gives the same results as in QED for the effective potential.

\subsubsection{Ward identity}
One of the important requirements in the field theory is to check the gauge invariance of the theory, as it is related to the conservation of the momentum. As to the self-energy of the photon, it states 
\begin{eqnarray}
\Pi^{\mu\nu}(p) p_\nu=0,\label{wardI}
\end{eqnarray}
and generates zero mass for the photon. 
Based on the discussions in Section \ref{photonreal},  the real part of the self-energy of the photon, $\Pi^{\mu\nu}$ is highly suppressed for $\beta\rightarrow \infty$ along the contour as shown in fig. \ref{Fig:threshold}.
 This means eq. (\ref{wardI}) is automatically satisfied for the real part. As to the imaginary part from the branch cut, examine how the dimensional regularization does to ensure the Ward identity.
The reason is that it provides an equality between the $d$-dimensional integrals: 
\begin{eqnarray}
\int \frac{d^d l_E}{(2\pi)^{{d}}} \frac{\left(-\frac{2}{d}+1\right)l^2_E}{(l^2_E+\Delta)^2}=
-\int \frac{d^d l_E}{(2\pi)^{d}} \frac{\Delta}{(l^2_E+\Delta)^2}.\label{wi}
\end{eqnarray}
Regardless of the real parts of the both integrals, for $d=4$, both of the imaginary parts are coming from the term, $\frac{\Delta}{(4\pi)^2}\log \Delta$, in the outcomes of above integrals, therefore eq. (\ref{wi}) is contented for the imaginary part. Thus the self-energy tensor of the photon from the real-time formalism satisfies eq. (\ref{wardI}). As for the imaginary-time propagators, we may know from the above one-loop calculations, it does contribute a small radiative correction to photon's mass. On the other hand, we know from the traditional QED, the Lorentz- and gauge-invariance result in zero correction to the mass for photons. In the imaginary-time formalism, the Lorentz invariance does not hold in the imaginary-time Lagrangian densities as shown in Section \ref{lagimag}.  Thus there is no wonder there are an appearance of quadratic divergence in photo's self-energy and a nonzero mass correction. Fortunately, the quadratic divergence can be removed by its counterpart at the high energy limit, $\beta_0$, and the nonzero mass correction is canceled by its chemical potential, which will be explained right below. Therefore after taking into account all of the factors, the self-energy correction to photon's propagator  can be expressed in the form of  $\Omega^{\mu\nu}(q)=\Omega(q) (g^{\mu\nu}-\frac{q^\mu q^\nu}{q^2})$ for both real-time and imaginary-time propagators, 
 and the Ward identity can be secured. 

\subsubsection{Chemical potential}\label{chempot}
The calculation presented here is generalized from Appendix \ref{appC}. In the 4-momentum space $(i\omega_n,{\bf p})$, the number operator of the vector field is proportional to  the product of the photon field operators, such as 
\begin{eqnarray*}
 \hat{N}(i\omega_n,{\bf p})\equiv :A^\mu(i\omega_n,{\bf p}) A_\mu(i\omega_n,{\bf p}):&=&\frac{1}{2}\sum^{}_{\lambda=1,2}:\left(a^\lambda_{\omega_n,{\bf p}}a^{\lambda\dagger}_{\omega_n,{\bf p}}+
a^{\lambda\dagger}_{\omega_n,{\bf p}}a^\lambda_{\omega_n,{\bf p}}\right):,\,\, 
\end{eqnarray*}
where $\lambda$ denotes the polarization states. The representation in space and imaginary-time: $\hat{N}(\tau,{\bf x})=A^\mu (\tau,{\bf x}) A_\mu (\tau,{\bf x})$. They are related by 
\begin{eqnarray*}
\frac{1}{\beta}\sum_n\int \frac{d^3{\bf  p}}{(2\pi)^3}\,\hat{N}(\omega_n,{\bf p})=\int^\beta_0 d\tau\int d^3{\bf x}\,\hat{N}(\tau,{\bf x}).
\end{eqnarray*}
 The interaction Hamiltonian in the imaginary-time and space, which is proposed in Section \ref{IntQED}, is $\hat{V}=e\int d^3{\bf x}\bar{\psi}(x)\gamma^\mu\psi(x)A_\mu=\int d^3{\bf x}H_{int}$. We may rewrite eq. (\ref{chemicalp}) in Appendix \ref{appC} in terms of the chemical potential density, $\mu_\gamma$, as 
 \begin{eqnarray}
\beta \,\mu\rightarrow\int^\beta_0d\tau\int d^3{\bf x}\,\mu_\gamma(\tau,{\bf x})&=&-\sum_{n=1}\frac{1}{n!}\langle \frac{\delta (\int^\beta_0d\tau\int d^3{\bf x}H_{int})^n}{\delta (A^\mu A_\mu)}\rangle.
\label{chemicalp1}
\end{eqnarray}
The first non-vanishing term is for $n=2$, the variation, $\delta$, is placed inside the integral on $H_{int}^2$. The leading order term of the right-hand-side becomes
 \begin{eqnarray}
\int^\beta_0d\tau\int d^3{\bf x}\,\mu_\gamma(\tau,{\bf x})&=&-\frac{e^2}{2!}\langle  \left(\int^\beta_0d\tau\int d^3{\bf x}\bar{\psi}\gamma^\mu \psi\right)^2\rangle.
\label{chemicalp1}
\end{eqnarray}
The Lorentz index, $\mu$, is contracted with the other one, so that the right-hand-side is kept as a scalar.  It
then can be computed 
 \begin{eqnarray*}
\hspace*{-1.cm}
{\rm RHS}&=&-\frac{e^2}{2!}\int^\beta_0d\tau\int d^3{\bf x}\int^\beta_0d\tau'\int d^3{\bf x'}{\rm Tr} \,S_F(x-x')\gamma^\mu S_F(x'-x)\gamma_\mu,\\
&=&-\frac{e^2}{2!}\int^\beta_0d\tau\int d^3{\bf x}\frac{1}{\beta}\sum_{n}\int \frac{d^3{\bf q}}{(2\pi)^3}{\rm Tr} \,\frac{1}{\slashed{q}_n-m_{\rm f}}\gamma^\mu \frac{1}{\slashed{q}_n-m_{\rm f}}\gamma_\mu . 
\end{eqnarray*}
The dependence of $x$ in the first line of the above 
disappears after integrating out the variable $x'$.
As compared with eq. (\ref{chemicalp1}), we may obtain the chemical potential density:
 \begin{eqnarray*}
\mu_\gamma&=&-\frac{e^2}{2\beta}\sum_{n}\int \frac{d^3{\bf q}}{(2\pi)^3}{\rm Tr} \,\frac{1}{\slashed{q}_n-m_{\rm f}}\gamma^\mu \frac{1}{\slashed{q}_n-m_{\rm f}}\gamma_\mu . 
\end{eqnarray*}
The above can be derived directly from eq. (\ref{omega}) by setting the incoming momentum, $p_n=0$ and contracting the two Lorentz indices, so in a similar way to define the renormalization condition for the chemical potential density we may obtain
\begin{eqnarray*}
\hat{\mu}_\gamma&=&-2{\hat{M}} . 
\end{eqnarray*}
The subtraction from the value at $\beta_0=0^+$ have been applied in the above. The corrected mass of photon from the self-energy is canceled by the chemical potential density, so that the effective mass, which is defined in eq. (\ref{meffphoton}), diminishes  to the order of $O(\alpha)$.


\subsection{Self-energy of fermions}
\subsubsection{Real-time}
In fig. \ref{Fig:feynSEelectron}, it shows that two situations of a fermion emitting and absorbing a photon during its propagation. The self-energy of the fermion can be expressed in the form of 
\begin{eqnarray*}
\Sigma(p)=\slashed{p}\Sigma_V(p^2)+m_{\rm f}\Sigma_S(p^2).
\end{eqnarray*}
The scalar quantity, $\Sigma_V(p^2)$, can be obtained by taking $\frac{1}{4}{\rm Tr}\slashed{p}$ on both sides,
\begin{eqnarray*}
-i\Sigma_(p)&=&-e^2\int_\otimes\frac{d^4k}{(2\pi)^4}\gamma^\mu  \frac{(\slashed{p}+\slashed{k})+m_{\rm f}}{[(k+p)^2-m_{\rm f}^2]k^2}\gamma^\nu (1-n_F((k+p)\cdot u))(1+n_B(-k\cdot u)).
\end{eqnarray*}
After projecting onto $\slashed{p}$, their coefficient functions are
\begin{eqnarray*}
\Sigma_V(p^2)
&=&2ie^2\int d\alpha_1 d\alpha_2\delta(\alpha_1+\alpha_2-1)(1-\alpha_1)\int_\otimes \frac{d^4 K}{(2\pi)^4}\frac{1}{[K^2-\Delta(p^2)]^2}\\
&&\hspace{5cm}\times
\frac{1}{e^{-\beta(K+(1-\alpha_1)p)\cdot u}+1}\frac{-1}{e^{\beta(K-\alpha_1p )\cdot u}-1},
\end{eqnarray*}
where $\Delta(p^2)=-\alpha_1\alpha_2p^2+\alpha_1m^2_{\rm f}$ 
and $a=e^{\beta(1-\alpha_1)p\cdot u}$ and  $b=e^{\beta\alpha_1p\cdot u}$. As for the scalar part of self-energy function,
\begin{eqnarray*}
\Sigma_S(p^2)
&=&-i4e^2\int_\otimes \frac{d^4 k}{(2\pi)^4}\frac{1}{[(k+p)^2-m^2_{\rm f}]k^2}
(1-n_F((k+p)\cdot u))(1+n_B(-k\cdot u)).
\end{eqnarray*}
In general, the 4-momentum integral, $I _{\rm SE, \lambda}$, where $\lambda$ is the power of the denominator,  with $n_F$ and $n_B$ can be written in the form of 
\begin{eqnarray*}
&&\hspace{-1.5cm}I_{\rm SE, \lambda}=\int_\otimes \frac{d^4 l}{(l^2-\Delta)^\lambda}\left\{1-n_{\rm F}((k+p)\cdot u)\right\}\left\{1+n_{\rm B}(-k\cdot u)\right\}\\
&&\hspace{-0.6cm}
=\int_\otimes \frac{d^4 l}{(l^2_0-{\bf l}^2-\Delta)^\lambda}\frac{1}{e^{-\beta l\cdot u-\log b}+1}\frac{-1}{e^{\beta l\cdot u-\log a}-1},
\end{eqnarray*}
where $l=k+x p$, $a=e^{\beta xp\cdot u}$ and $b=e^{\beta(1- x)p\cdot u}$. It can explicitly be written for the left and right contours of the imaginary axis as follows
\begin{eqnarray*}
&&\hspace{-0.7cm} I_{\rm SE, \lambda}=\oint_{\rm RC,LC} \frac{d l_{\rm 0}d^3 {\bf l}}{(l^2_{\rm 0}-{\bf l}^2-\Delta)^\lambda}\frac{-1}{e^{\beta l_{\rm 0}-\log a}-1}\frac{1}{e^{-\beta l_{\rm 0}-\log b}+1}\pm({\rm residues\,\,of\,\,\,n_B\,\,and\,\,n_F }),\\
&&\hspace{-1cm}=
\begin{cases}
i\oint_{\rm RC} d l_{\rm E}\int d^3 {\bf l} \frac{(-1)^\lambda}{(l^2_{\rm E}+{\bf l}^2+\Delta)^\lambda}\frac{-1}{e^{i\beta l_{\rm E}-\log a}-1}\frac{1}{e^{-i\beta l_{\rm E}-\log b}+1}+2\pi i\frac{ab}{(1+ab)\beta}\sum_{n\in {\rm I}}\int d^3{\bf l}\frac{(-1)^{\lambda+1}}{\left[\Delta+{\rm l}^2+(\frac{2n\pi}{\beta}-i\frac{\log a}{\beta})^2\right]^\lambda}\\
i\oint_{\rm LC} d l_{\rm E}\int d^3 {\bf l} \frac{(-1)^\lambda}{(l^2_{\rm E}+{\bf l}^2+\Delta)^\lambda}\frac{-1}{e^{i\beta l_{\rm E}-\log a}-1}\frac{1}{e^{-i\beta l_{\rm E}-\log b}+1}-2\pi i\frac{(-1)ab}{(ab+1)\beta}\sum_{n\in {\rm I}}\int d^3{\bf l}\frac{(-1)^{\lambda+1}}{\left[\Delta+{\rm l}^2+(\frac{(2n+1)\pi}{\beta}-i\frac{\log b}{\beta})^2\right]^\lambda},
\end{cases}
\end{eqnarray*}
where $il_{\rm E}=l_0$. Take the right contour for example, as $\beta\gg1$, the factor $\frac{ab}{ab+1}$ becomes unity; the integral and the summation can be expressed as 
\begin{figure}[t]
\begin{center}
   \subfigure[]{\includegraphics[width=5cm]{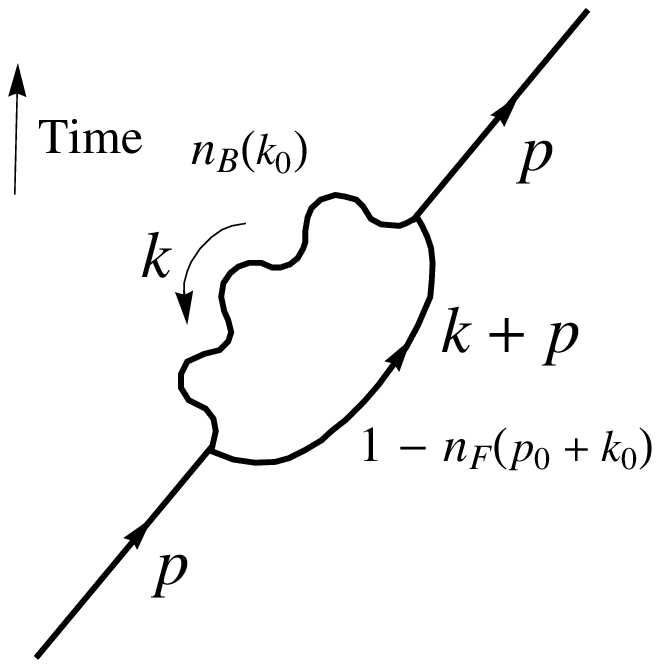}}
   \hspace*{0.07\textwidth}
   \subfigure[]{\includegraphics[width=5cm]{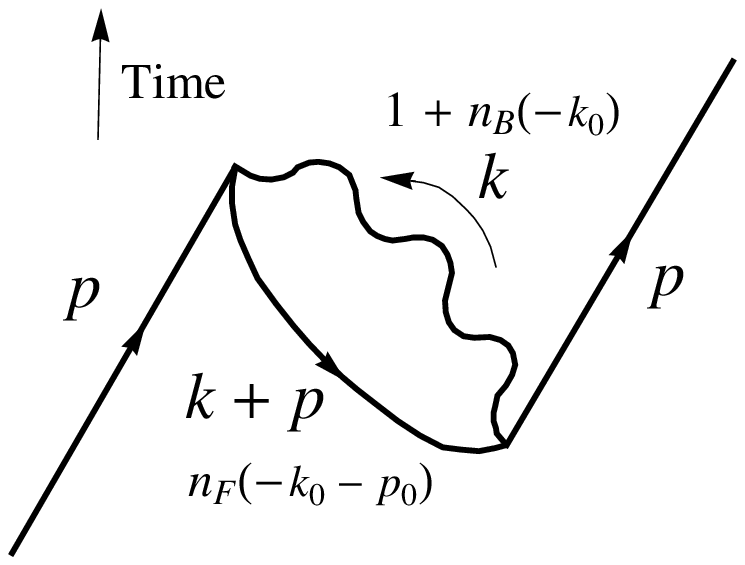}}
   \hspace*{0.0\textwidth}
\caption{\small   One-loop self-energy Feynman diagrams of an electron for different time ordering of the interaction vertices: (a) An electron annihilates first and creates another electron and a photon, later the two annihilate and an electron is created, the probability for this process is proportional to the product of $1-n_{\rm F}(k_0+q_0)$ and $n_{\rm B}(k_0)$. (b) An electron disappears due to the annihilation of a positron and a photon in the background, and its momentum has been carried away by another electron that is produced earlier. The probability is proportional to the product of $1+n_{\rm B}(-k_0)$ and $n_{\rm F}(-k_0-q_0)$.    } 
  \label{Fig:feynSEelectron}
\end{center}
\end{figure}
\begin{eqnarray*}
&&\hspace{0cm}=
i\int^{\Lambda_t}_{-\Lambda_t} d l_{\rm E}\int d^3 {\bf l} \frac{(-1)^\lambda}{(l^2_{\rm E}+{\bf l}^2+\Delta)^\lambda}+i\int^{\frac{2\pi N_{\rm max}}{\beta}}_{-\frac{2\pi N_{\rm max}}{\beta}}d l_{\rm E}\int d^3{\bf l}\frac{(-1)^{\lambda+1}}{\left[\Delta+{\rm l}^2+(l_{\rm E}-i\frac{\log a}{\beta})^2\right]^\lambda}.
\end{eqnarray*}
As discussed before, the cutoff  is chosen as $\Lambda_t=\frac{2\pi N_{\rm max}}{\beta}$, which is the same as the cutoff of 3-momentum, $\Lambda$. 
As for $\lambda=2$, the corresponding integral of the segments (A) and (B), similar to the contour in fig. \ref{Fig:threshold}, is
\begin{eqnarray*}
&&\hspace{-1cm}
=2\pi i\left[\left(-\frac{1}{\Lambda\Lambda_t}+\frac{2\Lambda\Lambda_t}{(\Lambda^2+\Lambda^2_t)^2}+\frac{\Lambda_t}{\Lambda(\Lambda^2+\Lambda^2_t)}-\frac{1}{\Lambda_t^2}\tan^{-1}\left(\frac{\Lambda}{\Lambda_t}\right)\right)\frac{\log^2a}{\beta^2}
+O\left(\frac{\log^4 a}{\beta^4}\right)\right]
\end{eqnarray*}
The result is suppressed by the cutoffs of the energy and the momentum, as well as (C) and (D), and approaches zero while taking their values to infinities; except the imaginary part if there is a branch cut.

\subsubsection{Imaginary-time}
The self-energy of an electron is calculated according to the Lagrangian of fermions in eq. (\ref{Lfermion}) and the interaction Lagrangian in eq. (\ref{Linteraction}). 
As in photon's self-energy, the fermion is assigned with a 4-momentum, $p^\mu_n\,\left(=(i\omega_n,{\bf p})\right)$, where $\omega_n=\frac{n\pi}{\beta}$ and $n$ is an odd integer. 
 Its self-energy takes the form:
\begin{eqnarray*}
\Xi \,(p^2_n)=\slashed{p}_n\Xi_V(p^2_n)+m_{\rm f}\,\Xi_S(p^2_n).
\end{eqnarray*} 
Similar to the procedures that are taken for the real-time case, as for the gauge, $\zeta=1$,
\begin{eqnarray*}
\Xi (p^2_n)=\frac{1}{\beta}\sum_m\int\frac{d^3{\bf k}}{(2\pi)^3}(e\gamma^\mu)  \frac{1}{(\slashed{p}_n+\slashed{k}_m)-m_{\rm f}}(e\gamma^\nu)\frac{-1}{k^2_m}\left.g_{\mu\nu}\right..
\end{eqnarray*}
After contracting with the vector,  $\slashed{p}_n$, and use the relation:
\begin{eqnarray*}
\Xi_V(p^2_n)=\frac{1}{4p^2_n}{\rm Tr}[\slashed{p}_n
\Xi] &{\rm and}&
\Xi_S(p^2_n)=\frac{1}{4m_{\rm f}}{\rm Tr}[
\Xi].
\end{eqnarray*}
The loop momentum is $k_m\,(=(i\omega_m,{\bf k}))$.
The function of the self-energy, $\Xi_V(p^2_n)$, can be obtained after taking the projection onto 4 momentum, $\slashed{p}_n$, and applying the Feynman parametrization: 
\begin{eqnarray}
\Xi_V(p^2_n)
&&\hspace{-0.0cm}=\frac{\alpha}{2\pi}\int^1_0 d\alpha_1 (1-\alpha_1)\frac{2\pi}{\beta}\sum_{m}\frac{1}{[(\frac{2\pi m}{\beta}+i\alpha_1p_n^0)^2+\Delta(p^2_n)]^{1/2}},
\end{eqnarray}
where the shifted loop momentum, $K^\mu_n=k^\mu_n+\alpha_1 p^\mu_n$, and $\alpha_i\,(i=1,2,3)$ are the Feynman parameters. 
As the factor $\beta$ gets large, the summation $\frac{2\pi}{\beta}\sum$ is approximated by the integral $\int d\omega$ from $-\Lambda_t$ to $\Lambda_t$. It becomes
\begin{eqnarray}\label{sigmaV}
&&\hspace{-1cm}\Xi_V(p^2_n)
\label{xiv}
=\frac{\alpha}{4\pi}\left(2\log \Lambda_t+2\log 2- \log \Delta(p^2_n) \right).
\end{eqnarray}
The self-energy at an infinite temperature can be expressed as
\begin{eqnarray}\label{BsigmaV}
\Xi_{0V}(p^2_n)&=&
\frac{\alpha}{2\pi}\int d\alpha_1 d\alpha_2\delta(\alpha_1+\alpha_2-1)(1-\alpha_1)\sum_{m}\frac{1}{|m|}\nonumber
=\frac{\alpha}{4\pi}\left(2\log N_{\rm max}+2\gamma_{\rm E}\right),
\end{eqnarray}
where the integer, $m$, goes from $\pm 1$ to $\pm N_{\rm max}$ and $\gamma_{\rm E}\,(=0.577216...)$ is the Euler-Mascheroni constant. As for the other function of the self-energy,  
\begin{eqnarray}
\Xi_{0S}(p_n^2)
 \label{xis}
&=&-\frac{\alpha}{\pi}\int^1_0 d\alpha_1\frac{2\pi}{\beta_0}\sum_{m}\frac{1}{[(\frac{2\pi m}{\beta_0}+i\alpha_1p^0_n)^2+\Delta(p^2_n)]^{1/2}}.
\end{eqnarray}
Similarly, the scalar part of the self-energy at infinite temperature can be expressed as
\begin{eqnarray*}
\Xi_{0S}(p^2_n)&=&-
\frac{\alpha}{\pi}\int^1_0 d\alpha_1 \sum_{m}\frac{1}{|m|}
=-\frac{\alpha}{\pi}\left(2\log N_{\rm max}+2\gamma_{\rm E}\right),
\end{eqnarray*}
when $N_{\rm max}$ is extremely large. From eq. (\ref{newIT}), we could have the renormalized self-energy functions for the imaginary-time as follows
\begin{eqnarray*}
\hat{\Xi}_V\equiv\Xi_V(\beta, p^2_n)-\Xi_{0V},\,\,{\rm and}\hspace{.4cm}
\hat{\Xi}_S\equiv\Xi_S\,(\beta, p^2_n)-\Xi_{0S}.
\end{eqnarray*} 
from eq. (\ref{sigmaV}) and (\ref{BsigmaV}), let $\Lambda_t=\frac{2\pi N_{\rm max}}{\beta}$,
\begin{eqnarray}
\hat{\Xi}_V(\beta, p^2_n)&=&\frac{\alpha}{4\pi}\left(2\log \frac{2\pi}{\beta}+2\log 2-2\gamma- \log \Delta(p^2_n) \right),\nonumber\\
\hat{\Xi}_S(\beta, p^2_n)&=&-\frac{\alpha}{\pi}\left(2\log \frac{2\pi}{\beta}+2\log 2-2\gamma- \log \Delta(p^2_n) \right).\label{xivxis}
\end{eqnarray} 
The dependence of $\beta$ in the above results is related to the renormalization group equations, and will be further explained in Section \ref{rgeq}. We combine the tree-level and the one-loop contribution 
\begin{eqnarray*}
&&\hspace{-0.8cm}\frac{1}{i\omega_n\gamma^0-\slashed{\bf p}-m_{\rm f}}+\frac{1}{i\omega_n\gamma^0-\slashed{\bf p}-m_{\rm f}}
(\hat{\Xi})\frac{1}{i\omega_n\gamma^0-\slashed{\bf p}-m_{\rm f}}+\dots
=
\frac{1}{i\omega_n\gamma^0-\slashed{\bf p}-m_{\rm f}-\hat{\Xi}}.
\end{eqnarray*}
We may approximate the self-energy as $\hat{\Xi} (\omega_n, |{\bf p}|) \approx  \left({i\omega_n}\gamma_0-\slashed{\bf p}\right)\hat{\Xi}_V(0,|{\bf p}|)+m_{\rm f}\hat{\Xi}_S(0,|{\bf p}|) $, since the self-energy functions $\hat{\Xi}_V(\omega_n,|{\bf p}|)$ and $\hat{\Xi}_S(\omega_n,|{\bf p}|)$ in eq. (\ref{xivxis}) have only weak dependence on the Matsubara frequency. The propagator for the imaginary-time may be derived to
\begin{eqnarray}
\approx\frac{1}{i\omega_n\gamma^0-\slashed{\bf p}-m_{\rm f}-(\left({i\omega_n}\gamma^0-\slashed{\bf p}\right)\hat{\Xi}_V+m_{\rm f}\hat{\Xi}_S)}
=\frac{1+\hat{\Xi}_V(0,|{\bf p}|)}{(i\omega_n\gamma^0-\slashed{\bf p})-m_{\rm f}(1+\hat{\Xi}_V+\hat{\Xi}_S)},
\label{effmassF}
\end{eqnarray}
where the effective mass of the fermion is defined as $m_{\rm f,\rm eff}=m_{\rm f}\left(1+(\hat{\Xi}_V+\hat{\Xi}_S)(0,|{\bf p}|)\right)$. One thing needs to emphasize is that this effective mass does not come into the play of the traditional propagator $1/(\slashed{p}-m_{\rm f})$ since in the summation of $1/(i\omega_n-\xi^{\rm eff}_{\bf p})$ over the Matsubara frequencies of eq. (\ref{diracprop}) the effective mass, $m_{\rm eff}$, would just add corrections to the density function. The pole of the traditional propagator is not shifted from the correction of the imaginary-time. From the previous section discussing the corrections for the real-time, no real part is generated, therefore a fermion mass defined by the pole of its propagator will always be a fixed value.  As we have seen from the approach of partition function in  eq. (\ref{Lagrangian}) and eq. (\ref{Lagrangian2}), such a replacement could be made, $\frac{1}{i\omega_n\gamma^0-\slashed{\bf p}-m_{\rm f}}\rightarrow \frac{1}{(i\omega_n-\xi_{\bf p})\gamma^0}$. The one-loop corrected propagator for the imaginary-time can be replaced by 
\begin{eqnarray*}
\hspace{0cm}&&\frac{1+\hat{\Xi}_V}{i\omega_n\gamma^0-\slashed{\bf p}-m_{\rm f,eff}}\rightarrow\frac{1+\hat{\Xi}_V}{(i\omega_n-\xi^{\rm eff}_{\bf p})\gamma^0}.
\end{eqnarray*}
The renormalized amplitude to a fermion field, $Z_2$, can be defined as 
\begin{eqnarray}
Z_2\equiv1+\hat{\Xi}_V(0,|{\bf p}|).\label{Z2}
\end{eqnarray} 
The amplitude will give rise to the renormalization group equation with respective to the variation of the variable, $\beta$, and will be discussed in Section \ref{rgeq}.

\subsection{Vertex correction}
\subsubsection{Real-time}
The computation of the vertex corrections for the real-time is similar to those in the previous sections.
The nonzero contribution is the imaginary part, while the real part is zero along the rectangular contour similar to that in fig. \ref{Fig:threshold} (b). Here the derivations of the loop integrals and the residues of the poles from the three density functions are shown. The corresponding Feynman diagram showing the incoming, outgoing and loop momenta are illustrated in fig. \ref{Fig:vertexC} (a). The correction to the vertex, $\delta \Gamma^\mu(p',p)$, is derived from
\begin{eqnarray*}
\hspace{-0.8cm}\delta \Gamma^\mu(p',p)
&=&\int_\otimes\frac{d^4k}{(2\pi)^4}\frac{-ig_{\nu\rho}}{(k-p)^2+i\varepsilon}\bar{u}(p')(-ie\gamma^\nu)
\frac{i(\slashed{k'}+m)}{k'^2-m^2_{\rm f}+i\varepsilon}\gamma^\nu \frac{i(\slashed{k}+m)}{k^2-m^2_{\rm f}+i\varepsilon}
(-ie\gamma^\nu)u(p) \\
&&\hspace{3.8cm}\times n_{\rm B}\left((k-p)\cdot u\right)\left\{1-n_{\rm F}((k+q)\cdot u)\right\}\left\{1-n_{\rm F}(k\cdot u)\right\},\\
&=&4ie^2\int_\otimes\frac{d^4l}{(2\pi)^4}\int^1_0dxdydz\delta(x+y+z-1)\\
&&\times\bar{u}(p')\left[\gamma^\mu\cdot\left(-\frac{1}{2D^2}-\frac{\Delta}{2D^3}+(1-x)(1-y)\frac{q^2}{D^3}+(1-4z+z^2)\frac{m^2_{\rm f}}{D^3}\right)\right.\nonumber\\
&&\left.\hspace{.5cm}+\frac{i\sigma^{\mu\nu }q_\nu}{2m_{\rm f}D^3}(2m^2_{\rm f}z(1-z))\right]u(p)\frac{-1}{e^{\beta (l\cdot u)-\log c}-1}\frac{1}{e^{-\beta l\cdot u-\log a}+1}\frac{1}{e^{-\beta l\cdot u+\log b}+1}.
\end{eqnarray*}
where $D=l^2-\Delta+i\varepsilon$, $\Delta=-xyq^2+(1-z)^2m^2_{\rm f}$ and $a=e^{\beta (1-\alpha_2)q\cdot u +\beta\alpha_3 (p\cdot u)}$, $b=e^{\beta \alpha_2 (q\cdot u )-\beta\alpha_3 (p\cdot u)}$ and $c=e^{\beta \alpha_2(q\cdot u) +\beta(1-\alpha_3 )p\cdot u}$. For a convenient purpose, we may write down their products: $ac=e^{\beta (p+q)\cdot u}$, $c/b=e^{\beta p\cdot u}$ and $ab=e^{\beta q\cdot u}$, which are constants without the dependence of the loop momentum $l$.
As $\beta \gg 1$, $ac$, $c/b$ and $ab\gg 1$.

A general form of the loop integrals for the vertex corrections, $I_{\rm VE,\lambda}$, of t\begin{figure}[t]
\begin{center}
   \subfigure[]{\includegraphics[width=5cm]{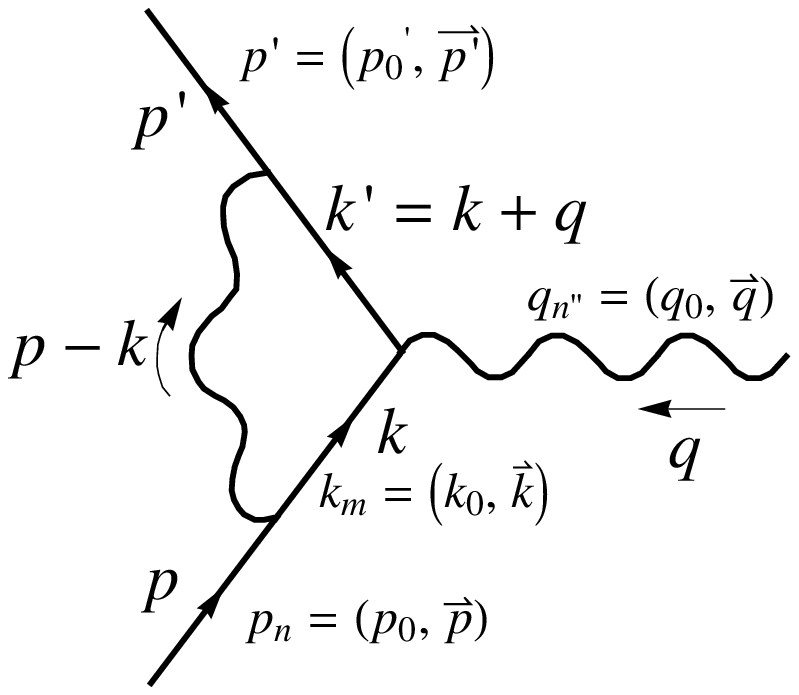}}
  \hspace*{0.0\textwidth}
   \subfigure[]{\includegraphics[width=5cm]{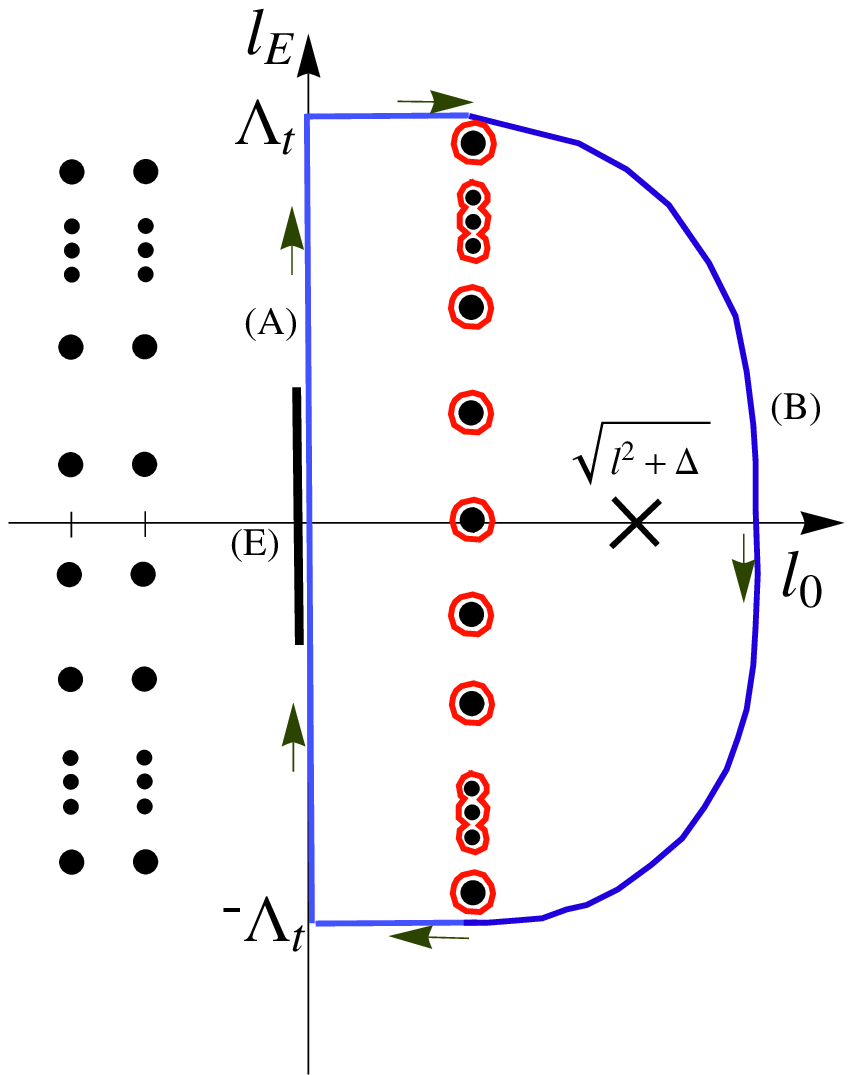}}
\caption{\small  (a) The Feynman diagram for the vertex correction is illustrated, and the momentum of each particle is shown for either the imaginary-time or the real-time. (b)
The contour for the vertex correction is represented by the blue lines with two segments (A) and (B). The branch cut, for $\Delta<0$, is denoted as (E). On the both side of the imaginary axis, the dots indicate the poles from  three density functions.  } 
  \label{Fig:vertexC}
\end{center}
\end{figure}he loop momentum, $k^\mu$, for the power, $\lambda$, of the denominator is written as
\begin{eqnarray*}
&&\hspace{-0cm}I_{\rm VE, \lambda}=\int_\otimes \frac{d^4 l}{(l^2-\Delta)^\lambda}n_{\rm B}\left((k-p)\cdot u\right)\left\{1-n_{\rm F}((k+q)\cdot u)\right\}\left\{1-n_{\rm F}(k\cdot u)\right\}\\
&&\hspace{-3cm}=
\begin{cases}
i\oint_{\rm RC} d l_{\rm E}\int d^3 {\bf l} \frac{(-1)^\lambda}{(l^2_{\rm E}+{\bf l}^2+\Delta)^\lambda}\frac{1}{e^{i\beta l_{\rm E}-\log c}-1}\frac{1}{e^{-i\beta l_{\rm E}-\log a}+1}\frac{1}{e^{-i\beta l_{\rm E}+\log b}+1}+2\pi i\frac{ac^2}{(b+c)(1+ac)\beta}\sum_{n\in {\rm I}}\int d^3{\bf l}\frac{(-1)^\lambda}{\left[\Delta+{\rm l}^2+(\frac{2n\pi}{\beta}-i\frac{\log c}{\beta})^2\right]^\lambda}\\
i\oint_{\rm LC} d l_{\rm E}\int d^3 {\bf l} \frac{(-1)^\lambda}{(l^2_{\rm E}+{\bf l}^2+\Delta)^\lambda}\frac{1}{e^{i\beta l_{\rm E}-\log c}-1}\frac{1}{e^{-i\beta l_{\rm E}-\log a}+1}\frac{1}{e^{-i\beta l_{\rm E}+\log b}+1}-2\pi i\frac{abc}{(b+c)(1-ab)\beta}\sum_{n\in {\rm I}}\int d^3{\bf l}\frac{(-1)^\lambda}{\left[\Delta+{\rm l}^2+(\frac{(2n+1)\pi}{\beta}-i\frac{\log b}{\beta})^2\right]^\lambda}\\
\hspace{8.65cm}-2\pi i\frac{ac}{(ac+1)(ab-1)\beta}\sum_{n\in {\rm I}}\int d^3{\bf l}\frac{1}{\left[\Delta+{\bf l}^2+(\frac{(2n+1)\pi}{\beta}-i\frac{\log a}{\beta})^2\right]^\lambda}.
\end{cases}
\end{eqnarray*}
Even though on the left-hand side there are twice number of the poles more than those on the right, the  coefficient of the second set of residues, $\frac{ac}{(ac+1)(ab-1)}$, diminishes as $\beta\gg1$. In fact, the sum of the two coefficients for the residues on the left is equal to the that on the right,
\begin{eqnarray*}
\frac{ac^2}{(b+c)(1+ac)}=\frac{abc}{(b+c)(1-ab)}+\frac{ac}{(ac+1)(ab-1)},
\end{eqnarray*}
so the two sets of residues on one side are comparable to the one on the other. As for the results of the above calculation, the same logic applies as those for the self-energy integrals, the only nonzero contribution is the imaginary part due to the branch cut, (E) in fig. \ref{Fig:vertexC}, under the condition $\Delta<0$. 
\subsubsection{Imaginary-time}\label{vertexIm}
The Feynman diagram for the vertex loop correction is shown in fig. \ref{Fig:vertexC} (a), including the related external and internal momenta.  In the calculations of the vertex correction, there is one thing needed to be clarified. As the Gordon decomposition is applied, the on-shell propertyties of the external momenta $p$ and $p'$ have to be used through $(\slashed{p}-m_{\rm f})u(p)=(\slashed{p'}-m_{\rm f})u(p')=0$. However, in the imaginary-time formalism, the external energy-momentum $p_n$ and $p'_{n'}$ are in the forms of ${p}^\mu_{n}=(i\omega_{n},{\bf p})$ and ${p'}^\mu_{n'}=(i\omega_{n'},{\bf p}')$ when they are in the loop. Thus, as the operator ${p}_n$ acts on the physical states $u(p)$, the analytic continuation has to be applied $p_n\rightarrow p$, as well as $p'_{n'}\rightarrow p'$, outside the loop. That means that the energy $p_0$ has to rotate from the imaginary-axis to the real axis after the loop integral is completed. It is similar to the Wick rotation. When the loop momentum is  rotated from the real axis to the imaginary one in the conventional way, the external momenta are in fact carried  by the shifted loop momentum along the Wick rotation implicitly. 
In addition, for a convenient reason in the calculations, the denominator, which is obtained after the Feynman parametrization, is denoted as  $D=(p_m+yq_{n''}-xp_n)^2-{\bf l}^2-\Delta(q^2_{n''})$, and the factor $\Delta(q^2_{n''})=-xy q^2_{n''}+(1-z)^2m^2_{\rm f}+zm_\gamma^2$, where $m_\gamma$ is the mass of photon. The variables $x$, $y$ and $z$ are the Feynman parameters.  The nonzero photon
mass $m_\gamma$ is given to avoid the infrared divergences at this stage. 
The correction to the vertex with a fermionic frequency $\omega_m=\frac{2\pi}{\beta}(m+\frac{1}{2})$ is
\begin{eqnarray}
&&\hspace{-.8cm}\delta \Gamma^\mu(\{p'_{n'}{\bf }\},\{p_n{\bf }\})\nonumber\\
&&\hspace{-1.5cm}=\frac{1}{\beta}\sum_{m}\int\frac{d^3{\bf k}}{(2\pi)^3}\frac{g_{\nu\rho}}{(k_m-p_n)^2}\bar{u}(p'_{n'})(e\gamma^\nu)
\frac{(\slashed{k}_m+\slashed{q}_{n''}+m)}{(k_m+q_{n''})^2-m^2_{\rm f}}\gamma^\mu \frac{(\slashed{k}_m+m)}{k^2_m-m^2_{\rm f}}
(e\gamma^\rho)u(p_n),\nonumber
\\
&&\hspace{-1.5cm}=\frac{4e^2}{\beta}\sum_{m}\int^1_0dxdydz\delta(x+y+z-1)\bar{u}(p'_n)\left[\gamma^\mu\cdot\left(-\frac{1}{16\pi\left[(\omega_m-iyq_{0''}+ixp_0)^2+\Delta(q^2_{n''}{\bf })\right]^{1/2}}\right.\right.\nonumber\\
&&\hspace{-1.0cm}\left.
-\left(\frac{\Delta(q_{n''}{\bf })}{2}-(1-x)(1-y)q^2_{n''}-(1-4z+z^2)m^2_{\rm f}\right)\frac{1}{32\pi \left[(\omega_m-iyq_{0''}+ix p_0)^2+\Delta(q^2_{n''}{\bf })\right]^{3/2}}\right)\nonumber\\
&&\hspace{-1.cm}\left.+\frac{i\sigma^{\mu\nu }q_{{n''}\nu}}{2m_{\rm f}}(2m^2_{\rm f}z(1-z)) \frac{1}{32\pi \left[(\omega_m-iyq_{0''}+ixp_0)^2+\Delta(q^2_{n''}{\bf })\right]^{3/2}}\right]u(p_n).\label{vertexqed}
\end{eqnarray}
Under the condition of $\beta\gg 1$, the summation over, $\frac{2\pi}{\beta}\sum_m$, is replaced by an integration, $\int d\omega$:\vspace{-.2cm}
\begin{eqnarray}
\delta \Gamma^\mu(\{p_{n'}\},\{p_n\})
&=&\frac{\alpha}{\pi}\int^1_0dxdydz\delta(x+y+z-1)\bar{u}(p'_n)\left[\gamma^\mu\cdot\left(-\log 2\Lambda+\frac{1}{2}\log \Delta(q^2_{n''})\right.\right.\nonumber\\
&&\left.-\left(\frac{\Delta(q^2_{n''})}{2}-(1-x)(1-y)q^2_{n''}-(1-4z+z^2)m^2_{\rm f}\right)\frac{1}{2\Delta(q^2_{n''})}\right)\nonumber\\
&&\hspace{0cm}\left.+\frac{i\sigma^{\mu\nu }q_{{n''}\nu}}{2m_{\rm f}}(2m^2_{\rm f}z(1-z)) \frac{1}{2\Delta(q^2_{n''})}\right]u(p_n).\label{vertex1}
\end{eqnarray}
Remember that we analytically continue $u(p_n)\rightarrow u(p)$ and $u(p'_{n'})\rightarrow u(p')$, while the Gordon decomposition is used in the above. We may retrieve the form factors from above by defining
\begin{eqnarray*}
 F_1(q_{n''})&=&-\frac{\alpha}{2\pi}\left(\log N_{\rm max}+\log 4\pi-\log \beta+\frac{5}{4}\right)\\&&+\frac{\alpha}{\pi}\int^1_0dz\int^{1-z}_0dx\left(\frac{1}{2}\log \Delta(q^2_{n''})\right.\left.
\hspace{0cm}+\frac{zq^2_{n''}+2(1-3z+z^2)m^2_{\rm f}}{\Delta(q^2_{n''})}\right),\\
 F_2(q_{n''})&=&\frac{\alpha}{\pi}\int^1_0 dz \int^{1-z}_0 dx \frac{m^2_{\rm f}z(1-z)}{\Delta(q^2_{n''})}.
\end{eqnarray*}
The corresponding form factors $F_{1}(p^2_{n''})$ and $F_{2}(p^2_{n''})$ in the limit of high temperature are  obtained by ignoring the terms that are not related to $1/\beta_0$ since they are negligible as compared to those with $1/\beta_0$, when $\beta_0 \rightarrow 0$. In addition, the Matsubara frequencies from the imaginary part of external 4-momenta, $\omega_n$ and $\omega_{n''}$, can also be ignored as the sum of loop frequency $\omega_m$ is going up to a very large number $N_{\rm max}$; it will be explained more by eq. (\ref{shiftingeffect}). The vertex correction at $\beta_0$ is
\begin{eqnarray}
&&\hspace{-2cm}\delta \Gamma_0^\mu(\{p_{n'}\},\{p_n\})=\frac{\alpha(2\pi)}{\pi\beta_0}\sum_{m}\int^1_0dxdydz\delta (x+y+z-1)\bar{u}(p'_n)\left[\gamma^\mu\cdot\left(-\frac{1}{2\left|\omega_m+y\omega_{n''}-x\omega_n\right|}\right.\right.\nonumber\\
&&\hspace{1cm}\left.-\left(\frac{\Delta(p^2_{n''})}{2}-(1-x)(1-y)q^2_{n''}-(1-4z+z^2)m^2_{\rm f}\right)\frac{1}{4 \left|\omega_m+y\omega_{n''}-x\omega_n\right]^{3}}\right)\nonumber\\
&&\hspace{1cm}\left.+\frac{i\sigma^{\mu\nu }q_{{n''}\nu}}{2m_{\rm f}}(2m^2_{\rm f}z(1-z)) \frac{1}{4 \left|\omega_m+y\omega_{n''}-x\omega_n\right|^{3}}\right]u(p_n),\nonumber\\
&&\hspace{.7cm}=-\frac{\alpha}{2\pi}\bar{u}(p'_n)\left.\gamma^\mu\cdot\left(\log N_{\rm max}+C_E\right)\right.
u(p_n).\label{vertex2}
\end{eqnarray}
It is obtained from the sum, $\sum_{m=1}^{N_{\rm max}}\frac{1}{m+\frac{1}{2}}\left(=\log N_{\rm max}+C_E\right)$.  
The terms with the denominator, $\left|\omega_m+y\omega_{n''}-x\omega_n\right|^3$, diminish due to the fact that, besides the constants, $\alpha$, $\pi$ and so on, it contains the factor, $\beta^2_0$, and the sum over the integer $m$, $\sum_{m=1}^{\infty}\frac{1}{(m+\frac{1}{2})^3}=0.414398...$, is a finite number. Thus, it becomes zero after taking the limit of $\beta_0\rightarrow 0$. This is similar to the case of $I_4$ explained in Section \ref{phi3}.
During the above derivation, the shift of the frequency, $\omega_m\rightarrow\omega_m+y\omega_{n''}-x\omega_{n'}$, is applied, then the summation range changes from the symmetric one, $\sum^{N_{\rm max}}_{-N_{\rm max}}$ to an asymmetric one, $ \sum^{N_{1}}_{-N_{2}}$, where $N_{1,2}=N_{\rm max}\pm\Delta N$, where the difference of the two cutoffs is $\Delta N=y\omega_{n''}-x\omega_{n'}$. The terms in eq. (\ref{vertex2}) involving the cutoffs is
\begin{eqnarray}
-\frac{\alpha}{4\pi}\left(\log N_{1}+\log N_{2}\right)=-\frac{\alpha}{2\pi}\left\{\log N_{\rm max}+\frac{1}{2}\log \left(1+\frac{\Delta N}{N_{\rm max}}\right)+\frac{1}{2}\log \left(1-\frac{\Delta N}{N_{\rm max}}\right)\right\}.\hspace{.5cm}\label{shiftingeffect}
\end{eqnarray}
 As long as $N_{\rm max}\gg\Delta N$, the shifting effect can be ignored. Therefore the form factors in the limit of small $\beta_0$ are 
\begin{eqnarray*}
F_{0,1}(q^2_{n''})=-\frac{\alpha}{2\pi}\left(\log N_{\rm max}+ C_E\right), \,\, {\rm and}\hspace{.5cm}F_{0,2}(q^2_{n''})=0.
\end{eqnarray*}
The renormalized form factors, which can be defined from eq. (\ref{newIT}), are given as below with a new reference points at $\beta_0=0^+$:
\begin{eqnarray*}
\hat{F}_{1}(q^2_{n''})&\equiv&F_{1}(q^2_{n''})-F_{0,1}(q^2_{n''}),\\
&=&-\frac{\alpha}{2\pi}(C_E+\log 4\pi-\log \beta+\frac{5}{4})+\frac{\alpha}{\pi}\int^1_0dz\int^{1-z}_0dx\left(\frac{1}{2}\log \Delta(q^2_{n''})\right.\\
&&\hspace{6cm}\left.
+\frac{zq^2_{n''}+2(1-3z+z^2)m^2_{\rm f}}{\Delta(q^2_{n''})}\right),\\
\hat{F}_{2}(q^2_{n''})&\equiv&F_{2}(q^2_{n''})-F_{0,2}(q^2_{n''})=\frac{\alpha}{\pi}\int^1_0 dz \int^{1-z}_0 dx \frac{z}{1-z}=\frac{\alpha}{2\pi}.
\end{eqnarray*}
 The form factor $\hat{F}_{1}(q^2_{n''})$ also possesses the infrared divergence.  From the result of $\hat{F}_{2}(q^2_{n''})$, the ratio $\frac{g-2}{2}=\frac{\alpha}{2\pi}$, which is consistent with what QED predicts for the one-loop level. It is known that there is an infrared divergence in the function $\hat{F}_{1}(q^2_{n''})$, and a tiny photon mass $m_\gamma$ is given to avoid the catastrophe temporarily. In a subsequent work \cite{huang14b} related to this issue, the infrared divergence can be regularized by giving a non-zero minimal Matsubara frequency $\big(\frac{2\pi}{\beta}\big)^2$ in the loop momentum. The added frequency is to emphasize the discreteness of the Matsubara frequency, since it is treated as a continuous variable in our calculations as $\beta\gg 1$. And in the case of QED presented here for the vertex correction, the loop frequency is not allowed to be zero for any fermionic propagator.

\section{Renormalization group equations}
\label{rgeq}
\subsection{Renormalized amplitude $Z_2$ and electron's effective mass $m_{\rm eff}$}
From one-loop corrections for the imaginary-time in the previous section, the corrections to the amplitudes of the fields are obtained. 
 The renormalized constant of a fermion is defined from eq. (\ref{Z2}) before being taken into the limit of $\beta\gg1$: 
\begin{eqnarray*}
Z_2&=&1+\hat{\Xi}_V|_{\omega_0=0},\\
&=&1+\frac{\alpha}{2\pi}\int^1_0 d\alpha_1 (1-\alpha_1)\frac{2\pi}{\beta}\sum_{m}\frac{1}{[(\frac{2\pi m}{\beta})^2+\Delta(0,{\bf p})]^{1/2}}-\frac{\alpha}{4\pi}\sum_m\frac{1}{|m|}.
\end{eqnarray*}
We may be interested in the change of the renormalized amplitude with respect to the temperature. Take derivative of the above formula:
\begin{eqnarray*}
\frac{\partial \log Z_2}{\partial \log \beta}
&=&-\frac{\alpha}{2\pi}\int^1_0 d\alpha_1 (1-\alpha_1)\left.\frac{2\pi}{\beta}\right.\sum_{m}\frac{\Delta(0,{\bf p})}{[(\frac{2\pi m}{\beta})^2+\Delta(0,{\bf p})]^{3/2}}.
\end{eqnarray*}
While $\beta\gg1$, the summation becomes an integration $\frac{2\pi}{\beta}\sum_m\rightarrow \int d\omega$:
\begin{eqnarray*}
\left.\frac{\partial \log Z_2}{\partial \log \beta}\right|_{\beta\gg 1}
&=&-\frac{\alpha}{2\pi}\int^1_0d\alpha_1(1-\alpha_1)\int^\infty_{-\infty} d\omega\frac{\Delta(0,{\bf p})}{[\omega^2+\Delta(0,{\bf p})]^{3/2}}=-\frac{\alpha}{2\pi}.
\end{eqnarray*}
We may change the variable $\beta$ to the temperature, $T\,(=\frac{1}{\beta})$, so that 
\begin{eqnarray*}
\left.\frac{\partial \log Z_2}{\partial \log T}\right|_{T\simeq 0}
&=&\frac{\alpha}{2\pi}.
\end{eqnarray*}
This is the same as the renormalization group equation that is obtained from $\overline{MS}$ scheme \cite{grozin} for the  renormalized amplitude of a fermion with the renormalization scale, usually denoted as $\mu$, is replaced by the temperature. As we shall see from below, all of the renormalization group equations to the next-to-leading order in the field theory are the same as those derived from the imaginary-time formalism. in the other limit, $\beta\rightarrow 0$, the term $\Delta(0,{\bf p})$ in the denominator can be ignored before the summation is done, so that
\begin{eqnarray*}
\left.\frac{\partial \log Z_2}{\partial \log \beta}\right|_{\beta\rightarrow 0}
&=&\lim_{\beta\rightarrow 0}-\frac{\alpha \zeta(3)}{2\pi}\frac{\beta^2\Delta}{(2\pi)^2}=0,
\end{eqnarray*}
and change the variable $\beta$ to temperature $T$
\begin{eqnarray*}
 \hspace{0.0cm} \left.\frac{\partial \log Z_2}{\partial \log T}\right|_{T\rightarrow \infty}
=0.
\end{eqnarray*}
In fact the above result for the limit of infinite value of $\beta$ can be obtained directly in eq. (\ref{xivxis}). 
Now consider the effective mass of a fermion, which is defined as $m_{\rm f, eff}=m\left(1+(\hat{\Xi}_V+\hat{\Xi}_S)(0,|{\bf p}|)\right)$, from eq. (\ref{xivxis}) and (\ref{effmassF}), take the derivative of the effective mass 
\begin{eqnarray*}
\frac{\partial \log m_{\rm eff}}{\partial \log \beta}&=&\frac{\alpha}{2\pi}\int^1_0 d\alpha_1 (1+\alpha_1)\left.\frac{2\pi}{\beta}\right.\sum_{m}\frac{\Delta(0,{\bf p})}{[(\frac{2\pi m}{\beta})^2+\Delta(0,{\bf p})]^{3/2}},
\end{eqnarray*}
so as $\beta \gg 1 $, transform the summation into an integral, we have 
\begin{eqnarray*}
\left.\frac{\partial \log m_{\rm eff}}{\partial \log \beta}\right|_{\beta\gg1}
&=&\frac{\alpha}{2\pi}\int^1_0 d\alpha_1 (1+\alpha_1)\int^\infty_{-\infty} d\omega\frac{\Delta(0,{\bf p})}{[\omega^2+\Delta(0,{\bf p})]^{3/2}}=6\frac{\alpha}{4\pi}.
\end{eqnarray*}
In the limit of $\beta\rightarrow 0$, the value is zero in a similar way to that of $Z_2$. We may change it into a function of temperature $T$,
\begin{eqnarray*}
\left.\frac{\partial \log m_{\rm eff}}{\partial \log T}\right|_{T\simeq 0}=-6\frac{\alpha}{4\pi},&{\rm and}\hspace{.5cm}&
\left.\frac{\partial \log m_{\rm eff}}{\partial \log T}\right|_{T\rightarrow \infty}=0.
\end{eqnarray*}
The result for the zero temperature is consistent as the renormalization equation, $\frac{\partial \log m_{\rm eff}}{\partial \log \mu}=-6\frac{\alpha}{4\pi}$, in field theory. As discussed in the previous section, this correction to the fermion's mass only goes to the corrected fermion density function and does not shift the pole of the propagator.  
\subsection{Renormalized amplitude $Z_3$ and charge renormalization constant $Z_\alpha$ }
In the limit of infinite $\beta$,  we may use the results from the previous sections for the discussions here.
For the photon's renormalized amplitude, it is defined from the one loop correction to the propagator in eq. (\ref{propcorr}),
\begin{eqnarray*}
Z_3&\equiv& 1-\hat{\Omega}_T(\beta,0,{\bf p}).
\end{eqnarray*}  
The corresponding renormalization group equations in the limit of infinite value of $\beta$, which can be easily obtained from eq. (\ref{pitpil}), as well as in the other limit of $\beta\rightarrow 0$, are
\begin{eqnarray*}
\hspace{-0cm}\left.\frac{\partial \log Z_3}{\partial \log \beta}\right|_{\beta\gg 1}=-\frac{8}{3}\frac{\alpha}{4\pi}, &{\rm and}\hspace{.5cm}& \hspace{-0cm}\left.\frac{\partial \log Z_3}{\partial \log \beta}\right|_{\beta\rightarrow 0}=0.
\end{eqnarray*}
Like the traditional renormalization constants defined in the field theory, we may also define the factor $Z_1=1+\hat{F}_{1}$ to describe the vertex correction. In a similar way, the correction to electron's charge
can be defined as $e_0=Z^{\mathsmaller{\frac{1}{2}}}_\alpha e$, and it is related to the other renormalized constants by $Z_\alpha=Z_1^2Z_2^{-2}Z_3^{-1}$. It is obvious that, in the framework of the imaginary-time, the Ward identity still holds in the same fashion, so that $Z_1=Z_2$ and  $Z_\alpha=Z^{-1}_3$. We may obtain
\begin{eqnarray*}
\hspace{-0cm}\left.\frac{\partial \log Z_\alpha}{\partial \log T}\right|_{T\simeq 0}
=-\frac{8}{3}\frac{\alpha}{4\pi},&{\rm and}\hspace{.5cm}&\hspace{-0cm}\left.\frac{\partial \log Z_\alpha}{\partial \log T}\right|_{T\rightarrow \infty}
=0.
\end{eqnarray*}
This is the same as the renormalization group equation, $\frac{\partial \log  Z_\alpha}{\partial \log \mu}=-\frac{8}{3}\frac{\alpha}{4\pi}$, from the $\overline{MS}$ scheme of renormalization. This may lead to a discussion of Landau pole \cite{gellmann,landau55}, which states that coupling constants become infinite in a finite momentum scale.
In the theory presented here, the renormalization group equations are identical with those predicted in the field theory, as the temperature of the vacuum is regarded as a parameter of scale, only in the low temperature limit.
The variation of couplings, with respect to a high $T$, vanishes, as shown in above. Therefore the electromagnetic coupling, $\alpha$, does not diverge over all range of temperature within this picture.
\section{Conclusion}\vspace*{-0cm}
In this paper, the vacuum is treated as a thermodynamical system, which is described by microscopic observables, and is applied on the calculations of radiative corrections as those in the field theory. The theory has been
constructed and compared with the conventional formalism of QED. It shows that, in a temperature close to zero, the mathematical formalisms coincide with those in the traditional theory, such as the propagators of the electron and the photon, and so on. The imaginary-time, $\tau$, is introduced with a finite range $(\beta_0,\beta)$, where $\beta_0(=0^+)$ is infinitesimal, in the construction of the partition functions; the field operators are expanded by the Matsubara frequencies and 3-momenta. The formalism can be separated into two parts for the real-time and the imaginary-time from the same partition function. The real-time evolution of the field is achieved by the analytic continuation of the variable $\tau$, by setting $\tau=it$, after Matsubara frequencies are summed. The one-loop radiative corrections have been calculated for the self-energy of the electron, the vacuum polarization of the photon, and the vertex correction. The results from the imaginary-time are always being real values, such as the self-energy; their UV divergences, instead of introducing counter terms, could be removed by subtracting the counterpart quantities, like corrections to the field amplitudes, masses, etc., at the reference point, $\beta_0= 0^+$. These values at a very high temperature lose the dependence of the external momenta, and are similar to the roles of the counter terms in the field theory. Those "counter terms" are automatically equipped in the theory from the lower integration bound of the imaginary-time, $\beta_0$, so the theory is intrinsically free of UV divergences.  
The correction to the self-energy of a fermion contribute to the propagator $1/(\slashed{p}_n-m)$, which only affects the density function and the field amplitude, and does not shift the pole of the propagator for the real-time. As for the self-energy of the photon, the imaginary-time corrections make a shift to photon's mass, which is defined as the pole of the propagator of the real-time. The one-loop mass correction to the photon is non-negligible and is canceled by its chemical potential, so that agreeable comparisons to the experimental results can be secured. The loop corrections from the real-time is found to be always finite and imaginary, because they comprise the conventional loop integrals and the residues from the density functions, which would cancel each other. 
The integrals along closed contours contribute nothing except the branch cuts, when the incoming momenta excesses the threshold. The resultant imaginary part is the same as those in the traditional Feynman integrals. The loop corrections from the real-time and the imaginary-time happen to account for the imaginary and the real parts of the calculations that are known in the field theory. As a result, it can be seen that this imaginary-time formalism in the low temperature limit is similar to the mathematical structure of the field theory. The consistency of the whole theory still needs more  efforts to check, but if there is any consistency problem in the theory it should not be like those in other imaginary-time theory for plasma because of their  mathematical similarities. 
Another important conclusion of this paper is the renormalization group effects. The Lagrangian densities are invariant under the scale transformation. From the Tolman-Ehrenfest relation that was discovered in the general relativity, the temperature determines the speed of time.  Similar discussions for all of four dimensions could also found in Wilson's approach in discussing the renormalization group in condensed  matter physics and the Hawking radiation in the curved space-time field theory. In the calculations provided here, the corrections to the renormalized amplitudes of the fields and fermion mass are given from the imaginary-time corrections; the renormalization group equations can be derived from them with respective to the variation of the temperature.   The amazing point of these computation is that it shows the the same renormalization coefficients at a temperature close to zero as those obtained in the field theory.  Besides, this theory for electrodynamic forces does not generate infinite values of couplings, the so-called Landau pole, as the variation of the coupling is found to vanish at a very high temperature. In addition to these derived results, the field theory in this imaginary-time formalism incorporates various concepts in the traditional field theory with those in the thermodynamics and the general relativity. The Einstein equation has been known before that it can be perceived in a thermodynamical perspective,  such as the Tolman-Ehrenfest relation. In the studies of the black hole, it is shown that the area of the event horizon  is proportional to the entropy of the black hole. Besides, the field theory in the curved space-time is usually studied in different vacuum states in which the space-time is under conformal transformations. They are combined in the construction of the imaginary-time field theory and make deep connections with the knowledge obtained in the particle physics.  To test its applicability on other physics of the vacuum, various effects, such as the Casimir effect, have been discussed in the following paper and many agreements can be proved. On the other hand, in the cosmology one important role that could be played by the vacuum is the dark energy. In another work, the cosmological constant can be computed without the interference of the divergence in the imaginary-time field theory through the DeWitt-Schwinger representation and the traditional calculation of the Casimir force. Both results give the same ratio for the equation of state, $w=-1$.

\appendix
\section{Functional approach}
\label{appA}
The following derivations are generalized from ref. \cite{shankar93}. The exponential factor of the above partition function can be divided into a product of $N$ components as follows 
\begin{eqnarray*}
e^{-\beta({\bf x})\mathcal{K}({\bf x})\Delta {\bf x}}&=&\lim_{N\rightarrow \infty}\left(e^{-(\beta({\bf x})/N) \mathcal{K}({\bf x})\Delta {\bf x}}\right)^N
=(1-\epsilon \mathcal{K} \Delta {\bf x})\cdot\cdot\cdot(1-\epsilon \mathcal{K} \Delta {\bf x}), 
\end{eqnarray*}
where $\epsilon=\beta/N$. Then, the partition function $Z$ becomes
\begin{eqnarray*}
Z&=&\prod_{\{{\bf x}\}}\int\langle-\psi^\dagger_1({\bf x})|(1-\epsilon \mathcal{K}({\bf x})\Delta {\bf x})|\psi({\bf x})_{N-1}\rangle
e^{-\psi^\dagger_{N-1}({\bf x})\psi_{N-1}({\bf x})\Delta {\bf x}}\langle\psi^\dagger_{N-1}({\bf x})|\\
&&\times(1-\epsilon \mathcal{K}({\bf x})\Delta {\bf x})|\psi_{N-2}({\bf x})\rangle
e^{-\psi^\dagger_{N-2}({\bf x})\psi_{N-2}({\bf x})}\langle\psi^\dagger_{N-2}({\bf x})| \cdot\cdot\cdot |\psi_2({\bf x})\rangle e^{-\psi^\dagger_{2}({\bf x})\psi_{2}({\bf x})\Delta {\bf x}}\langle\psi^\dagger_{2}({\bf x})|\\
&&\times(1-\epsilon \mathcal{K}({\bf x})\Delta {\bf x})|\psi_{1}({\bf x})\rangle
e^{-\psi^\dagger_{1}({\bf x})\psi_1({\bf x})\Delta {\bf x}}\prod_{i=1}^{N-1}d\psi^\dagger_id\psi_i,
\end{eqnarray*}
where we have used the definitions of the trace and the identity operator,
\begin{eqnarray*}
{\rm Tr}\,\Omega&=&\int \langle -\psi^\dagger({\bf x})|\Omega|\psi({\bf x})\rangle e^{-\psi^\dagger({\bf x})\psi({\bf x})\Delta {\bf x}}d\psi^\dagger({\bf x})d\psi({\bf x})\hspace{.3cm} {\rm and}\\
I&=&\int|\psi({\bf x})\rangle\langle\psi^\dagger({\bf x})|e^{-\psi^\dagger({\bf x})\psi({\bf x})\Delta {\bf x}}d\psi^\dagger({\bf x})d\psi({\bf x}).
\end{eqnarray*}
As $\epsilon$ is infinitesimal, an approximation can be made by the replacement
\begin{eqnarray*}
\langle\psi^\dagger_{i+1}({\bf x})|(1-\epsilon \mathcal{K}(\Psi^\dagger({\bf x}),\Psi({\bf x}))\Delta {\bf x})|\psi_i({\bf x})\rangle
&=&\langle\psi^\dagger_{i+1}({\bf x})|(1-\epsilon \mathcal{K}(\psi^\dagger_{i+1}({\bf x}),\psi_i({\bf x}))\Delta {\bf x})|\psi_i({\bf x})\rangle,\\
&=&e^{\psi^\dagger_{i+1}({\bf x})\psi_i({\bf x})}e^{-\epsilon \mathcal{K}(\psi^\dagger_{i+1}({\bf x}),\psi_i({\bf x}))\Delta {\bf x}}.
\end{eqnarray*}
 The additional conditions are given for the boundary values of the operators for $i$-index: $\psi^\dagger_N({\bf x})=-\psi^\dagger_1({\bf x})\,\,\, {\rm and }\,\,\, \psi_N({\bf x})=-\psi_1({\bf x})$.
With all of the formulas stated above, 
\begin{eqnarray*}
Z&=&\prod_{\{{\bf x}\}}\int\prod_{i=1}^{N-1}e^{\psi^\dagger_{i+1}({\bf x})\psi_i({\bf x})\Delta {\bf x}}e^{-\epsilon \mathcal{K}(\psi^\dagger_{i+1}({\bf x}),\psi_i({\bf x}))\Delta {\bf x} }e^{\psi^\dagger_{i}({\bf x})\psi_i({\bf x})\Delta {\bf x}}d\psi^\dagger_i({\bf x})d\psi_i(x),\\
&=&\prod_{\{{\bf x}\}}\int \prod_{i=1}^{N-1}\exp\left[\left(\frac{\psi^\dagger_{i+1}({\bf x})-\psi^\dagger_{i}({\bf x})}{\epsilon}\psi_i({\bf x})-\mathcal{K}(\psi^\dagger_{i+1}({\bf x}),\psi_i({\bf x}))\right)\epsilon\Delta {\bf x}\right]d\psi^\dagger_i({\bf x})d\psi_i({\bf x}).
\end{eqnarray*}
By making the transform, $\epsilon\rightarrow d\tau$, and changing the sum into an integral of $\tau$, then rewrite the field operators in 4 dimensions of the imaginary-time and space, $\psi_i({\bf x})\rightarrow \psi(\tau,{\bf x})$, we can obtain eq. (\ref{partitionZ2}) in Section \ref{fermionL}.
\section{Details for calculations of photon's self-energy}
\label{appB}
In derivation of Photon's self-energy with respect to the imaginary-time, the integration over the space is proceeded first. The sum is taken over Matsubara frequencies for the limit of $\beta\rightarrow 0$, and it becomes a continuous integral as for the other limit $\beta\rightarrow \infty$. Before the 3-dimension integration, the self-energy functions are
\begin{eqnarray}\label{omegaT1}
&&\hspace{-1.5cm}\Omega_T-\frac{M}{p^2_n}=-\frac{4e^2}{p^2_n}\int^1_0dx\frac{1}{\beta}\sum_{m}\int\frac{d^3{\bf l}}{(2\pi)^3}
\left\{ \frac{1}{2}\frac{1}{\left[(\omega_m+x\omega_n)^2+{\bf l}^2-x(1-x)p^2_n
+m^2_{\rm f}\right]}
\right.\nonumber
\\
&&\hspace{-1cm}+\frac{1}{{\left[(\omega_m+x\omega_n)^2+{\bf l}^2-x(1-x)p^2_n
+m^2_{\rm f}\right]^2}}
\left[\frac{1}{2}\left( x(1-x)p^2_n
-m^2_{\rm f}\right)
\left.
+(m^2_{\rm f}+x(1-x)p^2_n
) \right]\right\},\nonumber\\
\\
&&\hspace{-1.5cm}\Omega_L-\frac{M}{p^2_n}=-\frac{4e^2}{p^2_n}\int^1_0dx\frac{1}{\beta}\sum_{m} \int\frac{d^3{\bf l}}{(2\pi)^3}
\left\{ \frac{1}{2}\frac{1}{\left[(\omega_m+x\omega_n)^2+{\bf l}^2-x(1-x)p^2_n
+m^2_{\rm f}\right]}
\right.\nonumber
\\
&&\hspace{.4cm}+\frac{1}{{\left[(\omega_m+x\omega_n)^2+{\bf l}^2-x(1-x)p^2_n
+m^2_{\rm f}\right]^2}}
\left[\frac{1}{2}\left( x(1-x)p^2_n
-m^2_{\rm f}\right)\right.\nonumber\\
&&\hspace{6cm}\left.
\left.-2x(1-x)p^2_n+(m^2_{\rm f}+x(1-x)p^2_n
) \right]\right\},
\label{omegaL1}
\end{eqnarray}
where $\Delta(\omega_n,{\bf p})=m^2_{\rm f}-x(1-x){p}_n^2=m^2_{\rm f}+x(1-x)(\omega_n^2+{\bf p}^2)$. 
After integrating over the 3-momentum with a cutoff $\Lambda$, the transverse and longitudinal  parts are 
\begin{eqnarray}
\label{omegaT2}
&&\hspace{-1.5cm}
\Omega_T-\frac{M}{p^2_n}=
-\frac{4e^2}{p^2_n}\int^1_0dx\frac{1}{\beta}\sum_{m}
\left\{ \frac{1}{2}\left(\frac{\Lambda}{2\pi^2}-\frac{1}{4\pi}\sqrt{\left[(\omega_m+x\omega_n)^2+\Delta 
\right]} \tan^{-1}\left(\frac{\Lambda}{\sqrt{(\omega_m+x\omega_n)^2+\Delta}}\right)\right.
\right)\nonumber
\\
&&\hspace{1cm}\left.+\frac{1}{8\pi}\frac{1}{\sqrt{\left[(\omega_m+x\omega_n)^2+\Delta 
\right]}}
\left. \left[\frac{1}{2}\left( x(1-x)p^2_n
-m^2_{\rm f}\right)\right.
+(m^2_{\rm f}+x(1-x)p^2_n 
) \right]\right\},\\
&&\hspace{-1.5cm}\Omega_L-\frac{M}{p^2_n}=
-\frac{4e^2}{p^2_n}\int^1_0dx\frac{1}{\beta}\sum_{m}
\left\{ \frac{1}{2}\left(\frac{\Lambda}{2\pi^2}-\frac{1}{4\pi}\sqrt{\left[(\omega_m+x\omega_n)^2+\Delta 
\right]} \tan^{-1}\left(\frac{\Lambda}{\sqrt{(\omega_m+x\omega_n)^2+\Delta}}\right)\right.
\right)\nonumber
\\
&&\hspace{-.2cm}\left.+\frac{1}{8\pi}\frac{1}{\sqrt{\left[(\omega_m+x\omega_n)^2+\Delta 
\right]}}
\left. \left[\frac{1}{2}\left( x(1-x)p^2_n
-m^2_{\rm f}\right)\right.-2x(1-x)p^2_n
+(m^2_{\rm f}+x(1-x)p^2_n 
) \right]\right\}.\hspace{.5cm}
\label{omegaL2}
\end{eqnarray}
The above derivations lead to the results of eq. (\ref{betaInftyTps}) and eq. (\ref{beta0ps}) in Section \ref{photonimg}.

\section{Details for calculations of photon's chemical potential}\label{appC}
In the statistical mechanics, the thermodynamical potential is defined as 
\begin{eqnarray}
\Omega(T,V,\mu)=-\frac{1}{\beta}\ln Z,\label{thermoP}
\end{eqnarray}
where $T$ and $V$ is the temperature and the volume of the system. The relation between the thermodynamical potential and the Helmholtz free energy, $F(T,V,N)$ are related to each other through a Legendre transformation
\begin{eqnarray}
\Omega(T,V,\mu_{\rm ch})=F-\mu_{\rm ch} N,\label{legendreT}
\end{eqnarray}
where $N$ is the number of the particles and $\mu_{\rm ch}$ is the chemical potential. The partition function can be written as  
\begin{eqnarray*}
Z={\rm Tr}  \,e^{\beta\hat{ H}-\mu_{\rm ch} \hat{N}},
\end{eqnarray*}
where $\hat{H}$ and $\hat{N}$ are the Hamiltonian and the number operator of the system.  With a small perturbation in the Hamiltonian, it can be expressed as $\hat{H}=\hat{H}_0+\hat{V}$. The partition function can be perturbed in the following way
\begin{eqnarray*}
Z&=&{\rm Tr} \, e^{\beta\hat{ H}_0+\beta\hat{ V}-\mu_{\rm ch} \hat{N}}
={\rm Tr} \, \left(\sum_{n=0}\frac{1}{n!}\beta^n\hat{V}^n\right)e^{\beta\hat{ H}_0-\mu_{\rm ch} \hat{N}},\\
&=&{\rm Tr} \, e^{\beta\hat{ H}_0-\mu_{\rm ch}  \hat{N}}+\sum_{n=1}\frac{1}{n!} {\rm Tr} \, \left(\beta^n\hat{V}^n\right)e^{\beta\hat{ H}_0-\mu_{\rm ch}  \hat{N}}
=Z_0\left(1+ \sum_{n=1}\frac{1}{n!}\langle \beta^n\hat{V}^n\rangle\right).
\end{eqnarray*}
The above expression can be changed to thermodynamical potential according to (\ref{thermoP}).
\begin{eqnarray*}
\beta\left(\Omega-\Omega_0\right)&=&-\sum_{n=1}\frac{1}{n!}\langle \beta^n\hat{V}^n\rangle
\end{eqnarray*}
Similar to eq. (\ref{legendreT}), the difference of the Helmholtz free energy can be written as below
\begin{eqnarray}
\delta F&=&-\frac{1}{\beta}\sum_{n=1}\frac{1}{n!}\langle \beta^n\hat{V}^n\rangle-\mu_{\rm ch}  \delta N.\label{freeenergychange}
\end{eqnarray}
The symbol $\delta X$ denotes the total variation of the function, $X$.
As the equilibrium is reached and the free energy stays stable, so that $\delta F=0$. Therefore the change of the R.H.S. of eq. (\ref{freeenergychange}) with respect to the variation of particle's number is 
 \begin{eqnarray*}
\frac{\delta F}{\delta N}&=&-\frac{1}{\beta}\sum_{n=1}\frac{1}{n!}\langle \frac{\delta (\beta^n \hat{V}^n)}{\delta N}\rangle-\mu_{\rm ch} =0.
\end{eqnarray*}
In turn, the chemical potential can be obtained through the above equation
 \begin{eqnarray}
\mu_{\rm ch} &=&-\frac{1}{\beta}\sum_{n=1}\frac{1}{n!}\langle \frac{\delta (\beta^n \hat{V}^n)}{\delta N}\rangle.
\label{chemicalp}
\end{eqnarray}

\section{Summation of Matsubara frequency}\label{appD}
The summation formulas used in derivation of the fermion's propagator to sum over the Matsubara frequency, $\omega_n=\frac{\pi n}{\beta}$, are shown.
For  fermions, $n$ is an odd integer, and an even number for bosons. For fermionic frequencies,
\begin{eqnarray}
\hspace{-.5cm}\frac{1}{\beta}\sum_{n={\rm odd}}\frac{e^{-i\omega_n\tau}}{i\omega_n-\xi_{\bf p}}=-\frac{e^{\xi_{\bf p}(\beta-\tau)}}{e^{\beta \xi_{\bf p}}+1}\hspace{.3cm} {\rm for}\,\,\tau>0,\,\, {\rm and} \hspace{.3cm}
\frac{1}{\beta}\sum_{n={\rm odd}}\frac{e^{i\omega_n\tau}}{i\omega_n-\xi_{\bf p}}=-\frac{e^{\xi_{\bf p}(\beta+\tau)}}{e^{\beta \xi_{\bf p}}+1}\hspace{.3cm} {\rm for}\,\,\tau<0.\hspace{.8cm}\label{sumfermi}
\end{eqnarray}
For bosonic frequencies,
\begin{eqnarray}
\hspace{-.5cm}\frac{1}{\beta}\sum_{n={\rm even}}\frac{e^{-i\omega_n\tau}}{i\omega_n-\xi_{\bf p}}=-\frac{e^{\xi_{\bf p}(\beta-\tau)}}{e^{\beta \xi_{\bf p}}-1}\hspace{.3cm} {\rm for}\,\,\tau>0,\,\, {\rm and} \hspace{.3cm}
\frac{1}{\beta}\sum_{n={\rm even}}\frac{e^{i\omega_n\tau}}{i\omega_n-\xi_{\bf p}}=-\frac{e^{\xi_{\bf p}(\beta+\tau)}}{e^{\beta \xi_{\bf p}}-1}\hspace{.3cm} {\rm for}\,\,\tau<0,\hspace{.7cm}\label{sumbose}
\end{eqnarray}
where $\xi_{\bf p}=\sqrt{|{\bf p}|^2+m^2_{\rm f,b }}$ with a fermion or boson mass $m_{\rm f,b}$.

\section{Feynman rules for real-time}\label{E}
\label{appE}

Feynman rules corresponding to the fermionic and bosonic Lagrangian discussed in Section \ref{lagimag} are presented. The momentum, $p$, denoted the 4-momentum: $(p_0, {\bf p})$, where the boldface indicates the 3-momentum for spacial components. The density functions are  $n_F(p_0)=\frac{1}{e^{\beta p_0}+1}$ and $n_B(p_0)=\frac{1}{e^{\beta p_0}-1}$ for the fermionic and bosonic distributions respectively.
\vspace{-.8cm}
\begin{eqnarray}
\bullet\hspace{.5cm}
{\rm Dirac\,\, Propagator:}
&\raisebox{-13.5mm}{\psfig{figure=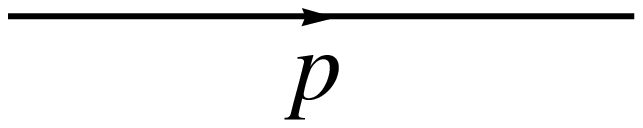, width=3cm}}
&=\frac{i}{\slashed{p}-m_{\rm f}+i\varepsilon}\left(1-n_F(p_0)\right);\nonumber
\end{eqnarray}
\vspace*{-2cm}
\begin{eqnarray}
\hspace{-.5cm}
\bullet\hspace{.5cm}
{\rm Photon\,\, Propagator:}
&\raisebox{-13.5mm}{\psfig{figure=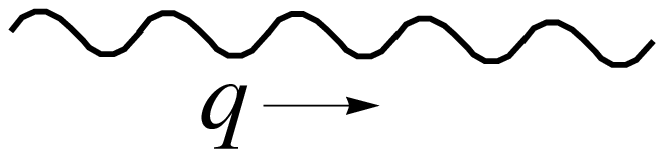, width=3cm}}
&=\frac{-ig_{\mu\nu}}{{q^2}+i\varepsilon}\left(1+n_B(q_0)\right);\nonumber
\end{eqnarray}
\vspace*{-1.5cm}
\begin{eqnarray}
\hspace{-2.9cm}
\bullet\hspace{.5cm}
{\rm QED\,\, vertex:}\hspace{.7cm}
&\raisebox{-13.5mm}{\psfig{figure=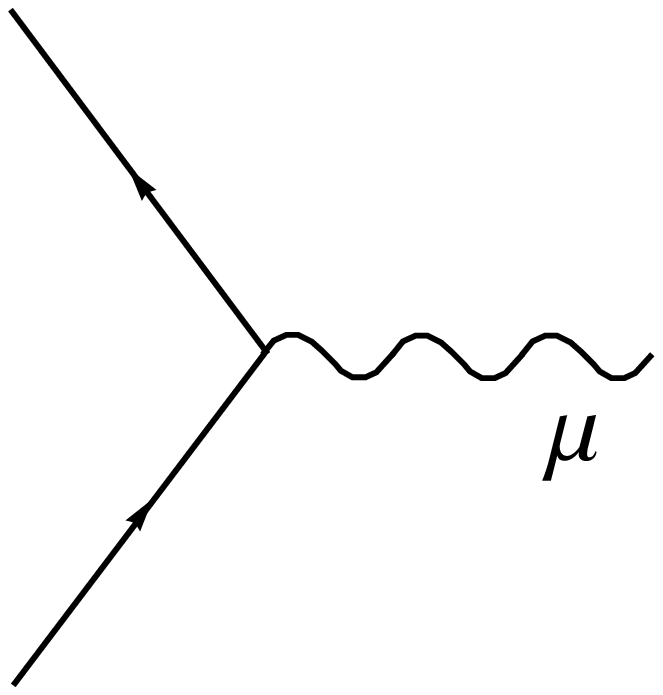, width=3cm}}
&\hspace{.5cm}
=-ie\gamma^\mu.\nonumber
\end{eqnarray}
where $e$ is the charge of the fermion and is $-|e|$ for an electron. The rest Feynman rules are the same as those in QED. 

\section{Feynman rules for imaginary-time}
\label{appF}
The Feynman rules for the imaginary-time is similar to those for the real-time except the zeroth component of the momentum is replaced by the Matsubara frequency, $\omega_n$. 
\vspace{-.5cm}
\begin{eqnarray}
\hspace{-2.4cm}\bullet\hspace{.5cm}{\rm Dirac\,\, Propagator:}
&\raisebox{-13.5mm}{\psfig{figure=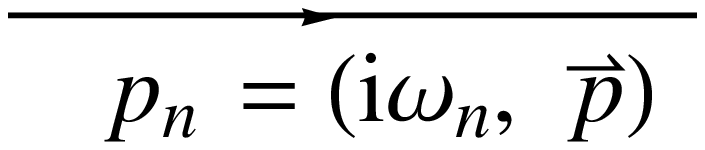, width=3cm}}
&\hspace{.5cm}=\frac{1}{\slashed{p}_n-m};\nonumber
\end{eqnarray}
\vspace*{-2cm}
\begin{eqnarray}
\hspace{-3cm}
\bullet\hspace{.5cm}
{\rm Photon\,\, Propagator:}
&\raisebox{-12mm}{\psfig{figure=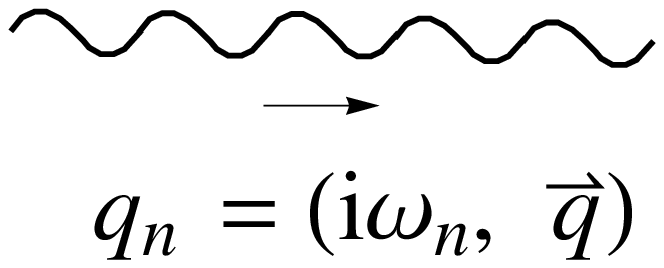, width=3cm}}
&=-\frac{g_{\mu\nu}}{{q^2_n}};\nonumber
\end{eqnarray}
\vspace*{-1.5cm}
\begin{eqnarray}
\hspace*{-.5cm}
&&\hspace{-0cm}
\bullet\hspace{.5cm}
{\rm QED\,\, vertex:}\hspace{1cm}
\raisebox{-13.5mm}{\psfig{figure=QEDvertex.eps, width=3cm}}
\hspace{.5cm}
=e\gamma^\mu;\nonumber\\
&&\bullet\hspace{.5cm} {\rm Impose\,\, conservations\,\, of\, \,  Matsubara\,\, frequency\,\, and\,\,momentum}\nonumber\\
&&\hspace{.6cm}{\rm \, \,at\, \,each\, \,vertex;}\nonumber\\
&&\bullet\hspace{.5cm} {\rm integrate\,\, over\,\, each\, \, loop\,\, frequency\,\,and\,\,momentum:}\,\,\frac{1}{\beta}\sum_n\int \frac{d^3{\bf p}}{(2\pi)^3}.\nonumber
\end{eqnarray}



%

\begin{thebibliography}{}
%
%

\bibitem{matsubara}
{}{{T.}~{Matsubara}},
{}{{ Prog. Theor. Phys.}}
{{}{14}}, {}\textbf{351} ({}{1955});\par
{}{{G.D.}~{Mahan}},
{}{{{\it Many-Particle Physics}. (Kluwer Academic/Plenum Publishers}},
{}{2000});
\par
{}{{A.L.}~{Fetter}},
{and}
{}{{J.D.}~{Walecka}},
{}{{{\it Quantum Theory of Many-Particle Systems}. (Dover Publications,}}
{}{2003}).


\bibitem{shankar93}
{}{{M.}~{Shankar}},
{}{{Rev. Mod. Phys.}}
{{}\textbf{66}}, {}{129} ({}{1994});\par

{}{{J.I.}~{Kapusta}},
{and}
{}{{C.}~{Gale}},
{}{{{\it Finite-Temperature Field Theory}. (Cambridge University Press,}}
{}{2006}).

\bibitem{ryden}
{}{{A.A.}~{Penzias}},
 {and}
{}{{R.W.}~{Wilson}},
{}{{Astrophys. J.}}
{{}\textbf{142}}, {}{419} ({}{1965});\par
{}{{B.}~{Ryden}},
{}{{{\it Introduction to Cosmology}. (Addison Wesley,}}
{}{2003}).


\bibitem{grozin}
{}{{A.}~{Grozin}},
{}{{{\it Lectures on QED and QCD}. (World Scientific,}}
{}{2007}).


\bibitem{tolman30}
  {}{{R.C.}~{Tolman}},
{}{{Phys. Rev.}}
{{}\textbf{35}}, {}{904}, {}{} ({}{1930});
\par
{}{{R.C.}~{Tolman}}, {and}
  {}{{P.}~{Ehrenfest}},
{}{{Class. Quant. Grav.}}
{{}\textbf{36}}, {}{1791}, {}{} ({}{1930}).\par

\bibitem{rovelli93}
  {}{{C.}~{Rovelli}},
{}{{Class. Quant. Grav.}}
{{}\textbf{10}}, {}{1549}, {}{} ({}{1993});
\par
{}{{A.}~{Connes}}, {and}
  {}{{C.}~{Rovelli}},
{}{{Class. Quant. Grav.}}
{{}\textbf{11}}, {}{2899}, {}{} ({}{1994});\par
  {}{{C.}~{Rovelli}}, {and}
  {}{{M.}~{Smerlak}},
{}{{Class. Quant. Grav.}}
{{}\textbf{28}}, {}{075007}, {}{} ({}{2011}).
\par
\bibitem{landsman87}
{}{{H.}~{Ezawa}},
{}{{Y.}~{Tomozawa}},
 {and}
{}{{H.}~{Umezawa}},
{}{{Nuovo Cim.}}
{{}\textbf{5}}, {}{810} ({}{1957});\par
{}{{D.A.}~{Kirzhnits}},
 {and}
{}{{A.}~{Linde}},
{}{{Phys. Lett. B}}
{{}\textbf{42}}, {}{471} ({}{1972});\par
{}{{R.L.}~{Bowers}},
 {and}
{}{{R.L.}~{Zimmermann}},
{}{{Phys. Rev. D}}
{{}{7}}, {}{296} ({}{1973});\par
{}{{C.}~{Bernard}},
{}{{Phys. Rev. D}}
{{}\textbf{9}}, {}{3312} ({}{1974});\par
{}{{L.}~{Dolan}},
 {and}
{}{{R.}~{Jackiw}},
{}{{Phys. Rev. D}}
{{}\textbf{9}}, {}{3320} ({}{1974});\par
{}{{S.}~{Weinberg}},
{}{{Phys. Rev. D}}
{{}\textbf{9}}, {}{3357} ({}{1974});\par
{}{{R.E.}~{Norton}},
 {and}
{}{{J.M.}~{Cornwall}},
{}{{Ann. Phys.}}
{{}\textbf{91}}, {}{106} ({}{1975});\par
{}{{N.P.}~{Landsman}},
 {and}
{}{{Ch.G.}~{van Weert}},
{}{{Phys. Rep.}}
{{}\textbf{145}}, {}{141} ({}{1987}).\par


\bibitem{bellac96}
{}{{M.L.}~{Bellac}},
{}{{{\it Thermal Field Theory}. (Cambridge University Press,}}
{}{1996}).


\bibitem{casimir}
{}{{H.B.G.}~{Casimir}},
{}{{ Proceedings of the Royal Netherlands Academy of Arts and Sciences}}
{{}\textbf{51}}, {}{793} ({}{1948});\par
{}{{P.W.}~{Milonni}},
{}{{{\it The Quantum Vacuum}. (Academic Press,}}
{}{1994}).

\bibitem{vanderwaals}
{}{{J.D.}~{van der Waals}},
{}{{ Nobel Lectures in Physics}}
{}{254} ({}{1910});\par
{}{{J.D.}~{van der Waals}},
{}{{ Verhand. Kon. Akad. V Wetensch. Amst. Sect. 1 (1893)}},
{}{{(English translation in J. Stat. Phys}}
{{}\textbf{20}}, {}{197} ({}{1979})).


\bibitem{unruh}
{}{{S.A.}~{Fulling}},
{}{{ Phys. Rev. D}}
{{}\textbf{7}}, {}{2850} ({}{1973});\par
{}{{P.C.W.}~{Davies}},
{}{{ J. Phys. A}}
{{}\textbf{8}}, {}{609} ({}{1975});\par
{}{{W.G.}~{Unruh}},
{}{{ Phys. Rev. D}}
{{}\textbf{14}}, {}{870} ({}{1976}).


\bibitem{hawking}
{}{{J.M.}~{Bardeen}},
{}{{B.}~{Carter}},
{and}
{}{{S.W.}~{Hawking}},
{}{{ Comm. Math, Phys.}}
{{}\textbf{31}}, {}{161} ({}{1973});\par
{}{{S.W.}~{Hawking}},
{}{{ Comm. Math, Phys.}}
{{}\textbf{43}}, {}{199} ({}{1975}).


\bibitem{huang13b}
{}{{Y.-C.}~{Huang}},
{}{{arXiv: 1311.5188 [gr-qc]}}. \par

\bibitem{dewitt75}
{}{{B.S.}~{DeWitt}},
{}{{{\it The dynamical theory of groups and fields in Relativity, groups and Topology}. {eds.}
{}{{B.S.}~{DeWitt}},
{and}
{}{{C.}~{DeWitt}} (New York:  Gordon \& Breach,}}
{}{1965});\par
{}{{B.S.}~{DeWitt}},
{}{{Phys. Rep.}}
{{}\textbf{19}}, {}{297} ({}{1975}).\par


\bibitem{huang13c}
{}{{Y.-C.}~{Huang}},
{}{{arXiv: 1311.6284 [gr-qc]}}. \par


\bibitem{huang13d}
{}{{Y.-C.}~{Huang}},
{}{{arXiv: 1312.4380 [gr-qc]}}. \par


\bibitem{difrancesco97}
{}{{P.}~{DiFrancesco}},
{}{{P.}~{Mathieu}},
{and}
{}{{D.}~{Senechal}},
{}{{{\it Conformal Field Theory}. (Springer-Verlag,}}
{}{1997});

{}{{J.}~{Zinn-Justin}},
{}{{{\it Quantum Field Theory and Critical Phenomena}. (Oxford University Press,}}
{}{2002}).


\bibitem{goldenfeld92}
{}{{N.}~{Goldenfeld}},
{}{{{\it Lectures On Phase Transitions And The Renormalization Group}. (Westview Press,}}
{}{1992}).

\bibitem{kadanoff66}
{}{{L.P.}~{Kadanoff}},
{}{{Physica}}, {{}\textbf{2}}, {}{263} 
({}{1966}).\par

\bibitem{wilson75}
{}{{K.G.}~{Wilson}},
{}{{Rev. Mod. Phys.}}
{{}\textbf{47}}, {}{4}, {}{773} ({}{1975}).\par




\bibitem{takesaki70}
{}{{M.}~{Takesaki}},
{}{{{\it Tomita’s theory of modular Hilbert algebras and its applications}. (Springer-Verlag Berlin,}}
{}{1970}).

\bibitem{kms}
  {}{{R.}~{Kubo}},
{}{{J. Phys. Soc. Jpn.}}
{{}\textbf{12}}, {}{570} {}{} ({}{1957});
\par
{}{{P.C.}~{Martin}}, {and}
  {}{{J.}~{Schwinger}},
{}{{Phys. Rev.}}
{{}\textbf{115}}, {}{1342} {}{} ({}{1959}).\par


\bibitem{peskin95}
  {}{{E.M.}~{Peskin}}, {and}
  {}{{D.V.}~{Schroeder}},
  {}{{\it An Introduction to Quantum Field Theory}}.
  ({}{{Addison-Wesley,}}  {}{1995});
\par
  {}{{C.}~{Itzykson}}, {and}
  {}{{J.B.}~{Zuber}},
  {}{{\it Quantum Field Theory}}.
 ({}{{McGraw-Hill, New York,}}  {}{1980});
\par
  {}{{J.D.}~{Bjoken}}, {and}
  {}{{S.D.}~{Drell}},
  {}{{\it Relativistic Quantum Mechanics}}.
  ({}{{McGraw-Hill, New York,}}  {}{1964}).\par




\bibitem{tu05}
{}{{C.}~{Amsler}}
{et al.},
{}{{Phys. Lett. B}},
{}\textbf{667}, 1 ({}{2008});\par
{}{{L.C.}~{Tu}},
{}{{J.}~{Luo}},
{and}
{}{{G.T.}~{Gillies}},
{}{{Rep. Prog. Phys.}}
{{}\textbf{68}}, {}{77} ({}{2005}).\par



\bibitem{huang14b}
{}{{Y.-C.}~{Huang}},
{}{{in preparation}}. \par



\bibitem{gellmann}
{}{{M.}~{Gell-mann}},
 {and}
{}{{F.E.}~{Low}},
{}{{Phys. Rev.}}
{{}\textbf{95}}, {}{1300} ({}{1954}).\par






\bibitem{landau55}
{}{{L.}~{Landau}},
{}{{in W. Pauli, ed., { Niels Bohr and the Development of Physics}. (Pergamon Press, London,}}
{}{1955});\par
{}{{L.}~{Landau}},
{}{{A.A.}~{Abrikosov}},
 {and}
{}{{I.M.}~{Khalatnikov}},
{}{{Dokl. Akad. Nauk SSSR}}
{{}\textbf{95}}, {}{497} ({}{1954});\par
{}{{L.}~{Landau}},
{}{{A.A.}~{Abrikosov}},
 {and}
{}{{I.M.}~{Khalatnikov}},
{}{{Dokl. Akad. Nauk SSSR}}
{{}\textbf{95}}, {}{773} ({}{1954});\par
{}{{L.}~{Landau}},
{}{{A.A.}~{Abrikosov}},
 {and}
{}{{I.M.}~{Khalatnikov}},
{}{{Dokl. Akad. Nauk SSSR}}
{{}\textbf{95}}, {}{1177} ({}{1954});\par
{}{{J.I.}~{Bogoliubov}},
{and}
{}{{C.}~{Shirkov}},
{}{{{\it Introduction to the Theory of Quantized Fields, 3rd ed. }. (Wiley, New York,}}
{}{1980}).




\end{thebibliography}
%

\end{document}